\documentclass[a4paper,11pt]{article}
\pdfoutput=1
\usepackage{jheppub}
\usepackage{amsmath,amssymb,amsfonts,bm,amscd,slashed,ytableau}
\usepackage{graphicx}
\usepackage{mathtools}
\usepackage{mathrsfs}
\usepackage{bbm}
\usepackage[bbgreekl]{mathbbol}

\DeclareSymbolFontAlphabet{\mathbbm}{bbold}
\DeclareSymbolFontAlphabet{\mathbb}{AMSb}

\newcommand{\Z}{{\mathbb Z}}
\newcommand{\R}{{\mathbb R}}
\newcommand{\C}{{\mathbb C}}

\newcommand{\cZ}{{\mathcal{Z}}}

\newcommand{\cN}{{\mathcal{N}}}

\newcommand{\dd}{{\rm d}}

\def\Tr{{\rm Tr \,}}

\def\tilde{\widetilde}
\def\hat{\widehat}
\def\bar{\overline}

\def\rQ{{\mathbbmtt{Q}\,}}

\def\gA{{\mathscr A}}

\def\cA{{\mathcal A}}
\def\cB{{\mathcal B}}
\def\cC{{\mathcal C}}
\def\cD{{\mathcal D}}

\def\cG{{\mathcal G}}

\def\cI{{\mathcal I}}

\def\cL{{\mathcal L}}

\def\cN{{\mathcal N}}
\def\cO{{\mathcal O}}

\def\cQ{{\mathcal Q}}

\def\cS{{\mathcal S}}
\def\cT{{\mathcal T}}
\def\cU{{\mathcal U}}

\def\cW{{\mathcal W}}

\def\cZ{{\mathcal Z}}

\DeclareMathOperator{\trace}{Tr}

\renewcommand{\bar}{\overline}
\renewcommand{\hat}{\widehat}

\title{Algebras, traces, and boundary correlators\\ in $\mathcal{N}=4$ SYM}

\author[1]{Mykola Dedushenko}
\author[2]{and Davide Gaiotto}

\affiliation[1]{Simons Center for Geometry and Physics,\\ Stony Brook University, Stony Brook, NY 11794-3636, USA}
\affiliation[2]{Perimeter Institute for Theoretical Physics, Waterloo, ON N2L 2Y5, Canada}

\abstract{We study supersymmetric sectors at half-BPS boundaries and interfaces in the 4d $\mathcal{N}=4$ super Yang-Mills with the gauge group $G$, which are described by associative algebras equipped with twisted traces. Such data are in one-to-one correspondence with an infinite set of defect correlation functions. We identify algebras and traces for known boundary conditions. Ward identities expressing the (twisted) periodicity of the trace highly constrain its structure, in many cases allowing for the complete solution. Our main examples in this paper are: the universal enveloping algebra $U(\mathfrak{g})$ with the trace describing the Dirichlet boundary conditions; and the finite W-algebra $\mathcal{W}(\mathfrak{g},t_+)$ with the trace describing the Nahm pole boundary conditions.}

\begin{document}
	\maketitle
	\flushbottom



\section{Introduction}
Boundary observables play especially important role in Quantum Field Theory (QFT) due to their direct practical relevance. Indeed, scattering processes in high energy physics take place on spacetime manifolds with asymptotic boundaries, while in condensed matter applications, any experiment involves ``probing'' a sample through some sort of a boundary, so boundary phenomena are ubiquitous and directly observable. Furthermore, the mathematical structure of boundary observables is different from bulk observables that have, until recently, attracted more attention in the literature, which makes them interesting subjects to explore in mathematical physics as well.

In this paper, we study aspects of boundary operators in the 4d $\cN=4$ super Yang-Mills (SYM) with gauge group $G$, subject to half-BPS boundary conditions. A rich class of boundary conditions preserving 3d $\cN=4$ SUSY, and often the full superconformal symmetry, are known in the literature \cite{Gaiotto:2008sa,Gaiotto:2008sd,Gaiotto:2008ak}. They are amenable to study via certain techniques originally developed for purely three-dimensional theories with the same amount of SUSY. More specifically, we will be looking at the supersymmetric sector in the cohomology of a chosen supercharge, which is described by the 1d theory often referred to as a topological quantum mechanics (TQM). The TQM is fully determined by the data of an associative algebra of observables and a twisted trace on this algebra that determines the $S^1$ partition function and correlators. These encode the $S^3$ partition functions and part of the OPE data of the 3d theory.

Each 3d $\cN=4$ theory has two TQMs associated to it: the Higgs and the Coulomb sector TQMs, determined by the algebras $\cA_H$, $\cA_C$ and twisted traces $T_H$, $T_C$ respectively. Such sectors were introduced and studied in \cite{Chester:2014mea,Beem:2016cbd,Dedushenko:2016jxl,Dedushenko:2017avn,Dedushenko:2018icp}, and in \cite{Gaiotto:2019mmf} a precise relation of the special traces $T_H$ and $T_C$ to traces over the Verma modules of $\cA_H$ and $\cA_C$ was conjectured. The algebras $\cA_H$, $\cA_C$ describe equivariant, short and even quantizations of the Higgs and Coulomb branches (the evenness can be broken by turning on the FI parameters and masses). Mathematical classification of such quantizations was recently studied in \cite{Etingof:2019guc}. The existence of nice traces $T_H$ and $T_C$ is what endows these quantizations with special properties listed above, and traces naturally follow from the $S^3$ partition function decorated by operator insertions, as we review momentarily.

The existence of $\cA_H$ and $\cA_C$ sectors follows from kinematics: the 3d $\cN=4$ SUSY implies that cohomology spaces of specially chosen supercharges produce algebras $\cA_H$ and $\cA_C$. This means that the half-BPS boundary of the 4d $\cN=4$ SYM should also carry similar algebras $\cA_H$ and $\cA_C$ of boundary local operators, whose constructions proceed along the same lines. The structure constants of these algebras, as well as traces on them, are part of the dynamical data. In purely 3d case, they require studying the $S^3$ partition function and correlators, while the 4d setting, as we will see, is related to the partition function and correlators on the hemisphere $HS^4$.

In the purely 3d case, the construction based on superconformal symmetry was discovered first in \cite{Chester:2014mea,Beem:2016cbd}, motivated by the similar construction of the chiral algebra in \cite{Beem:2013sza}, which we refer to as the ``$Q+S$'' type construction. In this approach, the operators from $\cA_H$ or $\cA_C$ live on a chosen line in spacetime. Later it was realized in \cite{Dedushenko:2016jxl} that the $S^3$ SUSY background provides a natural generalization away from conformal theories. In that case, the algebras $\cA_H$ and $\cA_C$ live on a great circle $S^1$ inside of $S^3$. We will see that the $\cA_H$ and $\cA_C$ structures at the boundary of 4d theory also admit two definitions: in terms of ``$Q+S$'' construction, in which case the operators live on a distinguished line, and in terms of hemisphere background $HS^4$, in which case the operators live on a distinguished great circle at the boundary, $S^1\subset S^3 = \partial(HS^4)$. In addition the Omega-background \cite{Nekrasov:2002qd,Nekrasov:2003rj,Nekrasov:2010ka} can also be used to give a variant of the definition, as we review below.

As we will see, both 1d sectors, $\cA_H$ and $\cA_C$, can be viewed as boundaries of certain 2d protected sectors of the 4d $\cN=4$ SYM. These sectors are special to $\cN=4$ SUSY and do not occur in $\cN=2$ theories. They again admit both ``$Q+S$'' definitions in flat space (the protected sector lives on a plane), and definitions in terms of the $S^4$ background (the protected sector lives on a great two-sphere $S^2$).\footnote{Note that by a conformal transformation, one can equivalently formulate the flat space construction in such a way that the protected sector would be supported on a two-sphere embedded in flat space. We do not do so, and prefer to have either two-plane in $\R^4$, or two-sphere in $S^4$.} We refer to the definition of $\cA_H$ as the ``electric construction'' and of $\cA_C$ -- as the ``magnetic construction'' for obvious reasons: they are exchanged by the electric-magnetic duality, and the electric construction is more manifest in the original Lagrangian description, while magnetic construction is more manifest from the S-dual point of view. 

\begin{figure}[h]
	\label{fig:HS4}
	\centering
	\includegraphics[scale=0.5]{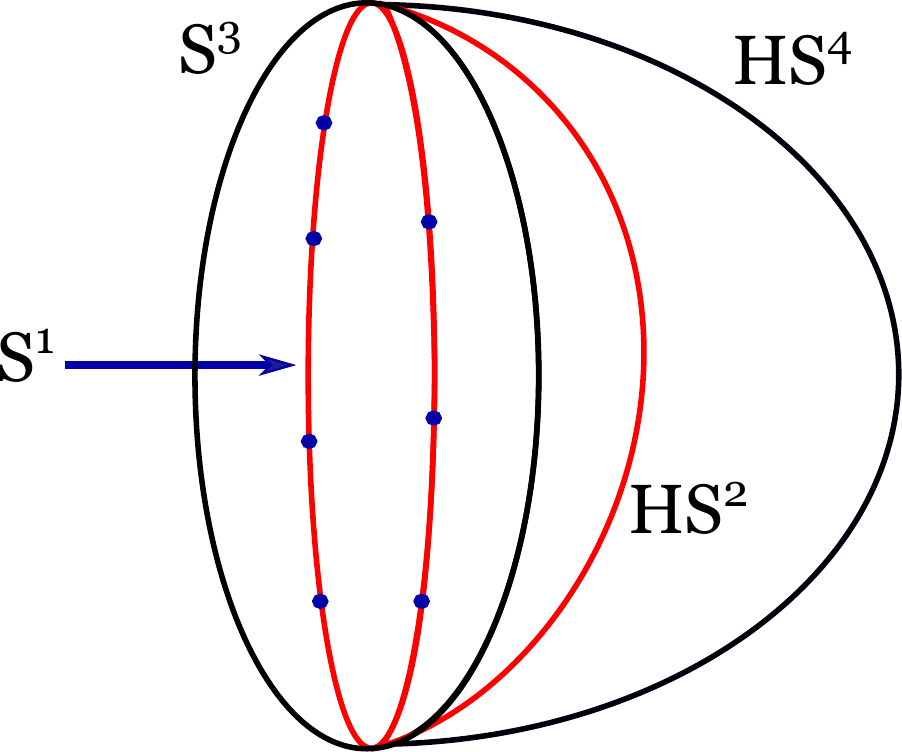}
	\caption{In the sphere background formulation, the 4d theory lives on $HS^4$, the 2d protected sector lives on a great $HS^2\subset HS^4$. The boundary of $HS^2$ is a great circle $S^1\subset S^3$ with some operator insertions.}
\end{figure}
The electric construction in the $S^4$ formulation was first discovered long time ago: it is the 2d constrained Yang-Mills sector discussed in \cite{Giombi:2009ds,Pestun:2009nn,Giombi:2009ek}. The word ``constrained'' refers to the fact that all instantonic contributions must be dropped, with the exact answer being defined by the perturbation series. The electric construction with boundary was also recently considered in \cite{Wang:2020seq,Komatsu:2020sup,Wang:2020jgh} from the spherical background point of view, and indeed these papers address questions closely related to our subject of study. On $HS^4$ we, therefore, have a distinguished locus given by a great 2d hemisphere $HS^2$, with the constrained 2d Yang-Mills (electric or magnetic) in the bulk, and either $\cA_H$ or $\cA_C$ TQM at the boundary, see Figure \ref{fig:HS4} for an illustration. Calling the 2d Yang Mills electric or magnetic means that the emergent 2d gauge field in the cohomology has a simple relation to either the electric or the magnetic variables of the 4d SYM theory.

In this paper, we determine the boundary correlators data, --- algebras $\cA_H$, $\cA_C$ and their twisted traces $T_H$, $T_C$, --- for a large class of boundary conditions. In the last section, we also provide some preliminary remarks on interfaces, which are the subject of a separate publication. We emphasize the algebraic approach to the problem: once, say, the algebra $\cA_H$ is known, the twisted trace property of $T_H$, which reads
\begin{equation}
T_H (xy) = T_H\left(\left( (-1)^{F_H} e^{-2\pi \ell m}\cdot y \right)x\right),\quad x,y\in\cA_H,
\end{equation}
can be thought of as a set of Ward identities, which often considerably simplify the problem of computing $T_H$. Here $(-1)^{F_H}$ is a $\Z_2$ grading on $\cA_H$ that originates in the $SU(2)_H$ R-charge of the 3d $\cN=4$ SUSY, and $m$ stands for various boundary masses that appear as twist parameters in the trace $T_H$. In the case of $\cA_C$ and $T_C$, the analogous property is very similar, except that $F_H$ is replaced by $F_C$ related to the $SU(2)_C$ R-charge, and the boundary masses $m$ are replaced by the boundary FI parameters. We now provide a more detailed summary.
\subsection{Technical summary}
One of the best understood examples here is the case of Dirichlet boundary conditions in a theory with gauge algebra $\mathfrak{g}$. While the $\cA_C$ algebra is just $\C$, the Higgs algebra is given by the universal enveloping algebra of the complexification of $\mathfrak{g}$,
\begin{equation}
\label{Ug_preview}
\cA_H = U(\mathfrak{g}_\C).
\end{equation}
This example already makes manifest one crucial distinction between the $\cA_H$, $\cA_C$ algebras in purely 3d theories and those in bulk-boundary systems. The 3d algebras describe quantizations of the 3d $\cN=4$ branches of supersymmetric vacua, which are hyper-K\"ahler cones or their resolutions/deformations. The algebra $U(\mathfrak{g}_\C)$, on the other hand, is a quantization of a complex Poisson manifold $\mathfrak{g}^*_\C$ (with its canonical Lie-Poisson structure), which is not symplectic (Poisson structure is not invertible). This is a general feature: the analog of moduli spaces of vacua in the bulk-boundary system with 8 conserved supercharges is not hyper-K\"ahler, but rather complex Poisson, and our algebras $\cA_H$, $\cA_C$ quantize the ``Higgs'' and ``Coulomb'' branches of such moduli spaces. In fact, we basically derive the result \eqref{Ug_preview} by quantizing $\mathfrak{g}^*_\C$: the constrained 2d Yang-Mills on $HS^2$ can be reformulated as a perturbative calculation in the BF theory on a disk with boundary insertions of $B$, which precisely gives such a quantization according to \cite{Cattaneo:1999fm}. A similar occurrence of the universal enveloping algebra at the boundary of BF theory can be found in \cite{Ishtiaque:2018str}, who basically applied the techniques of \cite{Cattaneo:1999fm}.

We also determine the trace on $U(\mathfrak{g}_\C)$. When the twist parameters (i.e., boundary masses) are turned off, the Ward identities (i.e., trace relations mentioned above) imply that $T_H$ is fully determined by its value on the center $\cZ[U(\mathfrak{g}_\C)]$. The latter can then be found if we start turning on boundary masses and differentiating with respect to them. Indeed, the untwisted trace $T_H^{m=0}$ and the twisted trace $T_H$ are related roughly by the insertion of a moment map for $\mathfrak{g}_\C$,
\begin{equation}
T_H(\dots) = T_H^{m=0}(\dots e^{-2\pi m\cdot B}),
\end{equation}
where $B\in \mathfrak{g}_\C$. The details are given in Section \ref{sec:tr_on_U(g)}. We also provide an explicit expression \eqref{verma_traces} for the trace $T_H$ as a continuous linear combination of traces on Verma modules of $U(\mathfrak{g}_\C)$, thereby generalizing the conjecture of \cite{Gaiotto:2019mmf} to the case of bulk-boundary systems. One important distinction of \eqref{verma_traces} is that the linear combination is continuous, while in \cite{Gaiotto:2019mmf} it is discrete, with the Verma modules being in correspondence with the massive vacua of the 3d theory.

We also emphasize the role of an algebra of bulk operators $\cB_H$, which can be described as local gauge-invariant operators in the 2d Yang-Mills. This algebra is commutative, and is isomorphic to the center of $U(\mathfrak{g}_\C)$: the bulk operators are simply given by gauge-invariant polynomials in curvature of the constrained 2d Yang-Mills. Such operators can be identified with $\mathfrak{g}$-invariant polynomials on $\mathfrak{g}$, or Weyl-invariant polynomials on $\mathfrak{t}$,
\begin{equation}
\cB_H = \C[\mathfrak{g}]^\mathfrak{g} = \C[\mathfrak{t}]^\cW.
\end{equation}

There is an important map, called the bulk-boundary map, which is obtained by colliding operators from $\cB_H$ with the boundary. The image of the bulk-boundary map is always in the center of $\cA_H$: indeed, we can always move such operators into the bulk and commute them past anything on the boundary. This map, denoted by
\begin{equation}
\rho_H: \cB_H \to \cZ[\cA_H]\subset\cA_H,\quad \text{where $\cZ$ means the center,}
\end{equation}
in the case when $\cA_H=U(\mathfrak{g}_\C)$, is a non-trivial isomorphism of commutative algebras, known as the Harish-Chandra isomorphism \cite{HarCha}. It identifies $\C[\mathfrak{t}]^\cW$ with the center of $U(\mathfrak{g}_\C)$, and this identification is not trivial: the Harish-Chandra map encodes the physics of the bulk-boundary map for the Dirichlet boundary conditions.

Next we study the Neumann boundary conditions enriched by a boundary theory $\cT$. This case is relatively straightforward: given the algebras $\cA_H(\cT)$ and $\cA_C(\cT)$ of the 3d theory, we show that the boundary algebra $\cA_H$ is given by the $\mathfrak{g}$-invariants:
\begin{equation}
\cA_H = \left(\cA_H(\cT) \right)^{\mathfrak g},
\end{equation}
while the $\cA_C$ algebra is obtained as a central extension, where the mass parameters of $\cA_C(\cT)$ at the boundary are promoted to dynamical fields:
\begin{equation}
0 \longrightarrow \C[\mathfrak{t}]^\cW \longrightarrow \cA_C \longrightarrow \cA_C(\cT) \longrightarrow 0.
\end{equation}
The bulk-boundary map is very simple for $\cA_C$: it is given by the injective arrow in the above short exact sequence, which identifies the bulk operators $\C[\mathfrak{t}]^\cW$ with polynomials in the mass parameters of $\cA_C(\cT)$. For $\cA_H$, the bulk-boundary map can be described as follows: if $\mu\in \cA_H(\cT)$ is the moment map for the action of $\mathfrak{g}_\C$ on $\cA_H(\cT)$, it determines a homomorphism
\begin{equation}
U(\mathfrak{g}_\C) \to \cA_H(\cT),
\end{equation}
which then gives the homomorphism $\cB_H \to \cZ[U(\mathfrak{g}_\C)] \to \left(\cA_H(\cT) \right)^{\mathfrak g}$, where the first arrow is again the Harish-Chandra map. In Sections \ref{sec:AH_Neum} and $\ref{sec:AC_Neum}$ we also describe the traces on $\cA_H$ and $\cA_C$ in terms of traces on $\cA_H(\cT)$ and $\cA_C(\cT)$, and then proceed to check the basic S-duality example of the Dirichlet boundary conditions and the Neumann boundary conditions enriched by $T[G]$. In the end, we also comment on how to go back from $\cA_H$, $\cA_C$ to $\cA_H(\cT)$ and $\cA_C(\cT)$.

After that we study the Nahm pole boundary conditions, which present several new challenges. In this case $\cA_C=\C$, so we only focus on $\cA_H$. In Section \ref{sec:Rmixing} we explain that the R-symmetry mixes with part of the gauge symmetry at the boundary. This happens because the Nahm pole breaks both the $SU(2)_H$ R-symmetry of the 3d $\cN=4$ SUSY, and some of the boundary symmetries present in the Dirichlet case (which is the limiting case of a trivial Nahm pole, $\varrho=0$). A certain combination of broken symmetries remains preserved and plays the role of ``boundary R-symmetry''. Using this property, we identify the boundary operators at the Nahm pole in Section \ref{sec:bdyOpNahm}. It turns out that the space of boundary operators is isomorphic to the space of regular functions on the Slodowy slice $t_+$, where $t_+\in\mathfrak{g}_\C$ is the nilpotent element associated to the embedding $\varrho: \mathfrak{su}(2)\to\mathfrak{g}$. 

The latter observation motivates our conjecture that the algebra $\cA_H$ of boundary operators at the Nahm pole is isomorphic to the finite W-algebra $\cW(\mathfrak{g}_\C, t_+)$. We provide a few checks of this conjecture in Section \ref{sec:Wconj}, and also write the trace on $\cA_H$ as a continuous linear combinations of traces on the Verma modules of $\cW(\mathfrak{g}_\C, t_+)$, similar to the Dirichlet case.

In the last section, we present a few computations of algebras on interfaces engineered by a single D5 or NS5 brane intersecting a stack of D3 branes on which our 4d theory lives. The cases where some of the D3 branes terminate on the fivebrane are also considered. In these examples, the answers for $\cA_H$ and $\cA_C$ are always simple and given by centers of the universal enveloping algebras (or two copies of those). The derivations are not completely trivial, and are in fact instructive exercises. In the case of $N$ D3 branes intersecting a D5, with additional $k$ D3 branes terminating on the right, the computation involves finding the $\mathfrak{gl}_N$-invariant subspace in the finite W-algebra $\cW(\mathfrak{gl}_{N+k}, t_+)$, and serves as an additional check of the finite W-algebra conjecture of Section \ref{sec:finiteW}.

\section{General constructions}\label{sec:general}
Maximal Super Yang-Mills (MSYM) in four dimensions has a rich class of well-known superconformal half-BPS boundary conditions and interfaces \cite{Gaiotto:2008sa,Gaiotto:2008sd,Gaiotto:2008ak}. Being invariant under a large 3d $\cN=4$ superconformal symmetry, they borrow a lot of their properties from pure three-dimensional theories with the same amount of supersymmetry, which were recently explored in great detail. The structures most relevant to us here are those of associative algebras with traces, encoding correlation functions of Higgs and Coulomb branch operators in these theories \cite{Chester:2014mea,Beem:2016cbd,Dedushenko:2016jxl,Dedushenko:2017avn,Dedushenko:2018icp}.

The algebras themselves can be identified either in the SCFT context \cite{Chester:2014mea,Beem:2016cbd}, or using the Omega-background \cite{Nekrasov:2002qd,Nekrasov:2003rj,Nekrasov:2010ka} applied to 3d $\cN=4$ theories \cite{Yagi:2014toa,Bullimore:2015lsa,Bullimore:2016hdc,Oh:2019bgz,Jeong:2019pzg} (which uses Omega-deformations of the A and B models \cite{Yagi:2014toa,Luo:2014sva,Nekrasov:2018pqq}), or from the $S^3$ supersymmetric background \cite{Dedushenko:2016jxl,Dedushenko:2017avn,Dedushenko:2018icp}. For some recent progress on the latter approach see \cite{Chang:2019dzt,Dedushenko:2019mzv,Pan:2019shz,Gaiotto:2019mmf,Dedushenko:2019mnd,Fan:2019jii,Chester:2020jay,Gaiotto:2020vqj,Feldman:2020dku}. In principle, this list may continue, as any background that is ``equivariant'' in the appropriate sense can be used for this purpose: for example, the 1d sector construction was recently extended to other backgrounds, such as $S^2\times S^1$, see \cite{Panerai:2020boq}. The structure of twisted traces is most manifest in the $S^3$ description (for recent mathematical constructions of traces, see \cite{Etingof:2020fls}). Below we will rely on results obtained using various combinations of these descriptions. We will briefly review the necessary facts about the 4d and 3d theories, and introduce our main players, -- half-BPS boundaries and interfaces in four dimensions, and their protected algebras. 

From the geometric point of view relevant to this work, one of the main differences between such objects and those in purely three-dimensional theories is that the analogs of Higgs and Coulomb branches are no longer hyper-K\"ahler manifolds, but rather complex Poisson. The corresponding associative algebras are quantizations of these complex Poisson manifolds.

\subsection{Half-BPS boundary conditions: a reminder}\label{sec:bndry_cond}
Consider 4d MSYM with gauge group $G$ on a half-space:
\begin{equation}
\R^3_{x_1,x_2,x_3}\times \R^+_{y},
\end{equation}
where the $\R^3$ is parametrized by $x_1, x_2, x_3$, and the half-line $\R^+$ --- by $y\geq 0$. At $y=0$ we impose some half-BPS superconformal boundary condition following \cite{Gaiotto:2008sa}. The R-symmetry algebra $\mathfrak{su}(4)$ of the 4d MSYM is broken at the boundary down to $\mathfrak{su}(2)_H\oplus \mathfrak{su}(2)_C$. The six scalars of the 4d vector multiplet, valued in the $\mathbf{6}$ irrep of $\mathfrak{su}(4)$, split into two groups, which are traditionally denoted by $\vec{X}$ and $\vec{Y}$. They are acted on by $\mathfrak{su}(2)_H$ and $\mathfrak{su}(2)_C$ respectively, that is we fix the following convention:
\begin{align}
(X_1, X_2, X_3)\quad &\text{form a triplet of } \mathfrak{su}(2)_H,\cr 
(Y_1, Y_2, Y_3)\quad &\text{form a triplet of } \mathfrak{su}(2)_C.
\end{align}
This is related to a convention we choose to follow in this paper (that slightly differs from \cite{Gaiotto:2008sa}): at the boundary, we always preserve the same 3d $\cN=4$ subalgebra of the 4d $\cN=4$. Under this subalgebra, gauge fields restricted to the boundary, together with $\vec{Y}$ and the appropriate fermions, transform as the 3d $\cN=4$ vector multiplet. Restriction of the normal component $A_y$ of the gauge field, together with $\vec{X}$ and the remaining fermions, form a boundary hypermultiplet. For convenience, we will use the now standard notation
\begin{equation}
\cO\big|
\end{equation}
for the bulk field or operator $\cO$ restricted to the boundary. 

The various boundary conditions are constructed, roughly, by imposing restrictions on one of these boundary multiplets as a whole. For example, the Neumann boundary condition basically eliminates the boundary hypermultiplet, leaving behind the boundary vector multiplet. Similarly, what is known as the Dirichlet boundary condition, eliminates the boundary vector multiplet, leaving the boundary hypermultiplet dynamical. The name Neumann and Dirichlet reflect the boundary conditions on gauge fields:
\begin{align}
\text{Neumann:}\quad &F_{iy}\big|=0,\ i=1,2,3,\cr
\text{Dirichlet:}\quad &A_{i}\big|=0,\ i=1,2,3.
\end{align}
SUSY implies the following boundary conditions on scalars in these two cases:
\begin{align}
\label{NeuSca}
\text{Neumann:}\quad  &\vec{X}\big|=0,\quad D_y \vec{Y}\big|=0,\\
\label{DirSca}
\text{Dirichlet:}\quad &\vec{Y}\big|=0,\quad \left(D_y \vec{X} -\frac{i}{2}[\vec{X}\times \vec{X}]  \right)\Bigg|=0,
\end{align}
where $[\vec{X}\times \vec{X}]$ stands for the commutator on the gauge indices and the vector product on the R-symmetry indices. Notice that the last boundary condition has the form of Nahm's equations \cite{Nahm:1981nb}, which are ubiquitous in extended SUSY in diverse dimensions (and also show up for the codimension-two defects, see e.g. \cite{Gukov:2006jk,Gukov:2008sn}).

Both Neumann and Dirichlet boundary conditions admit non-trivial modifications. For Neumann, they are given by placing extra boundary degrees of freedom that have a global $G$ symmetry gauged by the 3d $\cN=4$ vector multiplet formed by the boundary restrictions of the bulk fields. Such modification shifts the Dirichlet boundary conditions on $\vec{X}$:
\begin{equation}
\label{bdy_FI}
\vec{X}\big| = -\vec{\mu} + \vec{r},
\end{equation}
where $\vec\mu$ is the hyper-K\"ahler moment map of the boundary matter, and we also included the possibility of the ``boundary FI term'' given by $\vec{r}\in\mathfrak{c(g)}\otimes \R^3$, where $\mathfrak{c(g)}$ is the center of $\mathfrak{g}$. This $\vec{r}$ or course explicitly breaks conformal symmetry.

For Dirichlet, one modification is the boundary mass given by a commuting triple $\vec{m}$:
\begin{equation}
\label{bdy_mass}
\vec{Y}\big| = \vec{m},
\end{equation}
which similarly breaks the conformal symmetry. A more interesting modification is the Nahm pole. Namely, part of the Dirichlet boundary conditions imposes \eqref{DirSca} on the fields $\vec{X}$. The usual Dirichlet boundary conditions in addition require that all fields be regular at the boundary. The Nahm pole modification consists of choosing a homomorphism $\varrho:\mathfrak{su}(2)\to G$, and demanding instead a fixed singular behavior compatible with \eqref{DirSca},
\begin{equation}
\vec{X} \sim \frac{\vec t}{y},\quad \text{as } y\to0,
\end{equation}
where $(t_1, t_2, t_3)$ are images of the standard $\mathfrak{su}(2)$ generators under the homomorphism $\varrho$. We also use alternative notations for the $\mathfrak{sl}_2$ triple:
\begin{equation}
t_1+it_2 = t_+,\quad t_1 - i t_2 =t_-,\quad t_3.
\end{equation}
We sometimes denote the image of $\mathfrak{su}(2)$ under $\varrho$ as 
\begin{equation}
\varrho(\mathfrak{su}_2),
\end{equation}
to avoid excessive nested parentheses.

Notice also that the unmodified Dirichlet boundary conditions fully break the gauge symmetry at the boundary, leaving behind the boundary global symmetry $G$. Boundary masses can break $G$ global symmetry to a centralizer of $\vec{m}$. Likewise, the Nahm pole breaks this global symmetry to the centralizer of $\varrho$: the unbroken boundary global symmetry is $C_G(\varrho(\mathfrak{su}_2))$, a subgroup of $G$ that commutes with $\varrho(\mathfrak{su}_2)$. One can simultaneously have both the Nahm pole $\varrho$ and the boundary mass valued in the Lie algebra of $C_G(\varrho(\mathfrak{su}_2))$.

The most general boundary conditions are constructed as follows. We pick a subgroup $H\subset G$ that we wish to preserve as a gauge symmetry at the boundary. We give Neumann boundary conditions to the vector multiplets valued in $\mathfrak{h}={\rm Lie}(H)$, and couple them to some boundary theory $T$ that has an $H$ global symmetry. We may also include a boundary FI term for the abelian part of $H$. For the gauge fields valued in $\mathfrak{h}^\perp\subset \mathfrak{g}$, (where the orthogonal complemet is taken with respect to the Killing form on $\mathfrak{g}$,) we impose Dirichlet boundary conditions, possibly modified by the Nahm pole $\varrho: \mathfrak{su}(2) \to \mathfrak{g}$, such that $\varrho(\mathfrak{su}_2)$ commutes with $H$, and by the boundary mass commuting both with $H$ and $\varrho$.

Via the folding trick, these constructions admit an obvious generalization to interfaces.

\subsection{Protected sectors in the bulk}
There is a number of constructions of lower-dimensional theories emerging as sectors in supersymmetric quantum field theories in higher dimensions. They rely on a choice of equivariant supercharge that squares to a space-time rotation plus, possibly, an R-symmetry transformation. Passing to its cohmology localizes us to the fixed point locus of the said rotation, effectively reducing the number of spacetime dimensions.

In four-dimensional case, the fixed locus can either be zero-dimensional or two-dimensional. In the former case, it suggests that the theory localizes on a 0d QFT, i.e. a matrix model, and Pestun's localization result \cite{Pestun:2007rz} is essentially an example of this (see also \cite{Festuccia:2018rew}). The appearance of two-dimensional fixed point locus is best known in the context of connection to integrable systems \cite{Nekrasov:2009rc}, and for the chiral algebra construction in 4d $\cN=2$ SCFTs \cite{Beem:2013sza} (see \cite{Lemos:2020pqv} for a recent review of the latter). This clearly applies to 4d MSYM, whose chiral algebra is quite rich. Interestingly, the 4d MSYM admits other constructions with the two-dimensional fixed point locus, which we will now describe. 

It often happens that generic supersymmetric theories admit a holomorphic twist, while passing to the extended SUSY introduces new structures, such as topological twists, holomorphic-topological twists and Omega-deformations thereof (see, e.g., \cite{Eager:2018dsx,Elliott:2020ecf} for a general study of twists, \cite{Saberi:2019ghy,Saberi:2019fkq} for a recent study of holomorphic twists, \cite{Johansen:1994aw} for the first example of holomorphic twist, \cite{NikiThesis,Baulieu:1997nj} for studies on holomorphic and holomorphic-topological theories in 4d, and \cite{Closset:2017zgf,Closset:2017bse} for other examples of mixed twists). A morally similar phenomenon occurs in two-dimensional protected sectors in 4d SCFTs. While $\cN=2$ theories only possess 2d holomorphic sectors (i.e. chiral algebras), passing to $\cN=4$ introduces a new possibility: a 2d sector that is topological (or quasi-topological, as we will explain in a moment).

To be more precise, recall that the 2d holomorphic sector of \cite{Beem:2013sza} originates from the topological-holomorphic twist of 4d $\cN=2$ theories \cite{Kapustin:2006hi} admitting a specific Omega-deformation along the topological plane \cite{Oh:2019bgz,Jeong:2019pzg}. As it turns out, our 2d quasi-topological sector in 4d $\cN=4$ can be seen as originating from the Omega-deformation of the Kapustin-Witten \cite{Kapustin:2006pk} (also known as Marcus \cite{Marcus:1995mq}) twist. This statement, however, simply refers to the choice of a supercharge, not the background: indeed, we will work with the flat space and the ``physical'' four-sphere background, not the topological one. It would be interesting to explore possible connections to the topologically twisted theory  more systematically, but below we take a more hands-on approach and simply write down the corresponding supercharges that define the 2d sector.

This quasi-topological 2d sector comes in two guises: an ``electric'' one and a ``magnetic'' one. In fact, this sector has first appeared in the literature over a decade ago. Curiously, while the chiral algebra construction was first discovered in flat space (the ``$Q+S$'' construction of \cite{Beem:2013sza}), and only later reformulated using the spherical backgrounds \cite{Pan:2019bor,Dedushenko:2019yiw} (see also \cite{Pan:2017zie} for partial results on $S^4$), the 2d quasi-topological sector of 4d MSYM was first discovered in the context of localization on $S^4$. As some readers might have guessed by now, we are talking about the sector described by the 2d constrained Yang-Mills (cYM), as first conjectured in \cite{Giombi:2009ds,Giombi:2009ek} and then derived in \cite{Pestun:2009nn} from localization. This is also the reason we call it quasi-topological: the 2d Yang-Mills (and cYM is no different) is known to depend on the underlying geometry of space-time only through the 2d area. Correlators of Wilson loops are also only sensitive to area they enclose, while local operators do not feel the metric. See \cite{Migdal:1975zg,Rusakov:1990rs,Blau:1991mp,Witten:1991we,Witten:1992xu,Gross:1993hu,Douglas:1993iia,Gross:1994mr,Nunes:1995pv,Bassetto:1998sr} for some original references on 2d Yang-Mills and \cite{Cordes:1994fc} for the review.

Here we also give the ``$Q+S$'' style definition of the 2d quasi-topological sector, as it is useful to have several approaches at hand. Furthermore, as we will see later, this quasi-topological sector agrees with topological sector at the boundary (the one described by the associative algebra with a trace, as we mentioned before). This means that they are defined by the same supercharge in the bulk-boundary system. Indeed, this observation was the basis for the recent work \cite{Wang:2020seq,Komatsu:2020sup,Wang:2020jgh} on localization in 4d/3d systems. A related $\Omega$-deformation perspective lies behind the AAB-twisted topological string construction employed in \cite{Ishtiaque:2018str}.

The 4d $\cN=4$ superconformal algebra has Poincare supercharges $Q^A_\alpha$, $\tilde{Q}_{A\dot\alpha}$ transforming in the $\mathbf{4}$ and $\bar{\mathbf 4}$, and conformal supercharges $S_{A\alpha}$, $\tilde{S}^A_{\dot\alpha}$ transforming in the $\bar{\mathbf 4}$ and $\mathbf{4}$ of the R-symmetry group ${\rm Spin}(6)=SU(4)$ respectively. The (real anti-symmetric) generators of the latter are denoted $R_{IJ}$, $I,J=1\dots6$. The details on our conventions and the anti-commutation relations are given in the Appendix \ref{app:conv}.

Let us also pick an $\mathfrak{osp}(4|4)$ subalgebra, that is a 3d $\cN=4$ superconformal subalgebra, that will remain unbroken once we include a boundary in later subsections. The R-symmetry generators of this subalgebra are:
\begin{align}
R_{12}, R_{13}, R_{23} &\text{ generate } \mathfrak{su}(2)_C, \text{ with the chosen Cartan generator } R_C=iR_{31},\cr
R_{45}, R_{46}, R_{56} &\text{ generate } \mathfrak{su}(2)_H, \text{ with the chosen Cartan generator } R_H=iR_{56}.
\end{align}
Like \cite{Beem:2013sza}, we choose special linear combinations of supercharges denoted $\rQ_1^H$, $\rQ_2^H \in \mathfrak{osp}(4|4)$, and $\rQ_1^C$, $\rQ_2^C\in\mathfrak{osp}(4|4)$, which define the electric and the magnetic constructions respectively. From the point of view of the boundary 3d $\cN=4$ SUSY, these are the supercharges that define the Higgs and Coulmb branch constructions of \cite{Chester:2014mea,Beem:2016cbd,Dedushenko:2016jxl,Dedushenko:2017avn,Dedushenko:2018icp}. Below we describe them in our conventions, detailed in the Appendix \ref{app:conv}.

\subsubsection{Electric construction in the bulk}\label{sec:elec_bulk}
The defining supercharge is $\cQ^H=\rQ^H_1 + \rQ^H_2$, where
\begin{equation}
\label{elecQH}
\rQ_1^H = Q^2_1 - \tilde{Q}_{4\dot1} + \zeta(S_{31}+\tilde{S}^1_{\dot1}),\quad \rQ_2^H = Q^3_2 + \tilde{Q}_{1\dot2} + \zeta (S_{22}-\tilde{S}^4_{\dot2}),
\end{equation}
where $\zeta$ is a parameter of mass dimension one, which is related to the sphere radius $\ell$ by
\begin{equation}
\zeta = \frac1{2\ell}.
\end{equation}
These supercharges are nilpotent, and their anti-commutator is
\begin{equation}
\label{QQH}
\{ \rQ_1^H, \rQ_2^H \} = 8i\zeta (M_{12} + iR_{12}).
\end{equation}
Here $M_{12}$ generates rotations in the $(x^1,x^2)$ plane. The equivariant cohomology (on the space of local operators) of the $\rQ^H_1 + \rQ^H_2$ supercharge is supported at $x^1=x^2=0$. This is the plane parametrized by $(x^3, y)$, and we also sometimes write $y=x^4$ for the uniformity of notations. Define ``twisted translations'' in the $(x^3, x^4)$ plane:
\begin{equation}
\hat{P}_3 = P_3 + 2\zeta (R_{45}+i R_{46}),\quad \hat{P}_4 = P_4 + 2\zeta (R_{35}+iR_{36}).
\end{equation}
Their most important property is that
\begin{align}
\{\rQ_1^H, Q^4_2 - \tilde{Q}_{2\dot2} \}=\{\rQ_2^H, -Q^1_1 -\tilde{Q}_{3\dot1}\}&=-2\hat{P}_3,\cr
\{\rQ_1^H,Q^4_2 + \tilde{Q}_{2\dot2} \}=\{\rQ_2^H, Q^1_1 - \tilde{Q}_{3\dot1}\}&=2i \hat{P}_4,
\end{align}
that is both twisted translations are exact, implying that the sector of local operators in the cohomology of $\rQ^H_{1,2}$ is topological.

Local observables in the cohomology of $\rQ^H_{1,2}$ are constructed as gauge-invariant polynomials in a ``twisted-translated'' scalar operator, which is given by a linear combination:
\begin{equation}
\label{phiH}
\phi^H(x_3, x_4) = X_+ + \frac{ix_3}{\ell}X_3 + \frac{x_3^2+x_4^2}{4\ell^2}X_- - \frac{ix_4}{\ell}Y_1=e^{-ix_3\hat{P}_3 - ix_4\hat{P}_4}X_+ e^{ix_3\hat{P}_3 + ix_4\hat{P}_4},
\end{equation}
where we used the notation\footnote{If we use the six scalars $\Phi^{1,2,3,4,5,6}$ as in the Appendix \ref{app:MSYM}, such that $\Phi^{1,2,3}$ are acted on by $SU(2)_C$, and $\Phi^{4,5,6}$ are acted on by $SU(2)_H$, then $X_1=\Phi^5, X_2=-\Phi^6, X_3=-\Phi^4$ and $Y_1=\Phi^3, Y_2=-\Phi^1, Y_3=-\Phi^2$. The above formulas for observables are written in flat space, while those in the Appendix \ref{app:MSYM} are given on $S^4$, so comparison involves multiplication by a Weyl factor $1+\frac{x^2}{4\ell^2}$.}
\begin{equation}
X_\pm = X_1 \pm i X_2.
\end{equation}
Notice that at the origin we simply have:
\begin{equation}
\phi^H(0,0)=X_+.
\end{equation}
There are also the following complexified gauge fields in the cohomology:
\begin{align}
\gA_3^H&=A_3 - iY_1 + \frac{2ix_4(x_4Y_1 - x_3X_3 -2\ell X_2)}{4\ell^2+x_3^2+x_4^2},\cr
\gA_4^H&=A_4 - iX_3 -\frac{2ix_3(x_4Y_1 - x_3X_3 -2\ell X_2)}{4\ell^2+x_3^2+x_4^2},
\end{align}
where $A_\mu$ denotes the 4d gauge field. This $\gA^H_i$ can be used to construct arbitrary shape Wilson lines in the $(x^3, x^4)$ plane. The corresponding gauge field strength is not independent, and is in fact cohomologous to $\phi^H$ defined above (see Appendix \ref{app:coho}),
\begin{equation}
\mathscr{F}^H_{34}=\frac1\ell \phi^H + \{\cQ^H,\dots\}.
\end{equation}
Additionally, we identify the following angular gauge field in the cohomology:
\begin{equation}
\gA_\tau^H = x_1 (A_2 - iY_2) - x_2 (A_1 + iY_3) = A_\tau - i(x_1 Y_2 + x_2 Y_3),
\end{equation}
which is $\cQ^H$-closed for all values of $x_1,x_2,x_3,x_4$. It can be used to construct Wilson loops linking the $(x^3, x^4)$ plane:
\begin{equation}
\label{el_WL_H}
W_R^H = \trace_R {\rm Pexp}\, \left[ i\oint \gA_\tau^H \dd\tau \right],
\end{equation}
where $\tau$ parameterizes the circle $x_1^2 + x_2^2={\rm const}$, $x_3={\rm const}$, $x_4={\rm const}$. Furthermore, we will see that there also exists a $\rQ_{1,2}^H$-closed 't Hooft loop with the same support as \eqref{el_WL_H}. 

All these observables are nothing else but those of the 2d Yang-Mills sector of the 4d MSYM, which was discovered in \cite{Giombi:2009ds,Giombi:2009ek,Pestun:2009nn}, and recently considered in \cite{Wang:2020seq,Komatsu:2020sup}. In particular, the 't Hooft operators mentioned in the previous paragraph are familiar from \cite{Giombi:2009ek}, and are seen as instanton contributions in the 2d Yang-Mills.

\subsubsection{Magnetic construction in the bulk}\label{sec:mag_bulk} 
The dual construction goes along the same lines. We define the supercharges\footnote{ There is a family of possible pairs $\rQ^C_{1,2}$, and we chose those belonging to the same subalgebra as $\rQ^H_{1,2}$ that remains unbroken once we put our theory on $HS^4$ with half-BPS boundary. If we only studied the magnetic construction, we could use a simpler expression, e.g. $\rQ_1^C = Q^4_1 - \tilde{Q}_{2\dot1} + \zeta(S_{31}+\tilde{S}^1_{\dot1})$, $\rQ_2^C = Q^3_2 + \tilde{Q}_{1\dot2} + \zeta (S_{42}-\tilde{S}^2_{\dot2})$. However, one would not be able to preserve such $\rQ^C_{1,2}$ together with $\rQ^H_{1,2}$ from \eqref{elecQH} on $HS^4$. The choices in \eqref{elecQH} and \eqref{magnQC} agree with those in the Appendix A.2 of referemce \cite{Dedushenko:2017avn}.}
\begin{align}
\label{magnQC}
\rQ^C_1&=\frac12(Q^1_1-iQ^2_1+iQ^3_1-Q^4_1+i\tilde{Q}_{1\dot1}+\tilde{Q}_{2\dot1}+\tilde{Q}_{3\dot1}+i\tilde{Q}_{4\dot1})\cr &+\frac{\zeta}{2}(S_{11}-iS_{21}+iS_{31}-S_{41}+i\tilde{S}^1_{\dot1}+\tilde{S}^2_{\dot1}+\tilde{S}^3_{\dot1}+i\tilde{S}^4_{\dot1}),\cr
\rQ^C_2&=\frac12(Q^1_2+iQ^2_2-iQ^3_2-Q^4_2-i\tilde{Q}_{1\dot2}+\tilde{Q}_{2\dot2}+\tilde{Q}_{3\dot2}-i\tilde{Q}_{4\dot2})\cr &+\frac{\zeta}{2}(S_{12}+iS_{22}-iS_{32}-S_{42}-i\tilde{S}^1_{\dot2}+\tilde{S}^2_{\dot2}+\tilde{S}^3_{\dot2}-i\tilde{S}^4_{\dot2}),
\end{align}
which are also nilpotent, with the anti-commutator given by
\begin{equation}
\label{QQC}
\{ \rQ_1^C, \rQ_2^C \} = 8i\zeta (M_{12} + iR_{46}).
\end{equation}
The equivariant cohomology of local operators with respect to $\rQ^C_1 + \rQ^C_2$ is supported at $x^1=x^2=0$. The ``twisted translations'' in the $(x^3, x^4)$ plane are now defined according to
\begin{equation}
\hat{P_3}=P_3 + 2\zeta(R_{23}-iR_{12}),\quad \hat{P}_4=P_4 - 2\zeta(R_{35}+iR_{15}).
\end{equation}
As before, they are both exact:
\begin{align}
&\{\rQ_1^C, iQ^1_2-Q^2_2-Q^3_2+iQ^4_2-\tilde{Q}_{1\dot2}-i\tilde{Q}_{2\dot2}+i\tilde{Q}_{3\dot2}+ \tilde{Q}_{4\dot2} \}\cr
&=\{\rQ_2^C, -iQ^1_1-Q^2_1-Q^3_1-iQ^4_1-\tilde{Q}_{1\dot1}+i\tilde{Q}_{2\dot1}-i\tilde{Q}_{3\dot1}+\tilde{Q}_{4\dot1}\}=-4\hat{P}_3,\cr
&\{\rQ_1^C,iQ^1_2-Q^2_2-Q^3_2+iQ^4_2+\tilde{Q}_{1\dot2}+i\tilde{Q}_{2\dot2}-i\tilde{Q}_{3\dot2}-\tilde{Q}_{4\dot2} \}\cr
&=\{\rQ_2^C, iQ^1_1+Q^2_1+Q^3_1+iQ^4_1-\tilde{Q}_{1\dot1}+i\tilde{Q}_{2\dot1}-i\tilde{Q}_{3\dot1}+\tilde{Q}_{4\dot1}\}=4i \hat{P}_4,
\end{align}
implying that the sector of local operators in the cohomology of $\rQ^C_{1,2}$ is topological.

Local observables in the cohomology are similarly gauge-invariant polynomials in
\begin{equation}
\phi^C(x_3, x_4) = Y_+ + \frac{ix_3}{\ell}Y_3 +  \frac{x_3^2+x_4^2}{4\ell^2}Y_- -\frac{ix_4}{\ell}X_1=e^{-ix_3\hat{P}_3 - ix_4\hat{P}_4}Y_+ e^{ix_3\hat{P}_3 + ix_4\hat{P}_4},
\end{equation}
where
\begin{equation}
Y_\pm = Y_1\pm i Y_2.
\end{equation}
It appears that cohomology has no gauge field in the $(x^3, x^4)$ plane, unlike in the $\rQ^H_{1,2}$ case. This is misleading: S-duality implies that there actually must be one, since we found an emergent gauge field in the $\rQ^H_{1,2}$, or ``electric'' construction, and the $\rQ^C_{1,2}$ construction is simply related to it by S-duality. Therefore, we expect its magnetic dual, a 2d gauge field
\begin{equation}
\gA_3^C,\quad \gA_4^C,
\end{equation}
which is expressed through the magnetic gauge field of the 4d theory and scalars, to be in the cohomology. The magnetic gauge field is not manifest in the Lagrangian formulation, which is why naively we could not find the corresponding gauge field in the cohomology. However, Wilson lines built from the $\gA^C_{3,4}$, which are allowed line operators in the cohomology, do have a familiar description in the electric variables: they are the 't Hoofts operators. Thus the magnetic sector admits arbitrary shape 't Hooft lines supported on the $(x^3, x^4)$ plane.

We still easily find the angular $\rQ^C_{1,2}$-closed gauge field, though,
\begin{equation}
\gA_\tau^C = x_1 A_2 - x_2 A_1 + x_3 Y_3 - x_4 X_1 - i\ell Y_+ - i\frac{x_1^2 + x_2^2 + x_3^2 + x_4^2}{4\ell}Y_-,
\end{equation}
which can be used to construct circular Wilson loops linking the $(x^3, x^4)$ plane. Because we find such Wilson loops both in the $\rQ^H_{1,2}$ and $\rQ^C_{1,2}$ cohomology, S-duality implies that there also must exist 't Hooft loops with the same circular support there, as we claimed earlier. All these observables correspond to the magnetic dual version of the 2d Yang-Mills sector. It would be interesting to understand whether one can derive any of its properties via direct localization, but we do not take this route here.

\subsection{Protected sectors at the boundary}
In the presence of the boundary, we also have two S-dual constructions:
\begin{enumerate}
	\item The electric construction, as defined by the cohomology of $\rQ^H_{1,2}$. We will sometimes call it the H construction. At the boundary, this is the familiar 1d sector of Higgs branch operators \cite{Dedushenko:2016jxl}, and it is coupled to the 2d constrained Yang-Mills in the $\rQ^H_{1,2}$ cohomology of the 4d MSYM in the bulk. Some aspects of this 2d/1d coupled system were recently considered in \cite{Wang:2020seq,Komatsu:2020sup}. 
	\item The magnetic construction in the cohomology of $\rQ^C_{1,2}$. We sometimes call it the C construction. At the boundary, this defines the 1d protected sector of Coulomb branch operators like in \cite{Dedushenko:2017avn,Dedushenko:2018icp}, and it is coupled to the magnetic version of the 2d constrained Yang-Mills in the $\rQ^C_{1,2}$ cohomoloyg of the 4d MSYM in the bulk. Such a 2d/1d coupled system has not been considered before, and it would be somewhat interesting to perform its direct localization, but we do not address it here. Instead, we will use other methods, and sometimes study the magnetic construction using the electric construction in the dual 4d Yang-Mills (at the dual value of the 4d coupling constant).
\end{enumerate}

Our main focus in this work are boundary or interface local operators in the context of the above constructions. (We refer to them as boundary operators for brevity.) Their correlators are topological for familiar reason: the twisted-translation generator $\hat{P}_3$ is Q-exact and unbroken at the boundary. The boundary operators form certain interesting associative algebras, and their correlation functions are encoded in (twisted) traces on those algebras. We denote the boundary algebras in the H and C constructions by
\begin{equation}
\cA_H\quad \text{and} \quad \cA_C.
\end{equation}

Precise identification of the boundary operators of course depends on the boundary conditions. Yet, there are certain universal features, which we can address now. For one, boundary limits of the bulk operators in the cohomology, when non-zero, are also in the cohomology, and form the center of the boundary algebra. Let us call the bulk algebras $\cB_H$ and $\cB_C$. They are commutative and represented by gauge-invariant polynomials in $\phi^H$ and $\phi^C$ respectively. When we can take $\trace (\phi^H)^n$ and $\trace (\phi^C)^n$ as the generating sets of such polynomials, we simply have:
\begin{equation}
\cB_H \cong \C[\trace\phi^H, \trace(\phi^H)^2,\dots,\trace(\phi^H)^r],\quad \cB_C \cong \C[\trace\phi^C, \trace(\phi^C)^2,\dots,\trace(\phi^C)^r].
\end{equation}
More generally, we write\footnote{As a field, $\phi^H$ is a map to $\mathfrak{g}$. As an operator, it is an element of the dual space, which is why we find polynomial functions on $\mathfrak{g}$, denoted as $\C[\mathfrak{g}]$, rather than $\C[\mathfrak{g}^*]$.}
\begin{equation}
\cB_H \cong \C[\mathfrak{g}]^G\cong \C[\mathfrak{t}]^\cW,\quad \cB_C \cong \C[\mathfrak{g}]^G\cong\C[\mathfrak{t}]^\cW,
\end{equation}
where $\cW$ is the Weyl group of $G$. There exist natural homomorphisms from the bulk to boundary algebras defined via collision of the bulk operators with the boundary:
\begin{equation}
\rho_H: \cB_H \to \cA_H,\quad \rho_C: \cB_C \to \cA_C.
\end{equation}
The elements of $\rho_H(\cB_H)$ are necessarily in the center $\cZ(\cA_H)$ of $\cA_H$: any such local observable can be moved slightly into the bulk, and commuted past any other boundary observable without collisions. The same is true for $\rho_C(\cB_C)$. We actually conjecture that
\begin{equation}
\rho_H(\cB_H) \cong \cZ(\cA_H),\quad \rho_C(\cB_C) \cong \cZ(\cA_C).
\end{equation}
We will see in the examples that such isomorphisms can be non-trivial: e.g., for the Dirichlet boundary conditions, $\rho_H(\cB_H) \cong \cZ(\cA_H)$ is essentially the Harish-Chandra isomorphism.

As we will see, the boundary can also support other, non-commutative operators, such as non-gauge invariant polynomials in $\phi^H$ for the Dirichlet boundary conditions, or operators coming from the boundary matter in the Neumann case.

Because our bulk/boundary system is superconformal (to the extent it is possible with the boundary), it is quite convenient to use the usual unitarity bounds of the superconformal theories to identify the boundary operators in the cohomology. The relevant bounds have precisely the same form as in the purely 3d case of \cite{Chester:2014mea,Beem:2016cbd}. If we denote the scaling dimension of the boundary operator inserted at the origin $0$ by $E$, then
\begin{equation}
E\geq R_H + R_C,
\end{equation}
and in particular (since $E$ is the same throughout the $SU(2)_H$ and $SU(2)_C$ multiplets),
\begin{equation}
E\geq R_H,\quad E\geq R_C,
\end{equation}
where $R_H$ and $R_C$ are its $\mathfrak{su}(2)_H$ and $\mathfrak{su}(2)_C$ charges. The boundary operators in the $\rQ^H_{1,2}$ cohomology, when inserted at the origin $0$, are precisely those saturating the first bound:
\begin{equation}
\label{coho_H}
E=R_H,
\end{equation}
which also shows that they are $\mathfrak{su}(2)_H$-highest weights. They must also be neutral under the $\mathfrak{su}(2)_C$. Similarly, operators in the $\rQ^C_{1,2}$ cohomology are $\mathfrak{su}(2)_H$ neutral, and $\mathfrak{su}(2)_C$ highest weights saturating the other bound when inserted at the origin:
\begin{equation}
\label{coho_C}
E=R_C.
\end{equation}
Both types of operators are Lorentz scalars. This characterization will be quite useful when we discuss the Nahm pole boundary conditions.

Mathematically, characterizing the cohomology classes via \eqref{coho_H} or \eqref{coho_C} is just the familiar statement of Hodge theory. Indeed, the representatives obeying \eqref{coho_H} or \eqref{coho_C} are analogous to harmonic representatives in the de Rham cohomology.

As an example, we have seen before that 
\begin{equation}
X_+(0)
\end{equation}
is in the $\rQ^H_{1,2}$ cohomology (whenever non-zero), and indeed it has $R_C=0$, $E=R_H=1$. Similarly,
\begin{equation}
Y_+(0),
\end{equation}
when non-zero, is in the $\rQ_{1,2}^C$ cohomology, and it has $R_H=0$, and $E=R_C=1$.

\subsubsection{Special features of the $\cA_H$}
The electric boundary algebra $\cA_H$, in general, has a more direct description. Indeed, it is related to the Higgs sector at the boundary, which does not require non-perturbative correction, and is often straightforward to describe in the electric variables.

With little work, we will understand in the later sections how to describe boundary algebras for Dirichlet and Nahm pole boundary conditions. Inclusion of extra boundary matter also has an elegant algebraic interpretation in terms of quantum Hamiltonian reduction. Description of traces requires slightly more work, as we will see.

There are a few possible generalization involving extended operators that we do not address here in any details. In the case of electric construction, we can have bulk supersymmetric Wilson loops ending on the boundary. Their endpoints are charged and can be left un-screened for Dirichlet boundary conditions. For Neumann boundary conditions, gauge invariance forces us to place charged boundary operators there so that he total charge is zero. Another generalization would involve inserting circular Wilson and 't Hooft loops linking the 2d Yang-Mills plane in the bulk. Such insertions do not affect $\cA_H$, but are expected to modify the trace on $\cA_H$ (to be introduced soon). Surface operators are also possible \cite{Wang:2020seq} but not studied here, while codimention-1 objects (interfaces and boundaries) are possible, and of course are the main subject in this work.

Another important potential modification involves half-BPS line defects placed along the 1d locus in the boundary/interface. These modify directly $\cA_H$ or $\cA_C$  \cite{2020JHEP...02..075D}, depending on the type of line defect. Topological correlators supported on Wilson lines of the bulk 2d cYM, in the electric case, were also explored in \cite{Giombi:2018qox,Giombi:2018hsx,Giombi:2020amn}.

\subsubsection{Special features of the $\cA_C$ algebra}
From the general point of view, the boundary algebra $\cA_C$ can be seen as more challenging, for the same reason that the Coulomb branch is more challenging in 3d theories: it can have non-perturbative corrections in the form of monopole operators. In parctice, however, quite often we do not have to work too much. Local monopole operators that can appear are only those charged under gauge fields living at the boundary, not the restrictions of bulk fields. Bulk gauge fields cannot have ``boundary monopole operators'' in four dimensions (unlike in 3d \cite{Bullimore:2016nji,Witten:2011zz}).

This essentially reduces the problem to a hard but solved one: description of the Coulomb branch algebra $\cA_C$ in pure 3d theories \cite{Bullimore:2015lsa,Nakajima:2015txa,Braverman:2016wma} (see also \cite{Dedushenko:2018icp} for traces on such algebras). Masses enter this algebra as parameters. We claim that the $\cA_C$
algebra for enriched Neumann boundary conditions is obtained from the $\cA_C$ algebra for the 3d boundary degrees of freedom
by promoting their masses to dynamical fields identified with restrictions of the bulk scalars. This statement generalizes in a simple way to 
more general boundary conditions. Writing traces on this algebra is also straightforward once traces of the 3d algebras are known.

Again, there are generalizations involving extended operators that we do not explore here. We could include an open 't Hooft line in the $(x^3, x^4)$ plane with both of its ends supported at the boundary. These endpoints look like monopole operators on the boundary. We could also include boundary line defects 
modifying $\cA_C$ \cite{2020JHEP...02..075D}, circular Wilson and 't Hooft loops linking the $(x^3, x^4)$ plane resulting in different traces on $\cA_C$, and could also study surface operators in the cohomology. Our focus is on local operators supported on boundaries and interfaces only.

\subsection{Sphere partition function and the twisted trace}\label{sec:twisted_trace}
So far, we have discussed theories in flat space. It has been realized in \cite{Dedushenko:2016jxl} that it is useful to put 3d $\cN=4$ theories on $S^3$ to study the protected correlators on $S^1\subset S^3$. The 4d constructions of \cite{Giombi:2009ds,Giombi:2009ek,Pestun:2009nn} were also formulated in terms of MSYM on a sphere, the round four-sphere in that case. Fusing these two ideas, we should study the correlators of the 4d/3d system by putting it on a hemisphere $HS^4$, as was recently explored in \cite{Wang:2020seq,Komatsu:2020sup}. Since the system is superconformal, there is a canonical way to define it on the spherical background, -- indeed, $HS^4$ is related to a flat half-space by a Weyl transformation.

The protected $S^3$ correlators of \cite{Dedushenko:2016jxl} were fully encoded in a certain (twisted) trace on the algebra of operators in the cohomology. More specifically, the two such algebras, --- $\cA_H$ of Higgs operators and $\cA_C$ of Coulomb operators, --- come with a trace on each of them canonically determined by the theory. These traces play a prominent role in the conjecture of \cite{Gaiotto:2019mmf}, where it was proposed that in the special case of a 3d theory that can be made fully massive, they are given by specific linear combinations of traces over the Verma modules of $\cA_H$ and $\cA_C$, which are in one-to-one correspondence with massive vacua. 

Similarly, the protected $HS^4$ boundary correlators determine (and are encoded in) the (twisted) traces on the algebras of boundary operators in the H and C constructions. If we study interfaces, rather than boundaries, then the right setting is the full $S^4$ with an interface at the equator. The algebras of boundary (or interface) local operators in the electric and magnetic constructions were also denoted by $\cA_H$ and $\cA_C$ in the previous subsection. Let the corresponding twisted traces be
\begin{equation}
T_H: \cA_H\to\C,\quad T_C:\cA_C\to \C.
\end{equation}
Recall that the word ``twisted'' refers to the fact that the trace property is twisted by an automorphism $g\in {\rm Aut}(\cA_H)$, that is:
\begin{equation}
\label{twisted_def}
T_H(xy) = T_H(g(y)x),\ \forall x,y\in \cA_H,
\end{equation}
and similarly for $\cA_C$. We will return to the nature of this twist in a moment.

As said previously, the (twisted) trace determines correlation functions. For example, if $\cO_i\in\cA_H$ are the boundary operators in the electric construction, the relation between the $HS^4$ correlators and the trace, as mentioned in the introduction, is simply
\begin{equation}
\langle \cO_1(\varphi_1)\dots \cO_n(\varphi_n) \rangle_{HS^4} = T_H(\cO_1\dots\cO_n),\ \varphi_1>\varphi_2\dots>\varphi_n,
\end{equation}
where on the right, the operators are multiplied as elements of the associative algebra $\cA_H$. Here the correlators are not normalized, that is $T_H(1)$ is the $S^3$ partition function.

The twistedness, i.e. the choice of $g$ in \eqref{twisted_def} is determined by masses in the $\cA_H$ case, and by FI parameters in the $\cA_C$ case.\footnote{A big advantage of sphere backgrounds is that it is possible to turn on masses and FI parameters without breaking $\rQ^H_{1,2}$ and $\rQ^C_{1,2}$ \cite{Dedushenko:2016jxl}. Doing so in the flat space description would break these supercharges.} More precisely, we can have two types of masses and FI parameters that determine the relevant automorphism $g$. Suppose the boundary conditions are defined by coupling to some boundary degrees of freedom with a flavor symmetry $F\times G$, out of which $G$ is gauged by the bulk vector multiplet. Then we can still turn on the 3d twisted mass $m_F\in \mathfrak{f}={\rm Lie}(F)$ for $F$ at the boundary. This $m_F$ is going to act on $\cA_H$ as an automorphism, and appear in the twisted trace relations roughly as follows (assuming no other masses are turned on),
\begin{equation}
T_H\left(xy\right)= T_H\left(\left((-1)^{F_H} e^{-2\pi\ell m_F}\cdot y\right)x\right),\quad x,y\in\cA_H.
\end{equation}
Here $(-1)^{F_H}$ is the center of the $SU(2)_H$ R-symmetry, which twists the trace even when $m_F=0$. 
One way to understand the action by $e^{-2\pi\ell m_F}$ is that the 3d mass $m_F$ is reflected in the 1d localized sector of \cite{Dedushenko:2016jxl} living on $S^1\subset S^3$ as a holonomy $m_F$ of the global symmetry $F$ around a circle, which precisely implements such a twist in correlators.

Another kind of mass we can turn on is the ``boundary mass'', that was reviewed in Section \ref{sec:bndry_cond}, see equation \eqref{bdy_mass}. It can appear in situations when the bulk gauge symmetry is broken down to a global symmetry at the boundary, which happens when we impose Dirichlet boundary conditions on the bulk gauge multiplet. In this case we may turn on a mass at the boundary, which simply corresponds to deformong the boundary conditions according to \eqref{bdy_mass}. Only one of the three masses is compatible with the supercharges $\rQ^H_{1,2}$, $\rQ^C_{1,2}$, which corresponds to choosing\footnote{The component of $\vec{Y}$ that can have a vev on the sphere is $Y_1=\Phi^3$, since it commutes with $R_{12}$ appearing in \eqref{QQH}. Likewise, $X_1=\Phi^5$ is the component of $\vec{X}$ that commutes with $R_{46}$ appearing in \eqref{QQC}.}
\begin{eqnarray}
\label{bdy_m}
Y_1\big| =m,\quad Y_2\big|=Y_3\big|=0.
\end{eqnarray}
The same remark also holds for the purely three-dimensional masses: only one real component out of the three is allowed.

The situation is very similar for the FI terms. If we couple the bulk theory to some boundary degrees of freedom that include abelian 3d $\cN=4$ gauge multiplets, we can turn on their FI parameters (only one parameter in the $\mathfrak{su}(2)_H$-triplet of FI terms is consistent with SUSY on the $S^3$, see \cite{Dedushenko:2016jxl}). Also, as explained in the equation \eqref{bdy_FI}, when we impose Neumann boundary conditions on the bulk gauge multiplet, we can turn on the boundary FI parameter for abelian factors of the gauge group. Only one FI term is allowed; for example, in the case of pure Neumann boundary conditions, we can have:
\begin{equation}
\label{bdy_sphere_FI}
X_1\big|=r,\quad X_2\big|=X_3\big|=0.
\end{equation}
Both pure 3d and boundary FI parameters appear as twist parameters in the trace on $\cA_C$.

We sometimes refer to \eqref{twisted_def} as the Ward identities for the correlation functions. The space of solutions of these Ward identities is 
interesting. In general, we expect that 3d algebras $\cA_H$ or $\cA_C$ should always have a finite-dimensional space of solutions of the Ward identities, so that the infinite collection of protected $S^3$ correlators should be determined algebraically in terms of a finite generating collection. 
For $\cA_C$ in standard gauge theories, this expectation is part of the quantum Hikita conjecture \cite{kamnitzer2018quantum}.

Because of the large center in the 3d/4d case, the characterization of twisted traces is a bit more complicated. We will still see, though, how the infinite collection of protected $HS^4$ correlators can be reduced algebraically to a matrix integral.

\subsubsection{Reduction of $\cA_C$ to 3d}\label{sec:red_3d}
Consider a 4d $\cN=4$ theory on the cylinder $S^3\times I$, where $I=[0,\epsilon]$ is an interval. Suppose at $y=0\in I$ we impose our boundary conditions $\mathbf{B}$ of interest, and study the algebra $\cA_C[\mathbf{B}]$. At the opposite end $y=\epsilon\in I$, let us impose the Dirichlet boundary conditions with some generic boundary mass \eqref{bdy_m}. Let this system flow to the IR. In the limit, we will land at the 3d theory $T^{\mathbf{B}}_{\rm 3d}$, whose algebra $\cA_C[T^{\mathbf{B}}_{\rm 3d}]$ will be given by the central quotient:
\begin{equation}
\label{quot_3d}
\cA_C[T^{\mathbf{B}}_{\rm 3d}] \cong \cA_C[\mathbf{B}]/I,
\end{equation}
where $I$ is the ideal generated by the center $\cZ(\cA_C[\mathbf{B}])$ if boundary masses vanish, $m=0$. For non-zero masses, we take the natural deformation of this ideal, which basically sets $Y_3=m$.

This happens because the Dirichlet boundary conditions fix vevs of the bulk operators, and those belong to the center of the boundary algebra. Now the trace on the quotient \eqref{quot_3d}, according to the conjecture of \cite{Gaiotto:2019mmf}, is a linear combination of finitely many traces over the Verma modules. If we denote such $m$-dependent trace by
\begin{equation}
T_C^m:\cA_C[\mathbf{B}]/I \to \C,
\end{equation}
then the trace on the full algebra $\cA_C[\mathbf{B}]$ is given by gluing the cylinder to the hemisphere. The resulting trace on the full algebra can be written as
\begin{equation}
\label{tr_integral}
T_C(\cO) = \frac1{|\cW|}\int_{\mathfrak{t}} \dd m\, T_C^m(\cO)\, Z(m)\prod_{\alpha\in\Phi_+} 4\sinh^2\pi\langle\alpha,m\rangle, \quad \cO\in\cA_C[\mathbf{B}],
\end{equation}
where $Z(m)$ is the hemisphere partition function with Dirichlet boundary conditions:
\begin{equation}
Z(m)=e^{-\pi\,{\rm Im}(\tau)\trace(m^2)} \prod_{\alpha\in\Phi_+} \frac{\langle\alpha,m\rangle}{2\sinh\pi\langle\alpha,m\rangle}\equiv e^{-\pi\,{\rm Im}(\tau)\trace(m^2)}\frac{\Delta(m)}{\mathbbm{\Delta}(m)},
\end{equation}
where we also introduced the notations $\Delta(m)$ and $\mathbbm{\Delta}(m)$ for the ordinary Vandermonde and the $\sinh$-Vandermonde respectively:
\begin{equation}
\Delta(m)=\prod_{\alpha\in\Phi_+} \langle \alpha, m\rangle,\quad \mathbbm{\Delta}(m)=\prod_{\alpha\in\Phi_+} 2\sinh\pi\langle \alpha, m\rangle,
\end{equation}
with $m$ a boundary mass made dimensionless by absorbing $\ell$, and $\tau$ a 4d gauge coupling:
\begin{equation}
\tau=\frac{4\pi i }{g_{\rm YM}^2} + \frac{\theta}{2\pi},
\end{equation}
and $\Phi_+$ denotes the set of positive roots in the root system $\Phi$ of $\mathfrak{g}$.

The equation \eqref{tr_integral} reflects the statement that $\cA_C[\mathbf{B}]$ is obtained from $\cA_C[T_{\rm 3d}^{\mathbf B}]$ by promoting 3d masses in $T^{\mathbf B}_{\rm 3d}$ theory to dynamical variables. Indeed, in \eqref{tr_integral} we integrate over masses, and by including insertions given by $\cW$-invariant polynomials $P(m)$, we can compute correlators of bulk operators $P(Y_+)$ at the boundary. Indeed, $P(Y_+)$ reduces to an insertions of $P(m)$, as it follows from \eqref{bdy_m}.

In later sections we will also discuss similar statements for the $\cA_H$ algebra, see in particular Section \ref{sec:ungauge}.

\section{Dirichlet boundary conditions and $U(\mathfrak{g}_\C)$}

Let us now start examining concrete examples, and Dirichlet boundary conditions are among the most basic ones. Recall that the fields $\vec{Y}$ obey Dirichlet boundary conditions \eqref{DirSca}, which in the presence of boundary masses look like \eqref{bdy_m}. This means that $\vec{Y}$ is not dynamical, so the boundary observable $\phi^C$ is not available. There are no other potential candidates to contribute the $\rQ^C_{1,2}$ cohomology, and in fact we simply have
\begin{equation}
\cA_C=\C,
\end{equation}
with the unique trace on it, normalized to produce the $HS^4$ partition function with the Dirichlet boundary conditions and the boundary mass:
\begin{equation}
\label{HS4Dir}
T_C(1) = Z(m)=e^{-\pi\, {\rm Im}(\tau)\trace(m^2)} \frac{\Delta(m)}{\mathbbm{\Delta}(m)}.
\end{equation}
Again, we made the mass dimensionless by incorporating a factor of the radius $\ell$. The 4d derivation of \eqref{HS4Dir} can be found in \cite{Gava:2016oep,Dedushenko:2018tgx}, while \cite{Wang:2020seq} checked that the 2d perturbative YM reproduces the same answer.

The $\cA_H$ algebra is much more interesting, which is our next subject.

\subsection{$\cA_H$ and $U(\mathfrak{g}_\C)$}\label{sec:AH_Ug}
Recall that the 2d protected sector of the MSYM on $S^4$ is described by a perturbative complexified 2d Yang-Mills on $S^2\subset S^4$ \cite{Pestun:2009nn}. For the four-dimensional hemisphere $HS^4$, the same is true, with $S^2$ replaced by the disk $HS^2$. The gauge field of the 2d Yang-Mills is $\gA^H_i$, $i=3,4$ discussed in Section \ref{sec:elec_bulk}. The 4d Dirichlet boundary conditions \eqref{bdy_m} are translated into the Dirichlet boundary conditions in two dimensions \cite{Wang:2020seq} fixing the value of the boundary gauge field according to:
\begin{equation}
\label{YMDbc}
\gA^H\big|_{\partial(HS^2)}=-i\frac{m}{\ell}.
\end{equation}
We still keep $m$ dimensionless here.

The gauge symmetry is broken at the boundary, so the boundary observables are polynomials in $\phi^H$ that are not necessarily gauge-invariant. At the origin, $\phi^H$ coincides with $X_+=X_1+i X_2$, while away from the origin it becomes a more general complex linear combination of $\vec{X}$ \eqref{phiH}. Such observables are related to the curvature of $\gA^H_{3,4}$ because $\phi^H$ is proportional to it in the cohomology:
\begin{equation}
\mathscr{F}^H_{34} = \frac{1}{\ell}\phi^H + \{\cQ^H,\dots\}.
\end{equation}
So the boundary correlators in 4d reduce to perturbative correlation function in 2d Yang-Mills (YM) on the disk with boundary condition \eqref{YMDbc}, and with boundary insertions of the 2d gauge curvature. It is enough to consider separate insertions of $\mathscr{F}^H$, while more general polynomials can be obtained by collisions of such elementary boundary operators.

It is useful to write the 2d YM as a deformed BF theory:\footnote{The 2d gauge coupling constant $e$ is related to the 4d Yang-Mills coupling $g_{\rm YM}$ \cite{Pestun:2009nn} via $e^2=\frac{g^2_{\rm YM}}{8\pi\ell^2}$.}
\begin{equation}
\label{BF}
S = -i\int \trace BF + e^2\int \dd^2 x\, \trace B^2,
\end{equation}
where for the simplicity of notations, we renamed $\mathscr{F}^H$ as simply $F$. Here $B$ is a $\mathfrak{g}^*_\C$-valued 0-form, where $\mathfrak{g}^*$ is the dual of $\mathfrak{g}$, and $\mathfrak{g}^*_\C=\mathfrak{g}^*\otimes\C$, since all the fields are complexified. We also abuse notations, denoting by $\trace(ab)$ one of the three things: a Killing form if $a,b\in \mathfrak{g}$, a dual of the Killing form if $a,b\in\mathfrak{g}^*$, the canonical pairing if $a\in \mathfrak{g}$, $b\in\mathfrak{g}^*$ or vice versa.

Notice that the action \eqref{BF} is diagonalized by the field redefinition 
\begin{equation}
B = \tilde{B} + \frac{i}{2e^2}F_{34},
\end{equation}
upon which one gets back the 2d YM action plus $e^2\int\dd^2x\, \trace \tilde{B}^2$. Therefore, the field $\tilde{B}$ has an ultra-local Green function, given just by a contact term $\langle \tilde{B}(x)\tilde{B}(y)\rangle\propto\delta^{(2)}(x-y)$. From this we see that under the correlators, $B$ and $\frac{i}{2e^2}F$ are equivalent up to contact terms:
\begin{equation}
\left(\frac{i}{2e^2}\right)^n \langle F^{a_1}_{34}(\varphi_1)\dots F^{a_n}_{34}(\varphi_n)\rangle = \langle B^{a_1}(\varphi_1)\dots B^{a_n}(\varphi_n)\rangle + \text{contact terms}.
\end{equation}
Here $a_i$ are gauge indices corresponding to some choice of basis on $\mathfrak{g}$, and $\varphi_i$ are angular positions along $\partial D^2$. As long as we look at the insertions of $B$ at separate points, the contact terms do not contribute, and as we know, we can build more general operators by collisions. Therefore, we study correlators of $B$ now.

Another important simplification can be achieved by reinterpreting the boundary conditions \eqref{YMDbc} as a modification of the $m=0$ case,
\begin{equation}
\label{YMD0}
\gA^H\big|=0,
\end{equation}
by an insertion of the boundary deformation in the path integral:
\begin{equation}
\label{bdy_deform}
e^{-\oint_{\partial D^2}\dd\varphi\, \trace(m B)}.
\end{equation}
Indeed, such a modification, on boundary equations of motion, deforms \eqref{YMD0} into \eqref{YMDbc}. A better, perhaps more convincing, way to understand it is through the operator formalism. In canonical quantization of the BF theory on space $S^1$, the position variable is the holonomy $A_\varphi=u={\rm const}$ along the $S^1$, or rather its global version:
\begin{equation}
\cU = {\rm Pe}^{i\oint_{S^1} A}.
\end{equation}
The field $B$ plays the role of conjugate momentum. Denote the ``position basis'' vector with the holonomy $\cU=e^{2\pi i u}$ by $|u\rangle$. The hemisphere produces some state $|\Psi\rangle$, and the boundary condition \eqref{YMD0} corresponds to computing the overlap $\langle 0|\Psi\rangle$. A non-zero holonomy at the boundary corresponds to a more general overlap $\langle u|\Psi\rangle$. As is usual, one can obtain $\langle u|$ by applying a shift operator given by the exponential of the momentum operator, which in our case can be written as:
\begin{equation}
\langle u| = \langle 0| e^{-i\oint_{S^1}\trace(u B)\dd\varphi}.
\end{equation}
Passing to the path integral formulation, we see that indeed, acting with such an operator corresponds to including the deformation \eqref{bdy_deform} as a boundary action. In other words, correlators of operators $\cO$ at $m\neq 0$ are given in terms of those at $m=0$,
\begin{equation}
\label{mTrace}
\langle \cO\rangle_m = \langle \cO\, e^{-\oint_{\partial D^2}\dd\varphi\, \trace(m B)}\rangle_0.
\end{equation}

To determine the algebra of boundary observables, it is enough to both set $m=0$, and study the weak-coupling limit $e\to0$. An abstract, and at the same time very cheap argument for this can be made using the result found later in this subsection: the boundary algebra at $e=m=0$ is the universal enveloping algebra $U(\mathfrak{g}_\C)$. If the answer at finite $e, m$ were different, it would be some deformation of $U(\mathfrak{g}_\C)$ \emph{as an associative algebra}. It is known, however, that for semisimple $\mathfrak{g}$
\begin{equation}
HH^2(U(\mathfrak{g}), U(\mathfrak{g}))=0,
\end{equation}
implying that such deformations are trivial, that is at $e, m\neq 0$ we must find $U(\mathfrak{g}_\C)[[e,m]]$, if $e,m$ were treated as formal parameters. Since they are really numbers, we simply have $U(\mathfrak{g}_\C)$, perhaps with some $e$ and $m$-dependent renormalization happening along the way. 

\begin{figure}[h]
	\label{fig:alg_diag}
	\centering
	\includegraphics[scale=1]{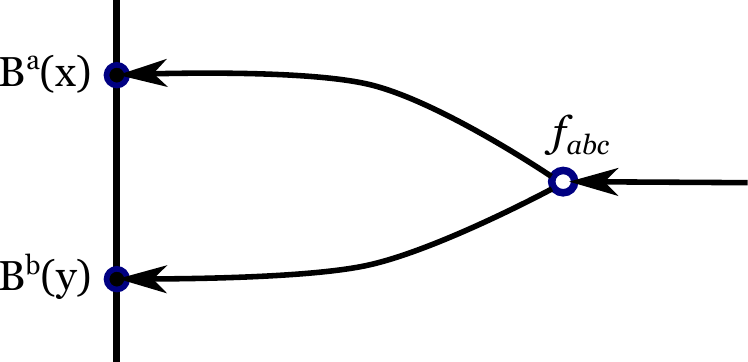}
	\caption{The boundary algebra is determined by this diagram. The two boundary insertions, $B^a(x)$ and $B^b(y)$, are connected to the bulk vertex $f_{abc}B^a A^b\wedge A^c$. A line with an arrow denotes the $\langle BA \rangle_{\rm free}$ propagator, with the arrow pointing at $B$. Similar Feynman rules were used in \cite{Cattaneo:1999fm}.}
\end{figure}

A more concrete argument is to compute boundary correlators perturbatively (we are only interested in the perturbative 2d YM anyways). We treat $\trace B[A,A]$ as an interaction, and account for its effect in perturbation theory. The propagators that follow from the free action are roughly
\begin{equation}
\langle BB\rangle_{\rm free}=0,\quad \langle BA\rangle_{\rm free}\sim\frac1\dd,\quad \langle AA\rangle_{\rm free} \sim \frac{e^2}{\dd * \dd}.
\end{equation}
As already mentioned, at $e=m=0$, we will find a non-trivial boundary algebra. The boundary correlators have jump discontinuities when operators collide, and such a discontinuous\footnote{In fact piece-wise constant, since the correlators are topological.} UV behavior encodes the algebra. The boundary algebra essentially follows from a single Feynman diagram on Figure \ref{fig:alg_diag}.
At $e\neq 0$, the non-zero $\langle AA \rangle_{\rm free}$ propagator allows to write many more non-trivial Feynman diagrams. However, this propagator is less singular than $\langle BA\rangle_{\rm free}$, and all such diagrams will not affect discontinuities of boundary correlators. We can easily see it from the dimensional analysis: $B$ is dimensionless, while both $e$ and $m$ have dimensions of mass. Therefore, positive powers of $e$ and $m$ in expressions for correlators can only appear with positive powers of coordinates, so that the total expression is dimensionless. Such contributions vanish at coincident points limit, not affecting the structure constants of the algebras. An example of such an expression is
\begin{equation}
(x_1-x_2)^{n+k}e^n m^k,
\end{equation}
where $x_1$, $x_2$ are positions of some insertions. In the $x_1-x_2\to 0$ limit, only the $n=k=0$, i.e. $e$ and $m$ independent, terms can survive.

Putting it in a more intuitive physical language, what we have just explained simply means that $e$ and $m$ are relevant perturbations. Therefore, they are immaterial in the UV limit, and meanwhile, the algebra of local operators is a question about the UV behavior of correlators. Thus, the algebra structure does not depend on $e$ and $m$, and in what follows, we simply compute it at $e=m=0$. Later we will determine the trace on it: unlike the algebra, it depends both on $e$ and $m$.

\paragraph{Algebra at $e=0$.} At $e=0$, the action \eqref{BF} simply describes the BF theory. It is well-known that the BF theory in 2d is equivalent to the Poisson sigma model with the dual Lie algebra, --- $\mathfrak{g}^*_\C$ in our case, --- taken as a target Poisson manifold. The Poisson structure is the canonical Lie-Poisson structure on $\mathfrak{g}^*_\C$. This equivalence is manifested by writing the action as
\begin{equation}
S=-i\int (A^a\wedge  \dd B^a -\frac12 f_{abc} B^a A^b\wedge A^c),
\end{equation}
where now $B^a$ play the role of coordinates on the target, and the Poisson structure is
\begin{equation}
\pi^{ab} = - f^{abc}B^c.
\end{equation}
We recognize this as the Poisson sigma model from \cite{Cattaneo:1999fm}, where it was shown perturbatively that the algebra of boundary operators is determined by the Konsevich's star product \cite{Kontsevich:1997vb}. Concretely, specializing to the case of $\mathfrak{g}^*_\C$ with the Lie-Poisson structure, we find that the star-product is
\begin{equation}
\label{star_prod}
B^a\star B^b = : B^a B^b :- \frac{i}{2} f^{abc}B^c.
\end{equation}
The Weyl-ordered product $:B^a B^b:$ on the right refers to the boundary operator we assign (via a chosen quantization map) to the usual product of smooth functions. This assumes a certain definition of composite operators adopted in \cite{Cattaneo:1999fm}.

We can avoid thinking about subtleties involved in defining composite operators as follows. Start with a collection of boundary operators $B^{a_1}$, $B^{a_2}, \dots$ inserted at separate points. We are allowed to move them through each other, which amounts to commuting them using the star-commutator that follows from \eqref{star_prod}:
\begin{equation}
[B^a, B^b]_\star=-if^{abc}B^c.
\end{equation}
Now we can define composite operators by colliding a collection of B's, that is, we bring them really close together without letting them pass through each other, i.e. the ordering is preserved. The commutator of $B^a$ with such an object is determined by consecutively commuting $B^a$ with its constituents. This is precisely the recipe we would use for computing commutators in $U(\mathfrak{g}_\C)$. Therefore, we see that with the standard definition of composite operators via collision of elementary fields, we simply find the universal enveloping algebra.  Any other definition of composite operators would produce the same algebra $U(\mathfrak{g}_\C)$, possibly written in a different basis.

\textbf{Remark:} observe that the dimensionless field $B$ is related to the dimension-1 field $\phi^H$ under the correlators via
\begin{equation}
\phi^H = \frac{g^2_{\rm YM}}{4\pi i\ell}B,
\end{equation}
up to Q-exact and contact terms. So in the natural 4d normalization, we have:
\begin{equation}
[\phi^H_a, \phi^H_b]_\star = -\hbar f^{abc}\phi^H_c,\quad \hbar=\frac{g_{\rm YM}^2}{4\pi\ell}.
\end{equation}
While in three dimensions, $\frac1{\ell}$ plays the role of a natural quantization parameter $\hbar$ \cite{Dedushenko:2016jxl}, in the four-dimensional problem with the Dirichlet boundary, $\hbar$ is also proportional to the 4d gauge coupling.

From now on, we will drop the $\star$ and simply write $B^a B^b$ for $B^a\star B^b$, etc. 

\subsection{The trace}\label{sec:tr_on_U(g)}
Now let us determine the trace on $U(\mathfrak{g}_\C)$ that encodes the boundary correlators. Assume from now on that there is no theta-angle:
\begin{equation}
{\rm Re}(\tau)=0.
\end{equation}
In the case of Dirichlet boundary conditions and their S-duals, analysis generalizes to $\theta\neq 0$ without too much work, and in particular the hemisphere partition function is $\theta$-independent.\footnote{This is only true for 4d $\cN=4$ theories. With $\cN=2$ SUSY, the $\theta$-dependence would appear through the non-trivial Nekrasov partition function \cite{Nekrasov:2002qd, Pestun:2007rz}.} We will not study $\theta\neq 0$ in the rest of this paper.

Consider again the trace $T_H$ with some twist parameter $m$, and let $T_H^{m=0}$ be the same trace with the twist parameter set to zero. Then we have
\begin{equation}
\label{TH(1)}
T_H(1) =T^{m=0}_H( e^{- 2 \pi m \cdot B} ) = e^{i\pi\tau\trace(m^2)} \frac{\Delta(m)}{\mathbbm{\Delta}(m)},
\end{equation}
where the first equality follows from \eqref{mTrace}. Decompose $U(\mathfrak{g}_\C)$ in irreps of $\mathfrak{g}_\C$. As $(-1)^{F_H}=1$, the trace relations for $T_H^{m=0}$,
\begin{equation}
T_H^{m=0}[B^a,\cO]=0,
\end{equation}
tell us that $T_H^{m=0}$ only takes non-zero values on $\mathfrak{g}_\C$-invariant operators, i.e. the center $\cZ[U(\mathfrak{g}_\C)]$ of $U(\mathfrak{g}_\C)$. There is a well-known Harish-Chandra isomorphism between $\cZ[U(\mathfrak{g}_\C)]$ and $\cB_H\cong \C[\mathfrak{g}]^G\cong \C[\mathfrak{t}]^\cW$, which will play role in the next subsection.

The left hand side of \eqref{TH(1)} is a generating function for Weyl-ordered operators: 
\begin{equation}
e^{- 2 \pi m \cdot B} = \sum_{n=0}^\infty \frac{(-2 \pi)^n}{n!} m_{a_i} \cdots m_{a_n} : B^{a_1} \cdots B^{a_n}:,
\end{equation}
which form a linear basis for $U(\mathfrak{t}_\C)$, and in particular include $\C[\mathfrak{t}]^\cW$ as a subspace. Thus \eqref{TH(1)} completely encodes the data of the trace. This means we already know, somewhat implicitly, all correlation functions. 

We can make the answer more explicit with a trick related to S-duality. First of all, notice the Fourier transform\footnote{We write the integration domain as $\mathfrak{t}^\vee$ anticipating the relation to S duality below.}
\begin{equation}
\label{Fourier}
\int_{\mathfrak t^\vee} [\dd a] e^{-\frac{i\pi}{\tau} \trace(a^2)}\Delta(a)e^{2\pi i m\cdot a}=\frac{\tau^{\dim(G)/2}}{i^{{\rm rk}(G)/2}}e^{i\pi\tau\trace(m^2)} \Delta(m).
\end{equation}
A way to prove it is to relate $\Delta(a)$ to the sinh-Vandermonde $\mathbbm{\Delta}(a)$ through
\begin{equation}
\Delta(a)=\lim_{\epsilon\to 0}\frac{\mathbbm{\Delta}(\epsilon a)}{(2\pi\epsilon)^{|\Phi_+|}},
\end{equation}
and apply the Weyl denominator formula for the sinh-Vandermonde:
\begin{equation}
\mathbbm{\Delta}(a)=\sum_{w\in\cW} (-1)^{l(w)} e^{2\pi\langle w(\rho), a \rangle},
\end{equation}
where $\rho=\frac12\sum_{\alpha\in\Phi_+}\alpha$. For each summand on the right-hand side of this formula, it is completely straightforward to evaluate the Fourier integral \eqref{Fourier}, after which we apply the Weyl formula again to recover $\mathbbm{\Delta}(\tau m\epsilon)$, and then take the $\epsilon\to 0$ limit to find \eqref{Fourier}.

Using \eqref{Fourier}, we can write:
\begin{equation}
T_H(1) = \frac{i^{{\rm rk}(G)/2}}{\tau^{\dim(G)/2}} \int_{\mathfrak t^\vee} [\dd a] e^{-\frac{i \pi}{ \tau} \trace(a^2)}\Delta(a)
\frac{e^{2\pi i m \cdot a}}{\mathbbm{\Delta}(m)}.
\end{equation}
Recognizing the character of the Verma module, we can also rewrite this as 
\begin{equation}
T_H(1) = \frac{i^{{\rm rk}(G)/2}}{\tau^{\dim(G)/2}} \int_{\mathfrak t^\vee} [\dd a] e^{-\frac{i \pi}{ \tau} \trace(a^2)}\Delta(a) \trace_{V_{-i a-\rho}} e^{- 2 \pi m \cdot B}
\end{equation}
in terms of (analytically continued) traces on Verma modules $V_{-i a-\rho}$ for $U(\mathfrak{g}_\C)$. 
Taking $m$ derivatives on both sides, we learn that 
\begin{equation}
\label{verma_traces}
T_H({\mathcal O}) = \frac{i^{{\rm rk}(G)/2}}{\tau^{\dim(G)/2}} \int_{\mathfrak t^\vee} [\dd a] e^{-\frac{i \pi}{ \tau} \trace(a^2)}\Delta(a) \trace_{V_{-i a-\rho}} e^{- 2 \pi m \cdot B} {\mathcal O},
\end{equation}
i.e. we have a decomposition of our trace as a continuous superposition of the twisted traces on $U(\mathfrak{g}_\C)$ 
associated to Verma modules. 

There is another useful way to rewrite this (recall that $\Delta(w\cdot a)=(-1)^{l(w)}\Delta(a)$):
\begin{equation}
\label{Dir_Sdual}
T_H({\mathcal O}) = \frac{i^{{\rm rk}(G)/2}}{|\cW|\tau^{\dim(G)/2}} \int_{\mathfrak t^\vee} [\dd a] \underbrace{e^{-\frac{i \pi}{ \tau} \trace(a^2)}\frac{\Delta(a)}{\mathbbm{\Delta}(a)}}_{\substack{\text{Dirichlet}\\\text{partition function}}}\cdot \underbrace{\mathbbm{\Delta}^2(a)}_{\substack{\text{gauging}\\ \text{factor}}}\cdot \underbrace{\left[\sum_{w \in W} (-1)^{l(w)} \frac{\trace_{V_{-i w(a)-\rho}} e^{- 2 \pi m \cdot B} {\mathcal O}}{\mathbbm{\Delta}(a)} \right]}_{T[G] \text{ at the boundary}}.
\end{equation}

This is consistent with S-duality: we recognize the localization expression for a hemisphere partition function with 
enriched Neumann boundary conditions and boundary degrees of freedom given by the 3d theory $T[G]$ (see \cite{Benvenuti:2011ga,Gulotta:2011si,Nishioka:2011dq} for the $T[G]$ partition function). 


\subsection{Bulk-boundary map and the Harish-Chandra isomorphism}\label{sec:HC}
As was mentioned before, the trace relations imply that $T_H^{m=0}$ is supported on the $\mathfrak{g}_\C$-invariants, i.e. the center $\cZ[U(\mathfrak{g}_\C)]$. Via the Harish-Chandra isomorphism, the latter is mapped to ${\rm Sym}(\mathfrak{t}_\C)^\cW\equiv \C[\mathfrak{t}^*]^\cW$, which we identify with the space of bulk operators $\C[\mathfrak{t}]^\cW$. We are going to argue here that the bulk-boundary map, given by bringing the Q-closed bulk operators to the boundary, is precisely the Harish-Chandra isomorphism.

In fact, the decomposition of $T_H$ in terms of traces on the Verma modules \eqref{verma_traces} is precisely what we need in order to show this. Recall that an element of the center, $z\in\cZ[U(\mathfrak{g}_\C)]$, acts by a scalar on a highest-weight module $V_\lambda$,
\begin{equation}
z\cdot v = \chi_\lambda(z) v,\quad \forall v\in V_\lambda.
\end{equation}
Furthermore, these scalars, considered as polynomial functions of $\lambda$ (polynomiality is obvious), are constant along the shifted Weyl orbits, which is part of the Harish-Chandra theorem. The latter means that considered as a function of $\tilde\lambda=\lambda + \rho$, $\chi_{\tilde\lambda}(z)$ is Weyl-invariant, and can in fact be regarded as an element of $\C[\mathfrak{t}]^\cW$. Thus we get the Harish-Chandra map $\cZ[U(\mathfrak{g}_\C)]\ni z \mapsto \chi_{\tilde\lambda}(z) \in \C[\mathfrak{t}]^\cW$.

Now suppose that we choose some $\mathfrak{g}_\C$-invariant
polynomial $\hat p(B)$ and insert it under the trace in \eqref{verma_traces}. By the above reasoning, it will evaluate to some Weyl-invariant polynomial $p(-ia)$ on the Verma module $V_{-ia-\rho}$. We then find:
\begin{equation}
\label{Verma_H-C}
T_H(\hat p(B) {\mathcal O} ) = \frac{1}{|\cW|} \frac{i^{{\rm rk}(G)/2}}{\tau^{\dim(G)/2}} \int [\dd a] e^{-\frac{i \pi}{ \tau} \trace(a^2)}\Delta(a)p(- i a) \left[ \sum_{w \in W} (-1)^{l(w)} \trace_{V_{-i w(a)-\rho}} e^{- 2 \pi m \cdot B} {\mathcal O} \right].
\end{equation}

In the S-dual picture, we recognize the factor $p(- i a)$ as the insertion of bulk BPS operator. Thus the Harish-Chandra isomorphism
tells us the boundary image of bulk operators that are S-dual to invariant polynomials in the bulk scalar fields. 

The Coulomb branch algebra of  $T[G]$ is expected to be the quantization of the regular nilpotent orbit in $\mathfrak{g}_\C$ 
or deformations thereof: it is the central quotient $U(\mathfrak{g}_\C)/I_a$, where $I_a$ is the ideal that fixes the value of the 
central elements $\hat p(B)$ in $\cZ[U(\mathfrak{g}_\C)]$ to $p(- i a)$. Here $a$ are the mass parameters in $T[G]$. 
The term in square brackets in the integral is ($\mathbbm{\Delta}(a)$ times) the trace $T_C$ in $T[G]$.

We see here the first example of a general principle: the algebra $\cA_C$ for a boundary condition has a center isomorphic to 
$\C[\mathfrak{t}]^\cW$, and the central quotient $\cA_C[a]$ of $\cA_C$ is a 3d Coulomb branch. The trace on $\cA_C$
is reconstructed as above by integrating the trace on $\cA_C[a]$ against an appropriate Gaussian measure. 

\paragraph{On $\mathfrak{g}_\C=\mathfrak{gl}_N$}
Let us give some more details in the case when the 4d gauge group is $U(N)$, so in particular $\mathfrak{g}_\C=\mathfrak{gl}_N$. The center of $U(\mathfrak{gl}_N)$ is formed by gauge-invariant polynomials that can be simply constructed from traces,
\begin{equation}
\trace(B^n).
\end{equation}
The Harish-Chandra isomorphism is {\it not} the naive identification of these with Weyl-invariant polynomials of the eigenvalues. It is a non-trivial deformation of that. To give a few examples of pairs $(\hat{p}, p)$, we write:
\begin{align}
\label{HC_examples}
\hat{p}(B) &\longmapsto p(-ia)\cr
\trace(B) &\longmapsto -i\sum_{j=1}^N a_j,\cr
\trace(B^2) &\longmapsto -\sum_{j=1}^N a_j^2 + \frac{N(N^2-1)}{12},\cr
\trace(B^3) &\longmapsto \sum_{j=1}^N \left(ia_j^3 +i\frac{N}{2}a_j^2 - i\frac{N^2-1}{4}a_j\right) - \frac{i}2 \left(\sum_j a_j \right)^2 - \frac{i N^2(N^2-1)}{24},\cr
\end{align}
which are all symmetric polynomials in $a_i$ as expected.

A simple way to derive $p(-ia)$ is to use the formula for $\mathfrak{g}_\C=\mathfrak{gl}_N$ (see, e.g. \cite{Umeda}):
\begin{equation}
\label{newtons}
1+i \Tr  \frac{1}{z-B} = \frac{\cC(z+i)}{\cC(z)},
\end{equation}
for a degree $n$ monic polynomial $\cC(z)$. The polynomial $\cC(z)$ itself can be computed as a {\it Capelli determinant} \cite{Capelli} of $z-B$,
which is a quantized analogue of the characteristic polynomial of $B$. It has the property that $\cC(B)=0$. Indeed, if we multiply the above expression by $\cC(z)$, we find 
\begin{equation}
\cC(z)+i \Tr  \frac{\cC(z)-\cC(B)}{z-B} +i \Tr  \frac{\cC(B)}{z-B}= \cC(z+i)
\end{equation}
so the coefficients $\Tr \cC(B) B^n$ of negative powers of $z$ must vanish. 

The important property of the Capelli determinant is that under the Harish-Chandra map, it has a simple expression, which in our conventions is $\prod_{j=1}^N (z-\lambda_j + i(j-1))$, where $\lambda_j$ parameterize the highest weight of the Verma module. For us, $\lambda=-ia -\rho$, and we find that that Harish-Chandra image of $\cC(z)$ is
\begin{equation}
\cC(z)=\prod_{j=1}^N (z + i a_j +\frac{i}{2}(N-1)).
\end{equation}
Combining this with \eqref{newtons} allows to write formulas like \eqref{HC_examples} straightforwardly.

\section{Neumann boundary conditions and their enrichment}\label{sec:Neum}
The half-BPS Neumann boundary conditions \eqref{NeuSca}, possibly deformed by the boundary FI term \eqref{bdy_sphere_FI}, and possibly enriched by the boundary 3d $\cN=4$ SCFT, form another well-known large class of boundary conditions. Without the boundary SCFT, it is quite easy to determine both $\cA_H$ and $\cA_C$. Because the boundary values of $\vec{X}$ are fixed, and the only operator obeying $E=R_H$ is $X_+ = X_1 + i X_2$, there are no non-trivial $\rQ^H_{1,2}$-closed boundary operators, and we simply have:
\begin{equation}
\cA_H=\C,
\end{equation}
with the trace given by the $HS^4$ partition function with Neumann boundary conditions:
\begin{equation}
T_H(1) = \frac1{|\cW|}\int_{\mathfrak t} [\dd a]\, \Delta(a) \mathbbm{\Delta}(a) e^{i\pi\tau\trace(a^2)}.
\end{equation}
Such boundary conditions are not compatible with the bulk $\theta$-term, except in the abelian case, which is why we set:
\begin{equation}
\theta=0,
\end{equation}
but still use the 4d coupling $\tau$ for convenience. Neumann boundary conditions allow for a boundary FI term $r$, which amounts to the insertion of $e^{2\pi i \trace(ra)}$ under the integral, where $r$ is valued in the abelian part of $\mathfrak{g}$, so $\trace(ra)$ picks out the components of $a$ in the abelian directions. Using the same trick based on the Weyl denominator formula, we find:
\begin{equation}
T_H(1) = \frac1{|\cW|}\int_{\mathfrak t} [\dd a]\, \Delta(a) \mathbbm{\Delta}(a) e^{2\pi i \trace(ra)} e^{i\pi\tau\trace(a^2)}=\frac{i^{{\rm rk}(G)/2}}{\tau^{{\rm rk}(G)/2}} e^{-\frac{i\pi}{\tau}\trace(r^2-\rho^2 -2ir\rho)}\Delta\left(\frac{i}{\tau}\rho-\frac{r}{\tau}\right),
\end{equation}
where $\trace(\rho^2)=\frac{h}{12}\dim(\mathfrak{g})$ by the Freudenthal-de Vries formula.

The nontrivial $\rQ^C_{1,2}$-closed boundary operators are built from the (twisted translations of) $Y_+=Y_1+iY_2$, which is dynamical at the boundary. Unbroken gauge invariance implies that the boundary operators are generated from building blocks of the form
\begin{equation}
\trace(Y_+^n)(0),
\end{equation}
and their twisted translations along the $x^3$ direction.
Such operators are well-defined both at the boundary and in the bulk, where they correspond to $\trace(\phi^C)^n$ from Section \ref{sec:mag_bulk}, thus form a commutative algebra, which, as we know by now, is the center of $U(\mathfrak{g}_\C)$. In other words, we can write
\begin{equation}
\cA_C = \C[\mathfrak{g}]^{\mathfrak{g}}\cong \C[\mathfrak{t}]^\cW,
\end{equation}
and a trace of some polynomial $P(Y_+)\in \C[\mathfrak{t}]^\cW$ is simply given by\footnote{$Y_+$ has dimension one, so it evaluates to $\frac1{\ell}a$, but we neglect obvious factors of $\ell$ for brevity.}
\begin{equation}
T_C(P(Y_+)) = \frac1{|\cW|}\int_{\mathfrak t} [\dd a]\, \Delta(a) \mathbbm{\Delta}(a) e^{2\pi i \trace(ra)} e^{i\pi\tau\trace(a^2)}P(a).
\end{equation}
This latter formula describes the one-point function of bulk operators.
\subsection{Adding boundary theory}
Now we enrich the Neumann boundary conditions by coupling to the boundary SCFT $\cT$ with a global symmetry $G$. Coupling is realized via gauging this global symmetry by the bulk vector multiplet. Sometimes it is useful to think about this boundary condition slightly differently: we start with the Dirichlet boundary condition (with global symmetry $G$), place theory $\cT$ at the boundary (that has another copy of global symmetry $G$), and gauge the diagonal subgroup of $G\times G$ by 3d vector multiplets living at the boundary. The corresponding hyper-K\"ahler moment map involved in this gauging is given in the equation \eqref{bdy_FI}. Suppose the theory $\cT$ itself has protected algebras
\begin{equation}
\cA_H(\cT)\quad \text{and}\quad \cA_C(\cT).
\end{equation}

\subsubsection{$\cA_H$ algebra and the trace}\label{sec:AH_Neum}
To describe the $\cA_H$ algebra of the coupled bulk-boundary system, the perspective of gauging the ${\rm Diag}(G\times G)$ is quite useful. Indeed, the algebra for [Dirichlet]$\otimes\cT$ is
\begin{equation}
\label{TotimesU}
\cA_H(\cT)\otimes U(\mathfrak{g}_\C),
\end{equation}
and the 3d gauging simply implements the quantum Hamiltonian reduction of \eqref{TotimesU}. From the localization, this comes about as follows: the bulk gives a 2d cYM on the hemisphere $HS^2$, the boundary theory $\cT$ produces a 1d topological quantum mechanics (TQM) living at the boundary $\partial(HS^2)$, and the 1d gauging couples these two systems, resulting in the quantum Hamiltonian reduction. The moment map constraint involved in this is
\begin{equation}
\mathbbm{m} \equiv \mu + X - r =0,
\end{equation}
where $\mu$ generates the $\mathfrak{g}_\C$-action on $\cA_H(\cT)$ and corresponds to the twisted translation of $\mu_+(0)=\mu_1(0) + i \mu_2(0)$ in the 3d theory $\cT$, and $X\in \mathfrak{g}_\C$ likewise corresponds to the twisted-translated version of $X_+(0)$ at the Dirichlet boundary of the 4d theory. The quantum Hamiltonian reduction reads
\begin{equation}
\cA_H = (\cA_H(\cT)\otimes U(\mathfrak{g}_\C)/(\mathbbm{m}))^{\mathfrak{g}_\C},
\end{equation}
where the quotient is over the left ideal generated by $\mathbbm{m}$, and then we pass to the $\mathfrak{g}_\C$-invariants (which is the same as $\mathfrak{g}$-invariants). Because $\mathbbm{m}$ is linear in $X$, taking the quotient is equivalent to eliminating the factor of $U(\mathfrak{g}_\C)$. Thus it only remains to take the subalgebra of invariants, and our answer is simply
\begin{equation}
\cA_H = (\cA_H(\cT))^{\mathfrak{g}}.
\end{equation}
We see that while gauging in 3d corresponds to the quantum Hamiltonian reduction of the protected algebra, gauging at the boundary of 4d is realized via a ``half'' of the same procedure, -- simply taking the subalgebra of invariants. One can compare this to the procedure in \cite{Costello:2020ndc}, where the boundary VOA for Neumann boundary conditions is computed by taking the derived invariants only, while the same problem in pure 2d setting requires performing Hamiltonian reduction (BRST reduction of the VOA) \cite{Dedushenko:2017osi}.

It is rather straightforward to write the twisted trace on $\cA_H$, given we know the twisted trace on $\cA_H(\cT)$ that follows from the sphere correlators for $\cT$. Let $T_H^{a}$ denote the twisted trace on $\cA_H(\cT)$, with masses $a$ for the $G$ symmetry turned on (and other parameters, such as masses $m$ for other flavor symmetries and, perhaps, FI terms kept implicit). We assume that it is normalized to the $S^3$ partition function of $\cT$:
\begin{equation}
T_H^{a}(1) = Z_{S^3}[\cT](a).
\end{equation}
From the 2d/1d viewpoint, $T^a_H$ describes partition function and correlators of the 1d TQM associated with the 3d theory $\cT$. We then couple it to the cYM on $HS^2$.
Again the perspective of gauging ${\rm Diag}(G\times G)$ turns out to be quite useful, and allows to write for $\cO\in (\cA_H(\cT))^{\mathfrak{g}}$:
\begin{equation}
T_H(\cO) = \frac1{|\cW|} \int_{\mathfrak t}[\dd a]\, \underbrace{T^{a}_H(\cO)}_{\substack{\text{trace on }\cA_H(\cT)}} \cdot \underbrace{e^{2\pi i \trace(ra)}\mathbbm{\Delta}(a)^2}_{\substack{\text{gauging}\\ {\rm Diag}(G\times G)}} \cdot \underbrace{e^{i\pi\tau\trace(a^2)}\frac{\Delta(a)}{\mathbbm{\Delta}(a)}}_{\text{Dirichlet partition function}}.
\end{equation}
We see that while $\cA_H$ does not, $T_H$ in general does depend on the boundary FI term $r$. Both $\cA_H$ and $T_H$ might also depend on FI parameters of the boundary theory $\cT$; the masses of $\cT$ also enter $T_H$ as the twist parameters, though we kept them implicit.

\subsubsection{$\cA_C$ algebra and trace}\label{sec:AC_Neum}
The above discussion makes it straightforward to describe the $\cA_C$ algebra at the boundary, especially since we have essentially alluded to the answer in Section \ref{sec:red_3d}, and later in \eqref{Dir_Sdual}-\eqref{Verma_H-C}. The operators in the Coulomb branch algebra $\cA_C(\cT)$ are neutral under the flavor symmetry $G$, so they remain in the algebra upon gauging. Additionally, we know that gauge-invariant operators built from $Y_+$, which form a copy of $\C[\mathfrak{t}]^\cW$, are also in the $\rQ^C_{1,2}$-cohomology, so we should extend $\cA_C(\cT)$ by such operators. Notice that $Y_1=a$ is the variable we integrate over as we couple bulk to the boundary, and meanwhile, this $a$ enters as a mass parameter in the algebra $\cA_C(\cT)$. Adjoining $\C[\mathfrak{t}]^\cW$ to the algebra $\cA_C(\cT)$ therefore simply means that we promote masses $a$ to dynamical fields. This is manifestly reflected in the fact that we integrate over $a$. Mathematically, this is a central extension:
\begin{equation}
0 \longrightarrow \C[\mathfrak{t}]^\cW \longrightarrow \cA_C \longrightarrow \cA_C(\cT) \longrightarrow 0.
\end{equation}
The first map here is just an inclusion, while the second map is a central quotient that sets $a$ to a constant value. This is (S-dual of) the same central quotient that was briefly discussed in the text after the equation \eqref{Verma_H-C}.

If we denote the trace on $\cA_C(\cT)$ as $T_C^a$, again only making the mass $a$ manifest in the notation, the trace on $\cA_C$ is written just like in Section \ref{sec:red_3d}:
\begin{equation}
\label{trace_AC}
T_C[\cO] =\frac1{|\cW|}\int_{\mathfrak t} [\dd a]\, T^a_C[\cO]\cdot  e^{2\pi i \trace(ra)}\mathbbm{\Delta}(a)^2 \cdot  e^{i\pi\tau\trace(a^2)}\frac{\Delta(a)}{\mathbbm{\Delta}(a)}.
\end{equation}
The boundary FI parameter $r$ can only affect the trace, not the algebra, but it does not introduce any new twists because the boundary monopoles (that would be charged under the corresponding ``topological symmetry'') do not exist in this setting. In the equation \eqref{trace_AC}, the insertion can be $\cO\in \cA_C(\cT)$, or it can be a central ``bulk operator'' from $\C[\mathfrak{t}]^\cW$ given by a Weyl-invariant polynomial
\begin{equation}
P(a)\in \C[\mathfrak{t}]^\cW.
\end{equation}

\subsubsection{Basic example of S-duality}
As an illustrative example, consider again the $T[G]$ boundary conditions. As a 3d theory, $T[G]$ has the nilpotent cone of $\mathfrak{g}$ for its Higgs branch, and the nilpotent cone of the Langlands dual $\mathfrak{g}^\vee$ for its Coulomb branch. Correspondingly, the TQM sectors describe quantizations of these branches, which are the central quotients $U(\mathfrak{g}_\C)/I_a$ and $U(\mathfrak{g}_\C^\vee)/I_b^\vee$ respectively.

As we couple $T[G]$ to the bulk $G$ gauge theory, we obtain algebras $\cA_H$ and $\cA_C$ at the boundary. They are computed according to our recipe, which gives
\begin{equation}
\cA_H \cong \left( U(\mathfrak{g}_\C)/I_a \right)^{\mathfrak{g}} \cong \C,
\end{equation}
because the $\mathfrak{g}$-invariants belong to the center, which is removed by taking the central quotient. The boundary algebra $\cA_C$ is obtained by promoting $b$ in $U(\mathfrak{g}_\C^\vee)/I_b^\vee$ to the dynamical field, which simply undoes the quotient, and we find
\begin{equation}
\cA_C\cong U(\mathfrak{g}^\vee_\C).
\end{equation}
The above answers for $\cA_H$ and $\cA_C$ of curse match those for $\cA_C$ and $\cA_H$ respectively at the Dirichlet boundary conditions in the $G^\vee$ gauge theory, as studied in the previous Section. The trace on $\cA_H=\C$ is given by the hemisphere partition function, and it obviously matches the dual answer. That the trace on $\cA_C= U(\mathfrak{g}^\vee_\C)$ matches the S-dual answer has already been observed in the equation \eqref{Verma_H-C} and the discussion after it.

To make things more explicit, consider the case of $T[SU(2)]$, i.e. SQED$_2$, coupled to the $SU(2)$ theory in the bulk. The Coulomb branch algebra is generated by $v_\pm, \varphi$, and it contains a mass parameter $m$, which we promote to the dynamical field upon coupling to the bulk. With respect to the bulk gauge symmetry, this is the \emph{abelianized} algebra, because $m$ is not invariant under the Weyl group of $SU(2)$, but $m^2$ is. The actual gauge-invariant algebra is obtained by throwing away $m$ but keeping $m^2$. While the classical relation in this algebra is $v_+ v_- = \varphi^2 - m^2$ \cite{Bullimore:2015lsa}, the quantum relations are:
\begin{equation}
v_+ v_- = \left(\varphi-\frac12 \epsilon\right)^2-m^2, \quad v_- v_+ = \left( \varphi + \frac12 \epsilon\right)^2 - m^2,\quad \text{where } \epsilon=\frac1{2\ell}.
\end{equation}
Either way, the relations simply eliminate $m^2$ from the algebra, and the remaining generators $v_\pm, \varphi$ only satisfy the following commutation relations
\begin{equation}
[v_+, v_-]=-2\epsilon\varphi,\quad [\varphi, v_\pm]=\pm\epsilon v_\pm.
\end{equation}
These are of course commutation relations of $\mathfrak{su}(2)$, so we indeed recover the algebra $\cA_C=U(\mathfrak{g}_\C)=U(\mathfrak{sl}_2)$ at the boundary.

\subsection{Ungauging and interval reduction}\label{sec:ungauge}
It is well-known that one can recover the boundary theory $\cT$ by an interval compactification \cite{Gaiotto:2008ak}, where on one end we use $\cT$ to define the boundary conditions, while the other end supports the Dirichlet boundary conditions. In the IR, this effectively ungauges the boundary, and we recover the $\cT$ SCFT.

Notice an interesting feature: while the algebra $\cA_H(\cT)$ decreases upon coupling to the bulk, the algebra $\cA_C(\cT)$ increases in size. The latter is related to the statement made in Section \ref{sec:red_3d}, which asserts that one can go back to $\cA_C(\cT)$ from $\cA_C$ by taking a central quotient, which is the same as the interval compactification with the Dirichlet boundary conditions on the other end. While going from $\cA_C(\cT)$ to $\cA_C$ involved promoting masses $a$ to dynamical variables, going back, obviously, corresponds to giving $a$ a fixed value, as already explained in Section \ref{sec:red_3d}, and this is exactly what the Dirichlet boundary conditions achieve.

Since $\cA_H$ is ``smaller'' than $\cA_H(\cT)$, understanding how ungauging works in this case is more interesting. As we reduce on the interval, we need to take into account the following:
\begin{enumerate}
	\item Algebra $\cA_H=\cA_H(\cT)^{\mathfrak g}$ on the left boundary;
	\item Algebra $U(\mathfrak{g}_\C)$ on the right (Dirichlet) boundary;
	\item Wilson lines stretched between the two boundaries constructed using the $\rQ^H_{1,2}$-closed gauge field $\mathscr{A}_4^H$ introduced in Section \ref{sec:elec_bulk}.
	\item Fermionic lines $\mathscr{L}_{1,2}$ sretched between the two boundaries that are cohomological descendants of the $\rQ^H_{1,2}$-closed bulk field $\phi^H$ introduced in Section \ref{sec:elec_bulk}.
\end{enumerate}
The fermionic lines from the item four are not $\rQ^H_{1,2}$-closed, but rather the descent equations imply (similar to \cite{Costello:2020ndc}) that
\begin{equation}
\rQ^H_{1,2} \mathscr{L}_{1,2} = \phi^H\big|_{\rm Left}^{\rm Right}.
\end{equation} 
On the Dirichlet boundary, the restriction of $\phi^H$ generates the boundary algebra $U(\mathfrak{g}_\C)$. On the other end, the Neumann boundary conditions imply that $\phi^H\big|=0$ there. The above equation therefore makes all operators in $U(\mathfrak{g}_\C)$ $\rQ^H_{1,2}$-exact, and completely eliminates the $U(\mathfrak{g}_\C)$ algebra.

What remains is the algebra $\cA_H=\cA_H(\cT)^{\mathfrak g}$ (item number one on the list) and Wilson lines (item number three). While the algebra $\cA_H$ only contains $\mathfrak{g}$-invariant operators from $\cA_H(\cT)$ due to gauge invariance, the Wilson lines are allowed to end on $\mathfrak{g}$-charged operators from $\cA_H(\cT)$. Such configurations of Wilson lines stretched between the two boundaries, with a charged operator from $\cT$ sitting at one end of the line, do contribute to the cohomology. They can be interpreted as giving non-trivial $\cA_H(\cT)^{\mathfrak g}$-modules. We therefore claim that $\cA_H(\cT)$ can be recovered from $\cA_H=\cA_H(\cT)^{\mathfrak g}$ as an extension by modules corresponding to such Wilson lines. This is an infinite extension, since charges of operators sitting at the endpoints of Wilson lines can be arbitrarily large:
\begin{equation}
\cA_H(\cT) = \text{infinite extension of } \cA_H.
\end{equation}
Alternatively, we could go to the S-dual picture, and recover the boundary algebra there using the $\cA_C$ prescription described above. The dual boundary theory is $T[G]\circ\cT$, which is obtained by gauging the diagonal $G$ symmetry in $T[G]\times \cT$. By specializing the center of the boundary algebra $\cA_H$ (which is the $\cA_C$ of the S-dual picture) to fixed values parametrized by $a$, we should recover $\cA_C(T[G]\circ \cT)$. This in particular implies that the trace on the boundary algebra $\cA_H$ of the original theory can be expressed through the trace on $\cA_C(T[G]\circ\cT)$ on the dual side according to:
\begin{equation}
\langle \cO \rangle_{{\rm Neumann }+\cT} =\frac1{|\cW|}\int[\dd a]\, \Delta(a)\mathbbm{\Delta}(a) e^{-\frac{i\pi}{\tau}\trace(a^2)}\langle \cO\rangle_{T[G]\circ\cT}(a).
\end{equation}
Notice that $Z_{T[G]\circ\cT}(a)$ is a character for the $\cA_H$ such that the center is given in terms of $a$. We do not pursue this approach any further.

\section{Nahm poles and finite W-algebras}\label{sec:finiteW}
To have answers for the whole class of boundary conditions introduced in \cite{Gaiotto:2008sa}, we must understand the algebras and traces for Nahm pole boundary conditions as well. Recall from Section \ref{sec:bndry_cond} that the Nahm pole is parameterized in terms of an embedding
\begin{equation}
\varrho: \mathfrak{su}(2) \to \mathfrak{g},
\end{equation}
and we fix a choice of such $\varrho$ in this section. Being a generalization of the Dirichlet boundary conditions, Nahm poles also break gauge symmetry at the boundary. While in the case of Dirichlet there is a remnant $G$ global symmetry at the boundary, the Nahm pole further breaks this global $G$ to a subgroup $F_\varrho$ that commutes with the pole, i.e. the centralizer of $\varrho(\mathfrak{su}_2)$:
\begin{equation}
\label{flavor_rho}
F_\varrho = C_G(\varrho(\mathfrak{su}_2)).
\end{equation}
For the global symmetry $F_\varrho$, like in the Dirichlet case, we can turn on the boundary mass, which then enters as the twist parameter in the trace.

Nahm poles also have a few completely novel features, such as modified R-symmetry at the boundary, as we discuss momentarily, and additional restrictions on the allowed boundary operators that basically follow from finiteness of the action.

\subsection{R-symmetry mixes with gauge symmetry at the Nahm pole}\label{sec:Rmixing}
Since the fields $\vec{X}$ form a triplet of what we previously called $SU(2)_H$, the Nahm pole 
\begin{equation}
\vec{X} \sim \frac{\vec{t}}{y},
\end{equation}
whenever non-trivial, explicitly breaks $SU(2)_H$. Superconformal invariance, on the contrary, requires this symmetry to be there. The resolution comes from the fact that
\begin{equation}
\varrho(\mathfrak{su}_2)\subset \mathfrak{g}
\end{equation}
integrates to $SU(2)_\varrho\subset G$ (or $SO(3)_\varrho$), which is also explicitly broken at the boundary. Crucially, there exists a subgroup of $SU(2)_H\times SU(2)_\varrho$ that remains unbroken. It plays the role of Higgs branch R-symmetry at the boundary, and we call it $\tilde{SU(2)}_H$. Let us choose a basis of the gauge algebra $\mathfrak{g}={\rm Lie}(G)$ as $(t_1, t_2, t_3, T_a)$, where the $T_a$ span the orthocomplement of $\varrho(\mathfrak{su}_2)$ in $\mathfrak{g}$. Then 
\begin{equation}
X_i = \sum_{j=1}^3 X_i^j t_j + \sum_{a} X_i^a T_a.
\end{equation}
The Nahm pole keeps $X_i^a$ finite, while $X_i^j$ is singular,
\begin{equation}
\label{NP_diag}
X_i^j \sim \frac{\delta_i^j}{y}.
\end{equation}
Here, the lower index is a triplet under the $SU(2)_H$, while the upper index is a triplet under the $SU(2)_\varrho$. The diagonal subgroup preserves the form of the singularity \eqref{NP_diag}, which makes it the unbroken subgroup we are looking for:
\begin{equation}
\tilde{SU(2)}_H \cong {\rm Diag}\left[ SU(2)_H\times SU(2)_\varrho \right].
\end{equation}
One might find a different global form if we only had the $SO(3)_\varrho$ to begin with, but its Lie algebra is really all we need here.

Besides $\tilde{SU(2)}_H$, there also exists the flavor symmetry group $F_\varrho$ defined in \eqref{flavor_rho}, which is preserved at the Nahm pole as well. It cannot further mix with the $\tilde{SU(2)}_H$ without breaking part of symmetry (as is always the case for non-abelian R-symmetry), thus $\tilde{SU(2)}_H$ must be the correct R-symmetry. 

One can also understand this fact from the SUSY algebra closure \eqref{closure}. Normally, the SUSY algebra closes up to a gauge transformation $\cG_\Lambda$, as written in \eqref{closure} (and up to equations of motion, which is not important to us here). This gauge transformation is trivial for gauge-invariant bulk operators, but not trivial for boundary operators that can be charged under $G$. This usually produces central charges that are proportional to boundary masses, and indeed this happens for the flavor symmetry $F_\varrho$. One can check, however, that the Nahm pole produces additional pieces in the boundary limit of $\cG_\Lambda$. They give a boundary gauge transformation with the gauge parameter proportional to
\begin{equation}
\varepsilon_{ijk} \lim_{y\to 0}X_k(y) y = \varepsilon_{ijk}t_k,
\end{equation}
whenever the right hand side of \eqref{closure} also includes an R-symmetry $R_{ij}$. Together these two pieces combine into a generator of $\tilde{SU(2)}_H$, which replaces $SU(2)_H$ when acting on boundary operators charged under $\varrho(\mathfrak{su}_2)\subset \mathfrak{g}$.

This redefinition of the R-symmetry group $SU(2)_H$ at the boundary has some interesting consequences. It means that the R-charges of boundary operators can be shifted from their naive values. Recall that the conformal dimensions and R-charges of operators must obey certain inequalities implied by unitarity, such as $E\geq R_H$ and $E\geq R_C$.\footnote{Superconformal primaries obey $E\geq R_H+R_C$ \cite{Chester:2014mea}, which is a stronger inequality if we consider the $SU(2)_H$ and $SU(2)_C$ highest weights.} It can happen that these inequalities are broken by the redefined R-charges. It is enough to consider the highest weight with respect to both $\tilde{SU(2)}_H$ and $SU(2)_C$: if it happens to violate the inequality
\begin{equation}
E \geq \tilde{R}_H + R_C,
\end{equation}
it means that the whole $\tilde{SU(2)}_H\times SU(2)_C$ multiplet is inconsistent with unitarity. What does it mean? It simply means that the corresponding multiplet is not part of the theory. In other words, the corresponding boundary operators do not exist. This situation is not unheard of: it was already pointed out in \cite{Gaiotto:2008sa} that the singular nature of the Nahm pole boundary conditions imposes certain additional constraints on the behavior of fields near the boundary, such that the action remains finite. These constraints must be precisely such that they remove all the operators that violate the unitarity bounds. This makes the identification of boundary operators subtle and interesting, which is what we address next.

\subsection{Boundary operators at the Nahm pole}\label{sec:bdyOpNahm}
The 4d action diverges in the presence of the Nahm pole. This is typical of disorder observables, and the standard cure is to excise a size-$\delta$ tubular neighborhood of the divergent locus, --- a boundary in our case, --- and include a boundary term that cancels the divergence.\footnote{See \cite{Mazzeo:2013zga}, where the boundary term was also included in the presence of the Nahm pole. For similar discussion in the case of monopole operators in 3d, see \cite{Dedushenko:2017avn}.} In our case, the leading divergence is of order $\delta^{-3}$, and it can be canceled by the boundary term that slightly modifies the action. The relevant terms in the action become:
\begin{equation}
\label{S_div}
S_{\rm div} \sim \trace \int\dd^4x \left( \left(\cD_y X_j + \frac{i}{2}\varepsilon_{jkl}[X_k, X_l]  \right)^2 - [X_j, Y_k] + {\rm fermions} \right).
\end{equation}
Now the action is less divergent, but still appears to have subleading $\delta^{-1}$ divergences. These cannot be removed, and in fact play a rather different role: they impose further restrictions on the behavior of fields at the boundary. It is clear that fields commuting with the pole are unconstrained, while others must obey certain additional restrictions.

Analyzing the field space in full detail is a cumbersome task, but fortunately we can almost avoid it using the results of the previous section. First we notice that the $\rQ^C_{1,2}$ cohomology at the Nahm pole is trivial, meaning that 
\begin{equation}
\cA_C=\C.
\end{equation}
Indeed, the only operator that obeys $E=R_C$ is $Y_+=Y_1 + i Y_2$, and it is not affected by the shift of R-charges. Because the Nahm pole boundary conditions, just like the Dirichlet, fix values of $Y_i$ at the boundary, we do not have any boundary operators in the $\rQ^C_{1,2}$ cohomology, besides the identity. The trace given by the hemisphere partition function will be addressed later.

To describe boundary operators in the $\rQ^H_{1,2}$ cohomology, we need to find those obeying
\begin{equation}
E= \tilde{R}_H.
\end{equation}
Furthermore, as we know, they must have $R_C=0$, be invariant under the boundary rotations that fix the $x^3$ axis, and be $\tilde{SU(2)}_H$ highest weights. This eliminates $Y_i$, fermions, and most of the gauge field components, except $F_{34}$. From the earlier analysis, we know that the $\rQ^H_{1,2}$-closed operator built from the $F_{34}$ and scalars is cohomologous to $\phi^H$, so we can simply ignore it. Therefore all the independent boundary operators can be constructed from the $X_i$ and derivatives.

Let us split $X_i$ into a singular and regular part:
\begin{equation}
X_i = \frac{t_i}{y} + \hat{X}_i.
\end{equation}
It is the regular part $\hat{X}_i$ that is used to construct boundary operators. Because they must be highest weight vectors of $\tilde{SU(2)}_H = {\rm Diag}[SU(2)_H \times SU(2)_\varrho]$, they are clearly highest weights both with respect to $SU(2)_H$ and $SU(2)_\varrho$. The former means that we are only interested in $\hat{X}_+ = \hat{X}_1 + i\hat{X}_2$, like before; the latter means that we should only consider gauge components of $\hat{X}_+$ valued in the subspace of highest weights with respect to $\varrho(\mathfrak{su}_2)$, which we denote as
\begin{equation}
P_+ \subset \mathfrak{g}.
\end{equation}
The subspace $P_+$ can be further decomposed into a subspace of zero weights with respect to $\varrho(\mathfrak{su}_2)$, weights $1/2$, $1$, etc. We write it as:
\begin{equation}
P_+ = \bigoplus_{m\in\frac12 \Z_{\geq 0}} P_{+,m}.
\end{equation}
The subspace $P_{+,0}$ of zero highest weights is clearly the same thing as the subalgebra $\mathfrak{f}_\varrho = {\rm Lie}(F_\varrho)$ commuting with $\varrho(\mathfrak{su}_2)$. Components of $\hat{X}_+$ valued in this space have $\tilde{R}_H = R_H = 1$, and because $\hat{X}_+$ has $E=1$, the equality $E=\tilde{R}_H$ is obeyed. Such components are indeed the allowed boundary local operators, as we can also see from the action: singularity does not affect the operators commuting with $\varrho(\mathfrak{su}_2)$.

For the highest weight vectors of positive weight, the inequality $E\geq \tilde{R}_H$ is broken. Indeed, if we focus on the subspace $P_{+,m}$ with $m>0$, then components of $\hat{X}_+$ valued in it have
\begin{equation}
\tilde{R}_H = R_H + m = 1+m>1,
\end{equation}
yet they still have $E=1$. Such components of $\hat{X}_+$ violate the unitarity bounds, thus the corresponding operators must not be part of the theory. In practice, it means that the Nahm pole must force such components to vanish at the boundary. Clearly all their derivatives along the boundary must vanish too.

Normal derivatives of $\hat{X}_+$, however, might not vanish. It is convenient to choose a partial gauge, in which
\begin{equation}
A_y=0
\end{equation}
in a neighborhood of the boundary. The reason is that $\hat{X}_+$ has unusual gauge transformations: indeed, it is defined by subtracting a gauge-noninvariant pole from a gauge-covariant quantity $X_+$. Thus taking covariant derivatives of $\hat{X}_+$ is a little cumbersome.\footnote{For a simple illustration, consider $\hat{X}_+$ itself, whose gauge transformation, due to the pole subtraction, is $\hat{X}_+ \mapsto \hat{X}_+ + i[\epsilon, \hat{X}_+] + i [\epsilon/y, t_+]$. Since $\epsilon\big|_{y=0}=0$, and its derivative is regular, we may write $\frac{\epsilon}{y}\big|_{y=0}=\partial_y\epsilon\big|_{y=0}$. Thus $\hat{X}_+\big|_{y=0}$ transforms according to $\hat{X}_+\big|_{y=0} \mapsto (\hat{X}_+ + i[\partial_y\epsilon,t_+])\big|_{y=0}$. The gauge-invariant boundary operator may be written as $(\hat{X}_+ - i A_y)\big|_{y=0}$, which coincides with $\hat{X}_+\big|_{y=0}$ in the gauge $A_y=0$. Taking normal derivatives of $\hat{X}_+$ is straightforward in this gauge, while trying to write them in a gauge-covariant way is inconvenient.} In the gauge $A_y=0$, though, we can take ordinary normal derivatives. In particular, if $m$ is an integer, then acting with the normal derivative $m$ times results in an operator that obeys $E=\tilde{R}_H$. Indeed,
\begin{equation}
(\partial_y)^m \hat{X}_+
\end{equation}
has dimension $E=1+m$. Thus its components valued in $P_{+,m}$ obey $E=\tilde{R}_H$, and may be valid boundary operators in the cohomology, if non-zero. This can be generalized to $m$ half-integral as well, if we write the answer in the following form. Let the boundary operator corresponding to the $\varrho(\mathfrak{su}_2)$ highest weight $v_\alpha\in P_+$ be denoted as $\hat{X}_+^{(\alpha)}$, and let its $\varrho(\mathfrak{su}_2)$ weight be $m_\alpha$. Then all such operators can be combined into the following $\mathfrak{g}$-valued and operator-valued function (not a field -- see below),
\begin{equation}
\label{X(y)_NP}
\hat{X}_+(y) = \sum_{v_\alpha\in P_+} \hat{X}_+^{(\alpha)} v_\alpha y^{m_\alpha},
\end{equation}
which makes it manifest that while $\hat{X}_+(y)$ has $E=1$, the coefficients $\hat{X}_+^{(\alpha)}$ have $E=\tilde{R}_H=1+m_\alpha$. Expansion \eqref{X(y)_NP} is simply a small piece of the more general expansion of $\hat{X}_+(y)$ in $y$, which has many more other terms, yet \eqref{X(y)_NP} contains the data of $\rQ^{H}_{1,2}$ cohomology at the boundary.

So far we have proven that all the boundary operators are generated by the coefficients in \eqref{X(y)_NP}. Logically, it is still a possibility that some of them vanish due to constraints imposed by the Nahm pole. We can actually do better and prove that all $\hat{X}_+^{(\alpha)}$ from \eqref{X(y)_NP} are non-trivial.

For that let us identify field configurations on which the operators from \eqref{X(y)_NP} are supported. In fact, we are in a rare situation where we have to distinguish \emph{operators} (we also occasionally call them \emph{observables}) and \emph{fields}, since these are elements of the dual spaces. For us the operators in the cohomology must be the $SU(2)_\varrho$ highest weights, and these are what we have in \eqref{X(y)_NP}. The dual fields on which such operators are supported, on the other hand, are the $SU(2)_\varrho$ lowest weights. The relevant field configurations, with the pole included, have the following form:
\begin{equation}
\label{NPsol}
X_+(y) = \frac{t_+}{y} + \sum_{v_\alpha \in P_-} x^{(\alpha)} v_\alpha y^{-m_\alpha},
\end{equation}
where now we sum over the lowest weights $v_\alpha\in P_-$. The succinct way to think about it is as follows: coefficients $\hat{X}_+^{(\alpha)}$ in \eqref{X(y)_NP} can be thought of as functions on the space of configurations of the form \eqref{NPsol}.

We can recall from \cite{Gaiotto:2008sa} that the moduli space of solutions of the Nahm's equations in the presence of pole has exactly the same description as \eqref{NPsol}. In \cite{Gaiotto:2008sa} this description was found by discarding one of the three Nahm's equations and complexifying the gauge group. Going backward, if we are given \eqref{NPsol}, we can perform a complex gauge transformation and find a solution of the usual Nahm's equations. Notice that the dangerous term in the action \eqref{S_div} (that constrains $\hat{X}_i$) is given by the square of Nahm's equations, and thus simply vanishes on their solutions. This shows that solutions of Nahm's equations are part of the physical space of fields on which the action is well-behaved despite the pole. In particular, coordinate functions on this space are valid independent observables that are not constrained in any way.

We therefore conclude that the operators $\hat{X}_+^{(\alpha)}$ from \eqref{X(y)_NP} have the interpretation of coordinates on the moduli space of solutions to the Nahm's equations, and all of them are independent nontrivial operators. Configurations of the form \eqref{NPsol} are known to be isomorphic to the Slodowy slice $\cS_{t_+}$ at $t_+$. Our conclusion therefore is the isomorphism of vector spaces:
\begin{equation}
\boxed{\text{Boundary operators } \cong \text{regular functions on } \cS_{t_+}}
\end{equation}
There is one algebraic structure on this space, which is that of a commutative point-wise multiplication of regular functions. It is expected to match the operator product on the sort of a ``chiral ring'' we would get if we dropped the ``S'' part in the ``Q+S'' construction of Section \ref{sec:general} (or took the commutative limit $\ell\to \infty$). Another algebraic structure on the space of operators is the one we actually considered in Section \ref{sec:general}. It is expected to match the appropriate equivariant deformation quantization of $\cS_{t_+}$ (with respect to its natural Poisson structure inherited from $\mathfrak{g}^*$), given by a star-product on the algebra of regular functions on $\cS_{t_+}$.

The latter is known to be the so-called finite W-algebra \cite{Kostant1978,de_Boer_1993,De_Vos_1995,PREMET20021,GanGinzburg,De_Sole_2006,Losev2007QuantizedSA} associated to the embedding $\varrho$ (in particular see \cite{GanGinzburg,Losev2007QuantizedSA} for the quantization). It only depends (up to isomorphism) on the conjugacy class of the nilpotent element $t_+$ (see, e.g. \cite[Theorem 1]{GoodGraPoly}), so we denote it as $\cW(\mathfrak{g}_\C, t_+)$. We largely follow \cite{brundan2005representations,brundan2008highest} where the highest weight theory of finite W-algebras was considered, and also use the results of \cite{BRUNDAN2006136} in the last section. We propose the following

\textbf{Conjecture: } the algebra of boundary operators in the $\rQ^H_{1,2}$ cohomology at the Nahm pole $\varrho$ is the finite W-algebra $\cW(\mathfrak{g}_\C, t_+)$.

Below we will provide some checks of this conjecture, but first let us briefly review the definition of finite W-algebras.

\subsection{Finite W-algebra at the boundary}\label{sec:Wconj}
\subsubsection{Definitions}
Within the limits of this subsection only, $\mathfrak{g}$ will be a finite-dimensional reductive Lie algebra over $\C$, equipped with a non-degenerate invariant symmetric bilinear form $\trace(\cdot,\cdot)$. This is what in the rest of paper is denoted by $\mathfrak{g}_\C$, a complexification of $\mathfrak{g}={\rm Lie}(G)$. For notational clarity, we drop the subscript $\C$ in this subsection only.

Let us briefly recall the definition of $\cW(\mathfrak{g},t_+)$ according to \cite{PREMET20021}. The embedding $\varrho: \mathfrak{su}(2) \to \mathfrak{g}$, as we know, gives a triple of $\mathfrak{su}(2)$ generators $(t_1, t_2, t_3)$ inside $\mathfrak{g}$. Choose $t_3$ as a semi-simple element, and $t_\pm = t_1 \pm i t_2$ as nilpotent elements, and use $t_3$ to define a $\frac12 \Z$-grading on $\mathfrak{g}$:
\begin{equation}
\mathfrak{g} \cong \bigoplus_{d\in \frac12\Z} \mathfrak{g}_d.
\end{equation}
This grading is an example of so-called \emph{good} gradings for $t_+$, and we do not have to define it using $t_3$, any good grading would do (as we mentioned earlier, see \cite[Therem 1]{GoodGraPoly} for this fact). That the grading is good means that $t_+\in \mathfrak{g}_1$, and the linear map
\begin{equation}
{\rm ad}_{t_+} : \mathfrak{g}_j \to \mathfrak{g}_{j+1}
\end{equation}
is injective for $j\leq -\frac12$ and surjective for $j\geq -\frac12$. In particular, the map ${\rm ad}_e: \mathfrak{g}_{-\frac12}\to \mathfrak{g}_{\frac12}$ is bijective, which implies that the skew-symmetric bilinear form $\langle\cdot,\cdot \rangle$, defined using the Killing form $\trace$ according to
\begin{equation}
\langle x,y \rangle = \trace(t_+ [x,y]),
\end{equation}
is non-degenerate. Pick a subspace $\mathfrak{l}\subset \mathfrak{g}_{-\frac12}$ Lagrangian with respect to $\langle\cdot,\cdot \rangle$, and define
\begin{equation}
\mathfrak{m} := \mathfrak{l}\oplus \bigoplus_{d\leq -1}\mathfrak{g}_d.
\end{equation}
This is a nilpotent subalgebra of $\mathfrak{g}$, and one can define its character $\chi: \mathfrak{m}\to \C$ as
\begin{equation}
\chi(x) = \trace(t_+ x).
\end{equation}
A finite W-algebra is then defined as the quantum Hamiltonian reduction of the universal enveloping algebra $U(\mathfrak{g})$ by this $\mathfrak{m}$, with $\chi$ treated as the moment map constraint. More precisely, $\chi$ extends to a homomorphism $U(\mathfrak{m})\to\C$, whose kernel $\ker\chi$ generates a left ideal $I_\chi$ of $U(\mathfrak{g})$, $I_\chi := U(\mathfrak{g}) \ker\chi$, and the finite W-algebra is then defined as
\begin{equation}
\cW(\mathfrak{g}, t_+) = \left(U(\mathfrak{g})/I_\chi\right)^{{\rm ad}\, \mathfrak{m}},
\end{equation}
where we take the subspace invariant under the adjoint action of $\mathfrak{m}$.\footnote{We took a few shortcuts, and made a few modifications compared to the standard mathematical literature treatment. First, we consider the $\frac12\Z$ grading with $t_+$ in degree one, instead of the $\Z$-grading with $t_+$ in degree two, as it is somewhat more natural physically. Second, the definition we reviewed is usually given second, being equivalent to the first definition, but such a shortened approach is enough for us. Finally, the algebra itself is often denoted $\cW(\chi)$, $\cW_\chi$, or $U(\mathfrak{g}, t_+)$, but we call it $\cW(\mathfrak{g}, t_+)$.}

Even though the definition depends on a few choices, such as a Lagrangian subspace $\mathfrak{l}$, an element $t_+$, and a good grading, it turns out that these choices are irrelevant. Up to an isomorphism, the algebra only depends on the conjugacy class of the nilpotent $t_+$.

Three more definitions of the same algebra $\cW(\mathfrak{g}, t_+)$ can be found reviewed in \cite{brundan2008highest}, where the relations (namely, equivalence) between all these definitions are given. The first definition there is very similar to the one we presented above, and can basically be obtained by dropping $\mathfrak{l}$. More precisely, define instead
\begin{equation}
\mathfrak{m} := \bigoplus_{d\leq -1}\mathfrak{g}_{d},\quad \mathfrak{n}:=\bigoplus_{d<0}\mathfrak{g}_d,
\end{equation}
and assume also $\mathfrak{t}\subset \mathfrak{g}_0$. The map $\chi$ is defined as before, it also is a character for $\mathfrak{m}$, and extends to $\chi: U(\mathfrak{m})\to\C$. Then again we define a left ideal $I_\chi:= U(g)\ker\chi$, and define
\begin{equation}
\cW(\mathfrak{g}, t_+) = (U(\mathfrak{g})/I_\chi)^{{\rm ad}\, \mathfrak{n}}.
\end{equation}
Repeating the same with the right ideal gives a canonically isomorphic algebra. Notice that this last definition is \emph{not} a quantum Hamiltonian reduction with respect to $\mathfrak{n}$, because $\chi$ does not extend to the character of $\mathfrak{n}$.

\subsubsection{Simple checks}
We now consider two simple checks of the proposal that $\cW(\mathfrak{g}_\C,t_+)$ describes the boundary algebra at the Nahm pole. 

First, suppose that $\varrho$ is a principal embedding. It is known that in this case the finite W-algebra is given by the center of $U(\mathfrak{g}_\C)$ \cite{Kostant1978},
\begin{equation}
\cW(\mathfrak{g}_\C, t_+) \cong \cZ\left[ U(\mathfrak{g}_\C) \right].
\end{equation}
So the proposal is that in this case $\cA_C=\C$ and $\cA_H=\cZ\left[ U(\mathfrak{g}_\C) \right]$, where the latter can be identified with the algebra of bulk operators,
\begin{equation}
\cB_H\cong \C[\mathfrak{t}]^\cW.
\end{equation}
The S-dual of the principal Nahm pole is given by the pure Neumann boundary conditions \cite{Gaiotto:2008ak}. According to Section \ref{sec:Neum}, in this case $\cA_H \cong \C$ and $\cA_C\cong \C[\mathfrak{t}^{\vee}]^\cW$, which agrees with the above up to exchange $\cA_C \leftrightarrow \cA_H$. We only have to mention that $\C[\mathfrak{t}]^\cW$ is a free polynomial ring with ${\rm rk}(G)$ generators, and the same is true for $\C[\mathfrak{t}^{\vee}]^\cW$, with ${\rm rk}(G^\vee)={\rm rk}(G)$ generators. Thus we see that the proposal for the principal Nahm pole agrees with S-duality.

Second, we can look at a more general Nahm pole $\varrho$, and its algebra $\cW(\mathfrak{g}_\C, t_+)$. It contains two obvious subalgebras:
\begin{equation}
\cZ\left[ U(\mathfrak{g}_\C) \right] \subset \cW(\mathfrak{g}_\C, t_+) \quad \text{and} \quad U(\mathfrak{f}_\varrho) \subset \cW(\mathfrak{g}_\C, t_+).
\end{equation}
The first one is just the subalgebra of bulk operators, which is also the center of $\cW(\mathfrak{g}_\C, t_+)$. This one is always part of the boundary algebra, as we have argued previously via the bulk-boundary map. The bulk-boundary map is not completely obvious in the presence of the Nahm pole: indeed, while the operators $\trace (X_+)^n$ are gauge-invariant, and so obey $E=R_H=\tilde{R}_H$ and must be in the cohomology, at the boundary they have to be expressed through $\hat{X}_+^{(\alpha)}$ appearing in \eqref{X(y)_NP}. In the process, there are some cancellations of powers of $y$ happening as we approach the boundary, between $\frac{t_+}{y}$ and $\hat{X}_+^{(\alpha)} v_\alpha y^{m_\alpha}$. The way it works is basically the finite W-algebra version of the Harish-Chandra map.

The second subalgebra corresponds to those components of $X_+$ that commute with $\varrho(\mathfrak{su}_2)$. They are essentially unaffected by the pole, and just like in the Dirichlet case, produce the universal enveloping algebra $U(\mathfrak{f}_\varrho)$ for $\mathfrak{f}_\varrho\subset \mathfrak{g}_\C$.

\begin{figure}[h]
	\centering
	\includegraphics[scale=1]{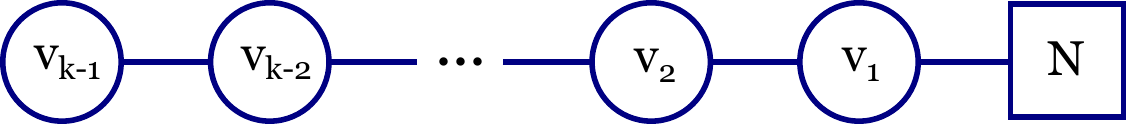}
	\caption{Quiver of the $T_\varrho[SU(N)]$ theory. The relation of ranks to $\varrho=(n_i)$ is: $v_{k-1}=n_k$, $v_i=v_{i+1} + n_{i+1}$ \cite{Gaiotto:2008ak,Nishioka:2011dq}.\label{fig:Trho}}
\end{figure}

\paragraph{A less trivial check}
Consider the case of gauge group $G=SU(N)$, so $\mathfrak{g}_\C=\mathfrak{sl}_N$. The boundary algebra $\cA_H$ is then expected to be $\cW(\mathfrak{sl}_N,t_+)$, where $\varrho = (n_1, n_2, \dots, n_k)$, with $\sum n_i = N$, $n_i\geq 1$, and $n_i\geq n_{i+1}$. The S-dual of the Nahm pole is given by the Neumann boundary conditions in the $SU(N)$ theory enriched with the boundary theory $T_\varrho[SU(N)]$ defined by the quiver in Figure \ref{fig:Trho}.
\\
To check the proposal, we need to know the algebra $\cA_C$ of this theory. The Coulomb branch algebras of such quiver theories have been identified in several works, see in particular \cite{Kamnitzer_2014,Bullimore:2015lsa,Braverman:2016pwk}, and the answer coincides with the one following from the methods of \cite{Dedushenko:2018icp}, implying that we get the same algebra on $S^3$. The algebra is given by a central quotient of the shifted truncated Yangian $Y_\mu(\mathfrak{sl}_k)$, and it appears in the representation in terms of shift operators like in \cite{Gerasimov_2005}. The shifted Yangians are related to finite W-algebras \cite{brundan2005representations}, and indeed in this case one finds precisely the right finite W-algebra, see e.g. Section 1.2.3 of \cite{Bullimore:2016hdc}, and \cite{Braverman:2010ef}.

\subsubsection{The trace}
Now we study the trace on $\cA_H$ at the Nahm pole. Its value on the identity, which also determines the trace on $\cA_C$, is given by the hemisphere partition function. We determine it using the S-duality, by computing the hemisphere partition function in the $G^\vee$ gauge theory, with Neumann boundary conditions enriched by $T_\varrho[G^\vee]$. The $S^3$ partition function of $T_\varrho[G^\vee]$ for $G^\vee=SU(N)$ can be found, e.g., in \cite{Nishioka:2011dq}, and we write it in the form
\begin{equation}
\label{Z_Trho}
Z_{T_\varrho[SU(N)]}(\zeta, m) = \frac{\sum_{w\in\cW}(-1)^{l(w)} e^{2\pi i m\cdot w(\zeta_\varrho)}}{\mathbbm{\Delta}(m)\mathbbm{\Delta}_\varrho(\zeta_\varrho)},
\end{equation}
where $m$ is in the Cartan of $\mathfrak{g}^\vee$, and the FI term $\zeta$ is in the Cartan of $\mathfrak{f}_\varrho\subset \mathfrak{g}$, that is if $\varrho=[n_1,n_2,\dots,n_k]$, then
\begin{equation}
\zeta = (\underbrace{\zeta_1,\dots,\zeta_1}_{n_1},\dots, \underbrace{\zeta_k,\dots,\zeta_k}_{n_k}),
\end{equation}
with $\zeta_1=0$, $\zeta_2=\alpha_1$, $\zeta_3=\alpha_1+\alpha_2$, \dots, $\zeta_k=\alpha_1+\dots +\alpha_{k-1}$, where $\alpha_i$ are the FI parameters associated with the gauge nodes of the quiver for $T_\varrho[SU(N)]$. The formula also includes the shifted parameter
\begin{equation}
\zeta_\varrho = \zeta-it_3,
\end{equation}
and the reduced sinh-Vandermonde $\mathbbm{\Delta}_\varrho(\zeta_\varrho)$, which is defined similar to the full sinh-Vandermonde, but with certain roots in the product omitted:
\begin{equation}
\label{mod_Vdm}
\mathbbm{\Delta}_\varrho(x) = \prod_{\alpha\in\Phi_+^\varrho} 2 \sinh\pi\langle \alpha, x \rangle.
\end{equation}
Here the restricted set of positive roots $\Phi^\varrho_+$ is determined according to the Young diagram encoding $\varrho$. Consider the Young diagram with column heights $n_1\geq n_2\geq \dots \geq n_k$ from the left to the right, and label boxes with the basis vectors $e_1, e_2, \dots, e_N$ ordered as follows: first top to bottom, then left to right. Then the roots belonging to $\Phi^\varrho_+$ are of the form $e_i - e_j$, with $i<j$, such that $e_i$ and $e_j$ belong to the same row of the Young diagram. (Here we identified the dual to Cartan of $\mathfrak{sl}_N$ with the subspace $\sum_{i=1}^N x_i=0$ of $\C^N$, where all the roots lie.)

This $\Phi^\varrho_+$ is in fact a positive part of the restricted root system $\Phi^\varrho$ introduced in \cite[Sections 2, 3]{GoodGraPoly}, see also \cite[Section 3.1]{brundan2008highest}. The roots in $\Phi^\varrho$ correspond to the non-zero weights of the adjoint action of $\mathfrak{t}^{t_+}$ on $\mathfrak{g}^{t_+}$, that is they are the roots of $\mathfrak{g}^{t_+}$. (Recall that the notation $x^{t_+}$ means ``centralizer of $t_+$ in $x$''.)

We are going to assume that for general $G^\vee$, the $S^3$ partition function of $T_\varrho[G^\vee]$ takes the same form \eqref{Z_Trho}, with the reduced sinh-Vandermonde in \eqref{mod_Vdm}, and with the restricted set of positive roots $\Phi_+^\varrho$ corresponding to $\mathfrak{g}^\vee$. This makes sense because the Nahm pole eliminates certain modes of the fields that would contribute to the one-loop determinants with the Dirichlet boundary conditions. Of course it would be desirable to have a derivation of this conjecture, but it seems to fit well with everything else in our story.

One can then determine the $HS^4$ partition function by integrating $Z_{T_\varrho[G^\vee]}$ against the Dirichlet partition function with the appropriate measure, which gives $T_H(1)$ on the original side of the duality:
\begin{align}
T_H(1) &= \frac1{|\cW|} \frac{i^{{\rm rk}(G)/2}}{\tau^{\dim(G)/2}} \int_{\mathfrak{t}^\vee} [\dd a]\, Z_{T_\varrho[G^\vee]}(\zeta, a)  \mathbbm{\Delta}(a)\Delta(a) e^{-\frac{i\pi}{\tau}\trace(a^2)}\cr
&=\frac{i^{{\rm rk}(G)/2}}{\tau^{\dim(G)/2}} \int_{\mathfrak{t}^\vee} [\dd a]\, e^{-\frac{i\pi}{\tau}\trace(a^2)}\Delta(a) \frac{e^{2\pi i a\cdot \zeta_\varrho}}{\mathbbm{\Delta}_\varrho(\zeta_\varrho)}
=\frac{\Delta(\zeta_\varrho)}{\mathbbm{\Delta}_\varrho(\zeta_\varrho)} e^{i\pi\tau\trace(\zeta_\varrho^2)}, 
\end{align}
where we included a factor $\frac{i^{{\rm rk}(G)/2}}{\tau^{\dim(G)/2}}$ by hands to compensate for the normalization term arising in S-duality due to the gravitational counterterms, -- this is the same factor we found in Section \ref{sec:tr_on_U(g)} by performing S-duality in the opposite direction, and since it cannot depend on the boundary conditions we can include it by hands here.

We can extract the following term from the second line,
\begin{equation}
\label{Verma_guess}
 \frac{e^{2\pi i a\cdot \zeta}}{\mathbbm{\Delta}_\varrho(\zeta_\varrho)},
\end{equation}
which can be similarly identified with the character (i.e. twisted trace of the identity) of the Verma module for $\cW(\mathfrak{g}, t_+)$. This deserves an explanation.

The highest weight theory for finite W-algebras was developed in \cite{brundan2008highest}, and they also introduced a \emph{good filtration} on $\cW(\mathfrak{g}, t_+)$, which was also called the \emph{loop filtration} in \cite{GOODWIN20102058}, such that the associated graded of $\cW(\mathfrak{g}, t_+)$ is $\mathfrak{t}^{t_+}$-equivariantly isomorphic, as a graded algebra, to $U(\mathfrak{g}^{t_+})$, where the latter is equipped with the good grading (see \cite[Theorem 3.8]{brundan2008highest}). This result allows to view $\cW(\mathfrak{g}, t_+)$ as a deformation of $U(\mathfrak{g}^{t_+})$, and as a vector space, $\mathfrak{g}^{t_+}$ is the space $P_+$ of highest weight vectors in $\mathfrak{g}$ with respect to $SU(2)_\varrho$ (which we found in Section \ref{sec:bdyOpNahm} to generate the boundary operators at the Nahm pole.)

Quite conveniently, \cite{GOODWIN20102058} found that Verma modules of $\cW(\mathfrak{g},t_+)$ can likewise be viewed as $\mathfrak{t}^{t_+}$-equivariant deformations of the Verma modules of $U(\mathfrak{g}^{t_+})$. In other words, each Verma module of $\cW(\mathfrak{g}, t_+)$ has a filtration, also called \emph{loop filtration} in \cite{GOODWIN20102058}, such that the associated graded is $\mathfrak{t}^{t_+}$-equivariantly and graded isomorphic to the Verma module of $U(\mathfrak{g}^{t_+})$. The character of Verma module can well be computed using such an associated graded module, as we are simply counting vectors weighted by their $\mathfrak{t}^{t_+}$-weights. 

The expression \eqref{Verma_guess} can be indeed recognized as the character of the $U(\mathfrak{g}^{t_+})$ Verma module $V_{-ia-\rho^{t_+}}$, where $\rho^{t_+}$ is the analog of Weyl vector in the case of $\mathfrak{g}^{t_+}$ (see \cite{GOODWIN20102058} for the precise definition of this shift). That we have $\zeta_\varrho$ instead of $\zeta$ in the denominator, i.e., that $\zeta$ is shifted by $t_+$, is related to the redefinition of R-symmetry discussed in Section \ref{sec:Rmixing}. Because our traces are also twisted by $(-1)^{2R_H}$ (see Section \ref{sec:twisted_trace}), this redefinition results in an extra $\Z_2$ twist in the trace, that is present in general, but was absent in the $U(\mathfrak{g})$ case. So we look at the $U(\mathfrak{g}^{t_+})$ character, with the additional $\Z_2$ twist that multiplies by $-1$ operators that have half-integral spin with respect to $t_3$.

We then further include operator insertions under the trace. We can also generate special insertions by taking derivatives with respect to the boundary masses, i.e. twist parameters of the trace, which are also identified as FI parameters $\zeta$ on the dual side. In the end of the day, we propose the formula very similar to the one in the Dirichlet case:
\begin{align}
T_H(\cO) &=\frac{i^{{\rm rk}(G)/2}}{\tau^{\dim(G)/2}} \int_{\mathfrak{t}^\vee} [\dd a]\, e^{-\frac{i\pi}{\tau}\trace(a^2)+2\pi a\cdot t_3}\Delta(a) \trace_{V_{-ia-\rho^{t_+}}}e^{-2\pi \zeta\cdot B}\cO,
\end{align}
where we use $V_{-ia-\rho^{t_+}}$ as a notation for the Verma module of the finite W-algebra that corresponds to the Verma module of $U(\mathfrak{g}^{t_+})$ with the same name. In this formula, $B$ denotes the element of $\mathfrak{f}_\varrho$, that corresponds to the $U(\mathfrak{f}_\varrho)$ subalgebra that couples to the boundary mass.

\section{Teaser on interfaces}
All our techniques can be straightforwardly applied to half-BPS interfaces of the 4d $\cN=4$ SYM. While many interesting aspects of such defects are subject of a separate paper, we describe some preliminary results here. We are interested in interfaces that can be engineered by D5 branes and their S-dual NS5 branes, and here we only consider examples with a single D5 or NS5 brane, while multiple fivebranes are treated in a separate article.

\subsection{Intersecting branes}
\paragraph{D5 frame.}
Suppose $N$ D3 branes intersect a single D5 brane. This system has a well-known description: it can be obtained by gluing two half-spaces with $U(N)$ SYM on them, each subject to the Dirichlet boundary conditions, with a fundamental hypermultiplet living at the interface. Each Dirichlet boundary supports a protected algebra $\cA_H=U(\mathfrak{gl}_N)$, according to our construction. The corresponding elementary boundary operators that generate the two copies of $U(\mathfrak{gl}_N)$ are denoted as $B_+$ and $B_-$. According to \cite{Beem:2016cbd,Dedushenko:2016jxl}, the interface hypermultiplet adds another ingredient: fields $(Q_\alpha, \tilde{Q}^\alpha)$, $\alpha=1..N$, valued in the fundamental of $U(N)$ and its dual, obeying
\begin{equation}
[Q_\alpha, \tilde{Q}^\beta] = \frac1{\ell}\delta_\alpha^\beta,
\end{equation}
that is they form $N$ copies of the Weyl algebra $W$.

Gluing is implemented via gauging ${\rm Diag}(U(N)\times U(N))$ on the interface, which can be derived from the localization \cite{Dedushenko:2018tgx}. At the algebra level, as we know, it corresponds to the quantum Hamiltonian reduction of $\cA_0=U(\mathfrak{gl}_N)\otimes U(\mathfrak{gl}_N)\otimes W^{\otimes N}$ with respect to the $\mathfrak{gl}_N$ action, whose moment map is
\begin{equation}
\mu_\alpha{}^\beta = (B_+)_\alpha{}^\beta + (B_-)_\alpha{}^\beta + Q_\alpha \tilde{Q}^\beta - r \delta_\alpha^\beta.
\end{equation}
Let us describe the resulting algebra. In quantum Hamiltonian reduction, we first take the quotient of $\cA_0$ over its left ideal generated by $\mu_\alpha{}^\beta$, which can be used to completely eliminate $B_-$, so that the quotient is represented by the subalgebra generated by $B_+$, $Q$ and $\tilde{Q}$. Next we pass to the subalgebra fixed by $\mathfrak{gl}_N$, and it clearly is generated by expressions of the form:
\begin{equation}
\trace (B_+)^n,\quad \tilde{Q} (B_+)^n Q, \quad n\geq0.
\end{equation}
This generating set is redundant, and not very convenient. First recall from the Section \ref{sec:HC} that the Capelli determinant $\cC(z)$, which is a degree-$N$ polynomial, acts as the characteristic polynomial, i.e. 
\begin{equation}
\cC(B_+)=0.
\end{equation}
Thus all the higher powers $(B_+)^n$, $n>N$, can be expressed through $(B_+)^n$ with $n\leq N$ and coefficients of $\cC(z)$, which are simply the center generators (Capelli invariants). The center can also be seen as generated by $\trace(B_+)^n$ with $n\leq N$ (these are called Gelfand invariants, whose relation to Capelli invariants was explained in Section \ref{sec:HC}). We therefore conclude that the true generating set is finite, and given by
\begin{equation}
\trace (B_+)^n,\quad \tilde{Q} (B_+)^n Q, \quad 0\leq n \leq N.
\end{equation}
This is still not the most convenient set, and one can easily show that it is equivalent to
\begin{equation}
\label{better_gen}
\trace(B_+)^n,\quad \trace (B_-)^n,\quad 0\leq n\leq N,
\end{equation}
which can be proven inductively by showing that all generators of the form $\tilde{Q} (B_+)^n Q$ can be expressed using \eqref{better_gen}, modulo $\mu$. Indeed, the moment map constraint implies
\begin{equation}
\trace (B_-)^n = \trace (\zeta\, {\rm Id} - B_+ - Q\tilde{Q})^n = (-1)^n n \trace \tilde{Q}(B_+)^{n-1}Q + \dots,
\end{equation}
where the ellipsis only involves products of expressions like $\trace(B_+)^m$ and $\tilde{Q} (B_+)^k Q$, with $k<n-1$. Taking $n=0$ as the base of induction, this proves that $\trace(B_-)^n$ can replace $\tilde{Q} (B_+)^n Q$ in the set of generators. Thus the algebra is generated by \eqref{better_gen},
which are simply the bulk operators on the two sides of the D5 brane, giving two copies of the center,
\begin{equation}
\cA_H = \cZ[U(\mathfrak{gl}_N)]\otimes \cZ[U(\mathfrak{gl}_N)].
\end{equation}

We can also easily determine the $\cA_C$ algebra. From the ``gluing by gauging'' perspective, each Dirichlet boundary supports a trivial algebra $\cA_C=\C$, and only the 3d vector multiplet involved in gauging contributes $\rQ^C_{1,2}$ closed operators of the type $\trace (Y_+)^n$ (in \cite{Dedushenko:2017avn}, these are called $\trace \Phi^n$), which generate the center. There are no monopole operators, because this 3d vector multiplet is not a purely 3d object: it lives at the interface and describes the ${\rm Diag}(G\times G)$ part of the 4d vector multiplet, and such setup does not admit local monopoles. We thus conclude:
\begin{equation}
\cA_C = \cZ[U(\mathfrak{gl}_N)].
\end{equation}

\paragraph{NS5 frame.} Now let us compare this to the S-dual configuration of $N$ D3 branes intersecting an NS5 brane. This can be described by Neumann boundary conditions on each side of the brane, coupled to a bi-fundamental hypermultiplet living at the interface. The algebra $\cA_H$ follows from the method of Section \eqref{sec:AH_Neum} immediately. We start with the algebra of free bifundamental hypers, which is $W^{N^2}$, i.e. $N^2$ copies of the Weyl algebra generated by
\begin{equation}
[Q_\alpha{}^{\dot\alpha}, \tilde{Q}^\beta{}_{\dot\beta}]=\frac1{\ell} \delta_\alpha^\beta \delta_{\dot\alpha}^{\dot\beta},
\end{equation}
where $\alpha, \beta$ are gauge index on the left, and $\dot\alpha, \dot\beta$ -- on the right of the brane. According to the Section \ref{sec:AH_Neum}, we pass to the $\mathfrak{gl}_N\times \mathfrak{gl}_N$-invariants, which are simply generated by
\begin{equation}
\trace(Q\tilde{Q})^n,\quad n\geq 0,
\end{equation}
where we use $(Q\tilde{Q})^{\dot\alpha}{}_{\dot\beta}=Q_\alpha{}^{\dot\alpha}Q^\alpha{}_{\dot\beta}$. Such $(Q\tilde{Q})$ obey the $\mathfrak{gl}_N$ relations, and we clearly obtain the center of $U(\mathfrak{gl}_N)$,
\begin{equation}
\cA_H=\cZ[U(\mathfrak{gl}_N)],
\end{equation}
which agrees with the $\cA_C$ algebra in the dual D5 brane description.

The $\cA_C$ algebra of the NS5 interface follows from Section \ref{sec:AC_Neum} without any work. Indeed, the Coulomb algebra of a free bi-fundamental hyper is trivial, that is $\C$. Extending it by the restrictions of bulk scalars $\trace(Y_+)^n$ on the left and on the right, we produce two copies of the center,
\begin{equation}
\cA_C=\cZ[U(\mathfrak{gl}_N)]\otimes\cZ[U(\mathfrak{gl}_N)],
\end{equation}
which of course agrees with $\cA_H$ in the D5 S-duality frame.

Using the known expression for the hemisphere partition function and the gluing rules, it is completely straightforward to write traces in all these cases, so we skip this. The S-duality, as usual, is implemented by the Fourier transform.

\subsection{Terminating and intersecting branes}
Another interesting configuration has different numbers of D3 branes on the two sides of the fivebrane. Namely, suppose we have $N$ D3 branes intersecting a fivebrane (D5 or NS5), and additionally $k$ D3 branes terminating on it from the right. The cases $k=1$ and $k>1$ are quite different, so we discuss them separately.

\paragraph{$k=1$, D5 and NS5 frames.} When the fivebrane is the D5 brane, the interface can be described as follows: take a left half-space with the $U(N)$ gauge group and Dirichlet boundary; take a right half-space with the $U(N+1)$ gauge group and Dirichlet boundary; pick an embedding $\iota: U(N) \to U(N+1)$; gauge the group ${\rm Diag}\big[U(N) \times \iota(U(N))\big]$ at the interface. Notably, there is no extra interface matter. As we now know, this procedure gives the $\cA_H$ algebra as a quantum Hamiltonian reduction of $U(\mathfrak{gl}_N)\times U(\mathfrak{gl}_{N+1})$ with respect to the diagonal $\mathfrak{gl}_N$, whose moment map is
\begin{equation}
\mu = B_- + B_+\big|_{\mathfrak{gl}_N} - r\, {\rm Id},
\end{equation}
where $B_+\big|_{\mathfrak{gl}_N}$ means restriction of $B_+\in\mathfrak{gl}_{N+1}$ to the $\mathfrak{gl}_N$ subalgebra, and $r$ is a possible boundary FI term. Again, taking quotient eliminates $B_-$, and taking the invariants results in the answer
\begin{equation}
\cA_H = U(\mathfrak{gl}_{N+1})^{\mathfrak{gl}_N}.
\end{equation}
To compute the invariant subalgebra, decompose $\mathfrak{gl}_{N+1}$ as a $\mathfrak{gl}_N$-module,
\begin{equation}
\mathfrak{gl}_{N+1}\cong \mathfrak{gl}_N \oplus \C \oplus \C^N \oplus \bar{\C}^N \ni (B, c, X, Y).
\end{equation}
It is easy to identify the $\mathfrak{gl}_N$-invariants as generated by $c$, $\trace B^k$, and $Y B^k X$, with $k\in \Z_{\geq 0}$. This makes it somewhat similar to the previous example, however there is an important difference: $X,Y$ do not form the Weyl algebra. Rather, the commutation relations are:
\begin{align}
\label{comm_NN1}
[B^\alpha_\beta, B^\gamma_\delta]&=\frac1{\ell}(\delta^\gamma_\beta B^\alpha_\delta - \delta^\alpha_\delta B^\gamma_\beta),\quad [B^\alpha_\beta, c]=0,\cr
[B^\alpha_\beta, X^\gamma]&=\frac1{\ell}\delta^\gamma_\beta X^\alpha,\quad [B^\alpha_\beta, Y_\delta] = -\frac1{\ell}\delta^\alpha_\delta Y_\beta,\cr
[X^\alpha, Y_\beta]&=\frac1{\ell}( B^\alpha_\beta - \delta^\alpha_\beta c),\quad [X^\alpha, c]=\frac1{\ell} X^\alpha,\quad [Y_\alpha, c]=-\frac1{\ell} Y_\alpha.
\end{align}
Because $B$ generates $\mathfrak{gl}_N$, commutativity of generators $\trace B^k, c$ is obvious, as well as vanishing of commutators of $\trace B^k, c$ with $Y B^k X$. The only non-obvious commutators are
\begin{equation}
[Y B^p X, Y B^q X]=0,
\end{equation}
which can be shown by a small computation using \eqref{comm_NN1}.

We want to claim that $c$, $\trace B^k$ and $Y B^k X$ generate $\cZ(\cU(\mathfrak{gl}_{N+1}))\otimes \cZ(\cU(\mathfrak{gl}_{N}))$, as expected from S-duality. However, to make things more manifest, it is again useful to define another generating set, which consists of operators that can be removed from the interface into the bulk. The moment map constraint tells us that
\begin{equation}
B_- = r - B,
\end{equation}
where $B_-$ is a generator on the $\mathfrak{gl}_N$ side of the interface. Then the generators of $\cZ(\cU(\mathfrak{gl}_N))$ are 
\begin{equation}
\trace(B_-)^m=\trace (\zeta - B)^m,
\end{equation}
which is an invertible change of generators from $\trace B^m$.
To identify the generators of $\cZ(\cU(\mathfrak{gl}_{N+1}))$, we consider the matrix of $\mathfrak{gl}_{N+1}$ generators $B_+$, which contains $B$, $X$, $Y$, $c$ as submatrices, and build the generators as
\begin{equation}
\trace (B_+)^m.
\end{equation}
For example
\begin{equation}
\trace B_+ = c + \trace B,\quad \trace (B_+)^2 = \trace B^2 + 2 YX + c^2 + \hbar \trace B - N\hbar c, \text{ etc.}
\end{equation}
An argument similar to that from the previous subsection proves that $\trace(B_+)^m$ are equivalent to the generators $c$, $YB^k X$. Thus we indeed find
\begin{equation}
\label{D5N_N1}
\cA_H = \cZ[U(\mathfrak{gl}_N)]\otimes \cZ[U(\mathfrak{gl}_{N+1})].
\end{equation}
This clearly matches the $\cA_C$ algebra for the S-dual NS5 interface. Indeed, in that case the treatment is not different from the previous subsection: there is a bi-fundamental hyper at the NS5 brane that has a trivial $\cA_C$ algebra, and its extension by the bulk fields produces product of bulk algebras on the left and on the right, exactly as in \eqref{D5N_N1}.

The $\cA_C$ algebra on the D5 interface is given by
\begin{equation}
\cA_C=\cZ[U(\mathfrak{gl}_N)],
\end{equation}
as is obvious from the gauging perspective, from the same reasons as in the previous section. This algebra also matches the S-dual algebra $\cA_H$ on the NS5 interface.

\paragraph{$k>1$, D5 and NS5 frames.} When we have more than one D3 brane terminating on the right, the description in the D5 frame changes quite dramatically, while the NS5 description is basically unchanged. The NS5 answer is still $\cA_H=\cZ[U(\mathfrak{gl}_N)]$ and $\cA_C=\cZ[U(\mathfrak{gl}_N)]\otimes \cZ[U(\mathfrak{gl}_{N+k})]$, with exactly the same derivation: we take $\mathfrak{gl}_N\times \mathfrak{gl}_{N+k}$ invariants pf the bi-fundamental hyper to find $\cA_H$, and we extend trivial algebra $\C$ by the bulk operators to obtain $\cA_C$.

The D5 frame now involves the Nahm pole. More precisely, we break $U(N+k)$ on the right into $U(N)\times U(k)$, give regular (or principal) Nahm pole boundary conditions to the $U(k)$-valued fields, and identify $U(N)$ with the $U(N)$ on the left via gauging ${\rm Diag}(U(N)\times U(N))$ as before. Again, there is no extra interface matter. The only non-trivial interface operators contributing to $\cA_C$ in this construction appear as $\trace(Y_+)^n$, where $Y_+$ corresponds to the $U(N)$ subgroup only. Thus we find $\cA_C=\cZ[U(\mathfrak{gl}_N)]$, matching the $\cA_H$ on the NS5 side of duality.

The most interesting case here is the $\cA_H$ algebra on the D5 side. It can again be obtained by gauging, i.e. the quantum Hamiltonian reduction. The left half-space (with the $U(N)$ theory) has Dirichlet boundary conditions, and contributes $U(\mathfrak{gl}_N)$. The right half-space has a $\varrho=[k,\underbrace{1,\dots,1}_{N}]$ Nahm pole, and so contributes the finite W-algebra $\cW(\mathfrak{gl}_{N+k}, t_+)$. Thus we have
\begin{equation}
\cA_0 = U(\mathfrak{gl}_N)\otimes \cW(\mathfrak{gl}_{N+k}, t_+),
\end{equation}
and we must perform the quantum Hamiltonian reduction of $\cA_0$ with respect to $\mathfrak{gl}_N$, where on the second factor, this $\mathfrak{gl}_N$ acts as the boundary symmetry commuting with the Nahm pole. Again, the first step in quantum Hamiltonian reduction is taking the quotient, and it simply eliminates $U(\mathfrak{gl}_N)$. The second step then results in
\begin{equation}
\cA_H = \cW(\mathfrak{gl}_{N+k}, t_+)^{\mathfrak{gl}_N},
\end{equation}
so we now proceed to compute it. To find this algebra, we need to understand the structure of the finite W-algebra sufficiently well. Fortunately, it has been studied in excruciating detail and related to shifted Yangians in \cite{BRUNDAN2006136}, so we will simply apply their result. We are not going to review them, rather only point at specific statements in that paper, so to understand the computation presented below, an interested reader will have to consult \cite{BRUNDAN2006136}. They use the classification of good gradings on $\mathfrak{gl}_N$ in terms of pyramids \cite{elashvili2003good}, and one can associate $\rho=[k,1,\dots,1]$ to various pyramids, the most convenient of which is:
\begin{center}
\ytableausetup{mathmode, boxsize=2em}
\begin{ytableau}
	\substack{1} \cr
	\substack{2} \cr
	.\cr
	.\cr
	\substack{N}\cr
	\substack{N+1} & \substack{N+2} & . & . & . & \substack{N+k}\cr
\end{ytableau}	
\end{center}
One easily reads off the shift matrix from the pyramid, which is a square matrix of size $N+1$:
\begin{equation}
\sigma = \left(\begin{matrix}
0 & 0 & \dots & 0 & k-1\\
0 & 0 & \dots & 0 & k-1\\
. & . & \dots & . & .  \\
0 & 0 & \dots & 0 & k-1\\
0 & 0 & \dots & 0 & 0
\end{matrix} \right).
\end{equation}
According to \cite{BRUNDAN2006136}, our finite W-algebra is given by the shifted Yangian (for $\mathfrak{gl}_{N+1}$) at level $k$, with the shift matrix $\sigma$ (recall that ``level $k$'' refers to a truncation of the Yangian by a certain ideal that will be described below):
\begin{equation}
\cW(\mathfrak{gl}_{N+k},t_+)^{\mathfrak{gl}_N} = Y_{N+1, k}(\sigma).
\end{equation}
Using the terminology of \cite{BRUNDAN2006136}, it is most convenient to describe this algebra in the parabolic presentation of shape
\begin{equation}
\nu = (N,1).
\end{equation}
With such a shape, the \cite[Corollary 6.3]{BRUNDAN2006136} identifies the generating set very explicitly as
\begin{equation}
\label{W_gen_NNk}
\left\{ \{D^{(1)}_{1;i,j}\}_{1\leq i,j\leq N},\quad \{D^{(r)}_{2;1,1}\}_{1\leq r \leq k},\quad \{E^{(k)}_{1;i,1}\}_{1\leq i \leq N},\quad \{F^{(1)}_{1;1,j}\}_{1\leq j \leq N}  \right\},
\end{equation}
where we used the same notations as in \cite{BRUNDAN2006136}. In fact, the Corollary 6.3 says more: ordered monomials in such elements form a basis of $Y_{N+1,k}(\sigma)$. The relations are recorded in \cite[(3.3)-(3.14)]{BRUNDAN2006136}: to establish closure of the algebra generated by \eqref{W_gen_NNk}, it is crucial to remember that one takes a quotient by the two-sided ideal generated by $\{D^{(r)}_{1;i,j}\}_{1\leq i,j\leq N, r>1}$ (this is what truncates the full Yangian to the finitely-generated algebra).

It is easy to see from \cite[(3.3)]{BRUNDAN2006136} that the first set $\{D^{(1)}_{1;i,j}\}_{1\leq i,j\leq N}$ in fact generates the $U(\mathfrak{gl}_N)$ of our interest; the second set $\{D^{(r)}_{2;1,1}\}_{1\leq r \leq k}$ is invariant under this $\mathfrak{gl}_N$ (i.e., commutes with it). The remaining two sets, $\{E^{(k)}_{1;i,1}\}_{1\leq i \leq N}$ and $\{F^{(1)}_{1;1,j}\}_{1\leq j \leq N}$, transform as a fundamental (defining) and an anti-fundamental irrep of $\mathfrak{gl}_N$. We then easily find that the $\mathfrak{gl}_N$-invariant subalgebra is generated by:
\begin{equation}
\label{glN_inv_NNk}
\{D^{(r)}_{2;1,1}\}_{1\leq r \leq k},\quad \trace (D^{(1)}_1)^m,\quad F^{(1)} (D^{(1)}_1)^m E^{(k)},
\end{equation}
where we suppressed some indices for brevity. At this point, the commutativity of $\trace (D^{(1)}_1)^m$ with all the other generators in \eqref{glN_inv_NNk} is obvious from $\mathfrak{gl}_N$-invariance. One can also use \cite[(3.3)]{BRUNDAN2006136} to prove by induction that
\begin{equation}
[D^{(r)}_{2;1,1},D^{(s)}_{2;1,1}]=0,
\end{equation}
where one assumes that $r>s$, and that commutativity holds for all $s\leq s_0$, and then proves that it must also hold for $s=s_0+1$.

The commutativity between $D^{(r)}_{2;1,1}$ and $F^{(1)} (D^{(1)}_1)^m E^{(k)}$ is slightly more challenging. First we have to show that, modulo the two-sided ideal $\cI$ generated by $\{D^{(r)}_{1;i,j}\}_{1\leq i,j\leq N, r>1}$,  the following relations hold: 
\begin{align}
E^{(k+r)}_{1;i,1} &= \sum_{j=1}^N(-D^{(1)}_1)^r_{ij}E^{(k)}_{1;j,1} \mod \cI,\cr
F^{(1+r)}_{1;1,j} &= \sum_{i=1}^N F^{(1)}_{1;1,i}(-D^{(1)}_1)^r_{ij} \mod \cI,
\end{align}
where $(D^{(1)}_1)^r$ means the $r$-th power of the matrix $D^{(1)}_{1;i,j}$. Then we can compute the commutator of $D^{(r)}_{2;1,1}$ with $F^{(1)} (D^{(1)}_1)^m E^{(k)}$ using \cite[(3.5), (3.6)]{BRUNDAN2006136} and these relations, to show that it vanishes. Commutativity of $F^{(1)} (D^{(1)}_1)^m E^{(k)}$ with $F^{(1)} (D^{(1)}_1)^n E^{(k)}$ can also be shown using similar considerations.

We can invoke the Capelli's determinant for $\mathfrak{gl}_N$ again to argue that only the following generators are independent,
\begin{equation}
\{D^{(r)}_{2;1,1}\}_{1\leq r \leq k},\quad \trace (D^{(1)}_1)^m,\quad F^{(1)} (D^{(1)}_1)^m E^{(k)},\quad 1\leq m \leq N.
\end{equation}
So we find that the algebra $\cW(\mathfrak{gl}_{N+k},t_+)^{\mathfrak{gl}_N}$ is generated by $k+2N$ free commuting variables, which agrees with the S-duality prediction
\begin{equation}
\cA_H = \cZ(U(\mathfrak{gl}_{N+k}))\otimes \cZ(U(\mathfrak{gl}_{N})).
\end{equation}
That this computation heavily relied on properties of finite W-algebras, and produced the expected answer, can be viewed as another check of our finite W-algebra proposal from the Section \ref{sec:finiteW}.

\section{Outlook}
In this work, we have mostly focused on boundary conditions, completely analyzing the Dirichlet case, and providing a lot of details on other standard classes of boundary conditions in 4d $\cN=4$ SYM, while other topics, such as interfaces, were only touched upon. A partial list of future directions, some of which we are planning to report on in the near future, includes:
\begin{itemize}
	\item The study of interfaces engineered by multiple D5 or NS5 branes. Such interfaces carry ineresting truncations of the Yangian, and provide new efficient connections with integrability.
	\item The holographic interpretation for many of these constructions is quite interesting, and the case of interfaces is closely related to the twisted holography setting studied in \cite{Ishtiaque:2018str}.
	\item While we have identified the $\rQ^H_{1,2}$ cohomology of boundary local operators (as a vector space) for the Nahm pole boundary conditions quite convincingly, that they form a finite W-algebra is more of a conjecture in general. The S-duality and some other computations we have done provide a very strong check of this statement, but it would be interesting to have a direct derivation of the operator product, like in the Dirichlet case, where the 2d constrained Yang-Mills perspective is quite useful.
	\item The construction of $\cA_C$ admits a lift to five dimensions, with boundary local operators lifting to boundary lines. Their operator algebras are also going to be certain associative algebras equipped with traces, which we expect to be ``quantum'' deformations of the structures studied in this paper, such as the quantum group $U_q(\mathfrak{g}_\C)$. It would be interesting to explore this further, taking some motivation from \cite{schrader2016cluster}.
	\item The construction of $\cA_H$ may also admit a lift to five dimensions, with boundary local operators lifting to a boundary chiral algebra, and the 5d half-index \cite{Gaiotto:2015una} in the appropriate limit playing the role of its character. It is unlikely that the flat five-dimensional half-space is the right setting for the 5d construction to work: the lack of conformal symmetry is one indication of it; not enough R-symmetry on the $S^4\times S^1$ background \cite{Kim:2014kta} is another. On the other hand, the 5d MSYM on $AdS_5$ is a better candidate to search for boundary chiral algebras, and indeed it was demonstrated to work in the simplest case \cite{Bonetti:2016nma}. It would be interesting to explore this further.
\end{itemize}
\acknowledgments

MD thanks Du Pei for asking a useful question during the talk. This research was supported in part by a grant from the Krembil Foundation. D.G. is supported by the NSERC Discovery Grant program and by the Perimeter Institute for Theoretical Physics. Research at Perimeter Institute is supported in part by the Government of Canada through the Department of Innovation, Science and Economic Development Canada and by the Province of Ontario through the Ministry of Colleges and Universities.

\appendix
\section{Conventions}\label{app:conv}
\subsection{Summary of notations}
Let us summarize some of the notations used throughout this work that could potentially be confusing:
\begin{align*}
G &- \quad \text{compact gauge group of 4d SYM}\cr
\mathfrak{g} &- \quad \text{Lie algebra of $G$}\cr
\mathfrak{g}^\vee &- \quad \text{Langlands dual of }\mathfrak{g}\cr
\mathfrak{g}^* &- \quad \text{vector space dual to }\mathfrak{g}\cr
\trace &- \quad \text{a unified notation for the Kiling form on $\mathfrak{g}$, the induced bilinear form on $\mathfrak{g}^*$,}\cr 
& \quad \quad\text{and the natural pairing of $\mathfrak{g}$ with $\mathfrak{g}^*$}\cr
\mathfrak{g}_\C, \mathfrak{g}^\vee_\C &- \quad \text{complexification of $\mathfrak{g}$ and $\mathfrak{g}^\vee$, that is $\mathfrak{g}\otimes\C$ and $\mathfrak{g}^\vee\otimes\C$}\cr
\Phi &- \quad \text{root system of }\mathfrak{g}\cr
\Phi_+ &- \quad \text{positive roots in }\Phi\cr
\cW &- \quad \text{Weyl group of $G$}\cr
\cS_{t_+} &- \quad \text{Slodowy slice to a nilpotent element }t_+\in\mathfrak{g}\cr
\cW(\mathfrak{g}_\C, t_+) &- \quad \text{finite W-algebra associated to a nilpotent element }t_+\in\mathfrak{g}_\C\cr
\Delta(a) &- \quad \text{Vandermonde determinant associated with $\Phi_+$}\cr
\mathbbm{\Delta}(a) &- \quad \text{sinh-Vandermonde associated with $\Phi_+$}\cr
\varrho &- \quad \text{embedding determining the Nahm pole}\cr
\rho &- \quad \text{the Weyl vector, i.e. the half-sum of roots in $\Phi_+$}\cr
\cC(z) &- \quad \text{Capelli determinant of $z-B$}\cr
\cA_H, \cA_C &- \quad \text{boundary algebras of local operators in the H and C constructions}\cr
\cA_H(\cT), \cA_C(\cT) &- \quad \text{Higgs and Coulomb branch algebras of a 3d theory }\cT\cr
\cA_C[\mathbf{B}] &- \quad \text{notation used in Section \ref{sec:red_3d} to specifically refer to the $\cA_C$ algebra}\cr
& \quad\quad \text{of the boundary condition }\mathbf{B}\cr
T_H, T_C &- \quad \text{twisted traces on $\cA_H$ and $\cA_C$ that encode physical correlators}\cr
\mathscr{A}^H, \mathscr{A}^C &- \quad \text{emergent gauge fields in the $\rQ^H_{1,2}$ and $\rQ^C_{1,2}$ cohomology respectively}\cr
A, F &- \quad \text{temporary notation for $\mathscr{A}^H$ and its curvature in Section \ref{sec:AH_Ug}}\cr
\cB_H, \cB_C &- \quad \text{commutative bulk algebras of local operators in the H and C constructions}\cr
E &- \quad \text{conformal dimension of local perators}\cr
R_H, R_C &- \quad \text{R-charges of local oeprators with respect to the choice}\cr 
& \quad\quad\text{of Cartan in $SU(2)_H$ and $SU(2)_C$}\cr
\end{align*}
\subsection{SUSY algebra}
The 4d $\cN=4$ superconformal algebra has the following anti-commutation relations:
\begin{align}
\{Q^{A}_{\alpha}, \tilde{Q}_{B\dot \alpha}\}&=\delta^A_B \gamma^\mu_{\alpha\dot\alpha} P_\mu,\cr
\{S_{A\alpha}, \widetilde{S}^{B}_{\dot{\alpha}}\}&=\delta_A^B \gamma^\mu_{\alpha\dot\alpha} K_\mu,\cr
\{Q^A_{\alpha},S_{B\beta}\}&=-\varepsilon_{\alpha\beta} (\bar\sigma^{IJ})^A{}_B R_{IJ} + \delta^A_B \gamma^{\mu\nu}_{\alpha\beta} M_{\mu\nu} - i\varepsilon_{\alpha\beta}\delta^A_B D,\cr
\{\tilde{Q}_{A\dot\alpha}, \widetilde{S}^B_{\dot\beta}\}&=-\varepsilon_{\dot\alpha\dot\beta}(\sigma^{IJ})_A{}^B R_{IJ} + \delta_A^B \gamma^{\mu\nu}_{\dot\alpha\dot\beta}M_{\mu\nu} -  i\varepsilon_{\dot\alpha\dot\beta}\delta_A^B D,
\end{align} 
and we do not need to write and of the remaining non-trivial commutators. Here $R_{IJ}$, $I,J=1..6$ are (real anti-symmetric) generators of the R-symmetry ${\rm Spin}(6)=SU(4)$, $\sigma^{IJ}$ are these generators in the representation $\mathbf{4}$ (the fundamental of $SU(4)$), while $\bar\sigma^{IJ}$ correspond to $\bar{\mathbf{4}}$. The supercharges $Q$ and $\widetilde{S}$ transform in $\mathbf{4}$, while $\tilde{Q}$ and $S$ -- in $\bar{\mathbf 4}$ of the R-symmetry. Matrices $\sigma^{IJ}$ are Hermitian, and we have
\begin{equation}
\bar\sigma^{IJ} = -(\sigma^{IJ})^*.
\end{equation}
We construct $\sigma^{IJ}$ as
\begin{align}
\sigma^{IJ}&=-\frac{i}{2}[\hat{\bar\gamma}^I, \hat\gamma^J],\quad \hat{\bar\gamma}^I=-(\hat\gamma^I)^*,
\end{align}
where $\hat\gamma_I^{AB}$ is an intertwiner between an irreducible component of $4\otimes 4$ and $6$ of $SU(4)$. The form of these matrices can be deduced from the isomorphism $\Lambda^2 \C^4 \cong \C^6$, and we take:
\begin{align}
\hat{\gamma}^1=\left(\begin{matrix}0&-i&0&0\\i&0&0&0\\0&0&0&i\\0&0&-i&0\end{matrix} \right), \quad \hat{\gamma}^2=\left(\begin{matrix}0&1&0&0\\-1&0&0&0\\0&0&0&1\\0&0&-1&0\end{matrix} \right),\quad
\hat{\gamma}^3=\left(\begin{matrix}0&0&-1&0\\0&0&0&1\\1&0&0&0\\0&-1&0&0\end{matrix} \right), \cr \hat{\gamma}^4=\left(\begin{matrix}0&0&i&0\\0&0&0&i\\-i&0&0&0\\0&-i&0&0\end{matrix} \right),\quad
\hat{\gamma}^5=\left(\begin{matrix}0&0&0&1\\0&0&1&0\\0&-1&0&0\\-1&0&0&0\end{matrix} \right), \quad \hat{\gamma}^6=\left(\begin{matrix}0&0&0&i\\0&0&-i&0\\0&i&0&1\\-i&0&0&0\end{matrix} \right).
\end{align}

For spinors, we work in conventions $\varepsilon^{12}=-\varepsilon_{12}=1$, and use gamma-matrices in Weyl representation (here $\sigma^i$ are Pauli matrices):
\begin{align}
(\gamma^i)_\alpha{}^{\dot\alpha} &= (\sigma^i)_\alpha{}^{\dot\alpha} \text{ for } i=1,2,3, \text{ and } (\gamma^4)_\alpha{}^{\dot\alpha}=i\delta_\alpha^{\dot\alpha},\cr
(\gamma^i)_{\dot\alpha}{}^\alpha &= (\sigma^i)_{\dot\alpha}{}^\alpha \text{ for } i=1,2,3, \text{ and } (\gamma^4)_{\dot\alpha}{}^\alpha=-i\delta_{\dot\alpha}{}^\alpha.
\end{align}

Introducing half-BPS boundary conditions, we break the R-symmetry down to $SU(2)_H \times SU(2)_C$, and choose
\begin{align}
&R_{12}, R_{13}, R_{23} \text{ generate } SU(2)_C, \text{ choose } R_C= iR_{12} \text{ as Cartan},\cr
&R_{45}, R_{46}, R_{56} \text{ generate } SU(2)_H, \text{ choose } R_H=iR_{56} \text{ as Cartan}.
\end{align}
Accordingly, we identify the 3d $\cN=4$ generators $Q^{a\dot b}_\alpha$ and $S_{a\dot{b}\alpha}$ preserved by the boundary:
\begin{align}
Q^{1\dot1}_\alpha&=Q^1_{\alpha} + \tilde{Q}_{3\dot\alpha=\alpha},\quad Q^{2\dot2}_\alpha=Q^3_{\alpha} + \tilde{Q}_{1\dot\alpha=\alpha},\quad Q^{1\dot2}_\alpha=Q^2_{\alpha}-\tilde{Q}_{4\dot\alpha=\alpha},\quad Q^{2\dot1}_\alpha=Q^4_{\alpha}-\tilde{Q}_{2\dot\alpha=\alpha},\cr
S_{1\dot1\alpha}&=S_{1\alpha}+\widetilde{S}^3_{\dot\alpha},\quad S_{2\dot2\alpha}=S_{3\alpha} +\widetilde{S}^1_{\dot\alpha},\quad S_{1\dot2\alpha}=S_{2\alpha}-\widetilde{S}^4_{\dot\alpha},\quad S_{2\dot1\alpha}=S_{4\alpha} - \widetilde{S}^2_{\dot\alpha},\ \dot\alpha=\alpha \text{ everywhere.}\cr
\end{align}

\subsection{4d $\cN=4$ Super-Yang Mills on $S^4$}\label{app:MSYM}
There are six scalars $\Phi^I$, $I=1..6$, and four Dirac spinors $(\lambda_{A\alpha},\bar\lambda^{A\dot\alpha})$, $A=1..4$, in the vector multiplet. The action and SUSY on a radius-$\ell$ four-sphere are given by
\begin{align}
S &= \frac{1}{g_{YM}^2}\trace\int_{S^4} \sqrt{g} \dd^4x \Big(\frac12 F_{\mu\nu}F^{\mu\nu} +  \cD_\mu\Phi^I \cD^\mu \Phi^I + 2i \bar\lambda^{A\dot\alpha} \gamma^\mu_{\dot\alpha}{}^\alpha \cD_\mu \lambda_{A\alpha} - i\hat\gamma_I^{AB} \lambda_A^\alpha [\Phi^I,\lambda_{B\alpha}]\cr 
&-i\hat{\bar\gamma}_{I AB} \bar\lambda^{A\dot\alpha}[\Phi^I, \bar\lambda^B_{\dot\alpha}] - \frac12 [\Phi^I, \Phi^J]^2 + \frac{2}{\ell^2} \Phi^I\Phi^I \Big),
\end{align}
\begin{align}
\delta A_\mu &= -i \left(\epsilon_A^{\alpha}\gamma_{\mu\alpha}{}^{\dot\alpha}\bar\lambda^A_{\dot\alpha} + \bar\epsilon^{A\dot\alpha}\gamma_{\mu\dot\alpha}{}^\alpha \lambda_{A\alpha}\right),\cr
\delta \Phi^I &= \hat\gamma^{IAB}\epsilon_A^\alpha \lambda_{B\alpha} + \hat{\bar\gamma}^I_{AB} \bar\epsilon^{A\dot\alpha}\bar\lambda^B_{\dot\alpha},\cr
\delta\lambda_{A\alpha}&=\frac12\sigma^{\mu\nu}F_{\mu\nu} \epsilon + i\cD_\mu\Phi^I \hat{\bar\gamma}_{IAB}\gamma^\mu_\alpha{}^{\dot\alpha}\bar\epsilon^B_{\dot\alpha} +\frac{i}{2} \Phi^I \hat{\bar\gamma}_{IAB}\gamma^\mu_\alpha{}^{\dot\alpha}\cD_\mu\bar\epsilon^B_{\dot\alpha} - \frac12[\Phi^I,\Phi^J](\sigma^{IJ})_A{}^B \epsilon_{B\alpha},  \cr
\delta\bar\lambda^A_{\dot\alpha}&=\frac12 \bar\sigma^{\mu\nu} F_{\mu\nu}\bar\epsilon + i\cD_\mu\Phi^I \hat\gamma_I^{AB}\gamma^\mu_{\dot\alpha}{}^\alpha \epsilon_{B\alpha} + \frac{i}{2} \Phi^I \hat\gamma_I^{AB}\gamma^\mu_{\dot\alpha}{}^\alpha \cD_\mu \epsilon_{B\alpha} - \frac12 [\Phi^I,\Phi^J](\bar\sigma^{IJ})^A{}_B \bar\epsilon^B{}_{\dot\alpha},\cr
\end{align}
where the SUSY parameters obey:
\begin{align}
\nabla_\mu\epsilon &=\gamma_\mu \tilde\epsilon,\quad \nabla_\mu\tilde\epsilon = -\frac{1}{4\ell^2}\gamma_\mu\epsilon,\cr
\nabla_\mu\bar\epsilon&=\gamma_\mu \tilde{\bar\epsilon},\quad \nabla_\mu \tilde{\bar\epsilon} = -\frac{1}{4\ell^2} \gamma_\mu\bar\epsilon.
\end{align}
A chosen SUSY closes on-shell as follows,
\begin{align}
\label{closure}
\delta^2 = 2i \cL_v + 2i \rho \Delta + 2 w^{IJ} R_{IJ} + \cG_\Lambda,
\end{align}
where $\cL_v$ is the Lie derivative with respect to the vector field
\begin{equation}
v^\mu = \bar\epsilon^{A\dot\alpha}\sigma^\mu_{\dot\alpha}{}^\alpha \epsilon_{A\alpha},
\end{equation}
$\Delta$ is the dilation with parameter
\begin{equation}
\rho = \bar\epsilon^{A\dot\alpha}\widetilde{\epsilon}_{A\dot\alpha}+\epsilon_A^\alpha \widetilde{\bar\epsilon}^A_\alpha,
\end{equation}
$R_{IJ}$ is the R-symmetry rotation with parameter
\begin{equation}
w^{IJ} = -\bar\epsilon^{A\dot\alpha}(\sigma^{IJ})_A{}^B \widetilde{\epsilon}_{B\dot\alpha}-\epsilon_A^\alpha (\bar\sigma^{IJ})^A{}_B \widetilde{\bar\epsilon}^A_\alpha, 
\end{equation}
and $\cG_\Lambda$ is the gauge transformation with parameter
\begin{equation}
\Lambda=(\epsilon_A^\alpha \hat\gamma_I^{AB}\epsilon_{B\alpha} + \bar\epsilon^{A\dot\alpha}\hat{\bar\gamma}_{IAB}\bar\epsilon^B_{\dot\alpha})\Phi^I-2\bar\epsilon^{A\dot\alpha}\sigma^\mu_{\dot\alpha}{}^\alpha \epsilon_{A\alpha}A_\mu,
\end{equation}
which acts on fields according to $\cG_\Lambda X = -i[\Lambda, X]$ and $\cG_\Lambda A_\mu = \cD_\mu\Lambda$.

For now, let us work in stereographic coordinates $x^\mu$ on $S^4$. The metric is
\begin{equation}
g_{\mu\nu}=e^{2\Omega}\delta_{\mu\nu},\quad e^{2\Omega}=\frac1{\left( 1+\frac{x^2}{4\ell^2} \right)^2},
\end{equation}
the vielbein and the Spin connection are
\begin{equation}
e_\mu^a = \delta_\mu^a e^\Omega,\quad \omega_\mu^{ab}=\frac1{2\ell^2}(x^a e_\mu^b -x^b e_\mu^a).
\end{equation}
Conformal Killing spinors are given by
\begin{align}
\epsilon &=   e^{\Omega/2} (\hat\epsilon + x^a \bar\sigma_a \hat\eta),\quad \tilde{\epsilon} = e^{\Omega/2}\left(\hat\eta -\frac1{4\ell^2}x^a\sigma_a \hat\epsilon\right), \cr
\bar\epsilon&=e^{\Omega/2} (\hat{\bar\epsilon} + x^a\sigma_a \hat{\bar\eta} ),\quad \tilde{\bar\epsilon} = e^{\Omega/2}\left(\hat{\bar\eta} - \frac1{4\ell^2} x^a\bar\sigma_a\hat{\bar\epsilon}\right).
\end{align}
We choose conventions where 
\begin{equation}
\delta_{\epsilon,\bar\epsilon} \cO = [\hat\epsilon_A^{\alpha}Q^A_{\alpha} + \hat{\bar\epsilon}^{A\dot\alpha}\widetilde{Q}_{A\dot\alpha} + \hat\eta_A^{\dot\alpha}\widetilde{S}^A_{\dot\alpha} + \hat{\bar\eta}^{A\alpha} S_{A\alpha}, \cO].
\end{equation}
Our two supercharges correspond to
\begin{align}
\cQ^H = \rQ_1^H + \rQ_2^H,\quad \text{which has:}\cr
\hat\epsilon_{22}=1,\quad \hat\epsilon_{31}=-1,\quad \hat{\bar\epsilon}^1_{\dot1}=-1,\quad \hat{\bar\epsilon}^4_{\dot2}=-1,\cr
\hat\eta_{1\dot2}=\frac1{2\ell},\quad \hat\eta_{4\dot1}=\frac1{2\ell},\quad \hat{\bar\eta}^2_1=-\frac1{2\ell},\quad \hat{\bar\eta}^3_2=\frac1{2\ell},
\end{align}
and
\begin{align}
\cQ^C=\rQ_1^C + \rQ_2^C,\quad \text{which has:}\cr
\hat\epsilon_{11}=-\frac12,\  \hat\epsilon_{12}=\frac12,\  \hat\epsilon_{21}=-\frac{i}2,\ \hat\epsilon_{22}=-\frac{i}2,\ \hat\epsilon_{31}=\frac{i}2,\ \hat\epsilon_{32}=\frac{i}2,\ \hat\epsilon_{41}=\frac12,\ \hat\epsilon_{42}=-\frac12,\cr
\hat{\bar\epsilon}^1_{\dot1}=\frac{i}2,\  \hat{\bar\epsilon}^1_{\dot2}=\frac{i}2,\  \hat{\bar\epsilon}^2_{\dot1}=-\frac12,\  \hat{\bar\epsilon}^2_{\dot2}=\frac12,\  \hat{\bar\epsilon}^3_{\dot1}=-\frac12,\  \hat{\bar\epsilon}^3_{\dot2}=\frac12,\  \hat{\bar\epsilon}^4_{\dot1}=\frac{i}2,\  \hat{\bar\epsilon}^4_{\dot2}=\frac{i}2,\cr
\hat\eta_{1\dot1}=\frac{i}{4\ell},\  \hat\eta_{1\dot2}=\frac{i}{4\ell},\  \hat\eta_{2\dot1}=-\frac{1}{4\ell},\  \hat\eta_{2\dot2}=\frac1{4\ell},\  \hat\eta_{3\dot1}=-\frac1{4\ell},\  \hat\eta_{3\dot2}=\frac1{4\ell},\  \hat\eta_{4\dot1}=\frac{i}{4\ell},\  \hat\eta_{4\dot2}=\frac{i}{4\ell},\cr
\hat{\bar\eta}^1_1=-\frac1{4\ell},\  \hat{\bar\eta}^1_2=\frac1{4\ell},\  \hat{\bar\eta}^2_1=-\frac{i}{4\ell},\  \hat{\bar\eta}^2_2=-\frac{i}{4\ell},\  \hat{\bar\eta}^3_1=\frac{i}{4\ell},\  \hat{\bar\eta}^3_2=\frac{i}{4\ell},\  \hat{\bar\eta}^4_1=\frac1{4\ell},\  \hat{\bar\eta}^4_2=-\frac1{4\ell}.
\end{align}

\subsubsection{Observables in the cohomology}\label{app:coho}

\paragraph{The H case. } Let us describe observables in the $\cQ^H$ cohomology. One can check that the following combinations are $\cQ^H$ closed:
\begin{align}
\phi^H(x_3, x_4) &= \Phi^5 -i\Phi^6 \frac{4\ell^2-x_3^2 - x_4^2}{4\ell^2 + x_3^2 + x_4^2} - \Phi^4 \frac{4i\ell x_3}{4\ell^2 + x_3^2 + x_4^2} - \Phi^3 \frac{4i\ell x_4}{4\ell^2 +  x_3^2 + x_4^2}\Bigg|_{x_1=x_2=0},\cr
\gA^H_3&=A_3 -i\Phi^3 +\frac{2ix_4(x_4\Phi^3 + x_3\Phi^4 + 2\ell\Phi^6)}{4\ell^2 + x_3^2 + x_4^2}\Big|_{x_1=x_2=0},\cr
\gA^H_4&=A_4+i\Phi^4 -\frac{2ix_3(x_4\Phi^3 + x_3\Phi^4 + 2\ell\Phi^6)}{4\ell^2 + x_3^2 + x_4^2}\Big|_{x_1=x_2=0},
\end{align}
where recall that $x_3$ and $x_4$ parametrize the $S^2$ at the fixed locus $x_1=x_2=0$. We therefore have a complex scalar $\phi^H$ and a complexified gauge field $\gA_\mu^H$ on $S^2$. One can check that the following shifted gauge field,
\begin{equation}
\label{effectiveA}
\tilde{\gA}^H_3 = \gA^H_3 + \frac{x_4}{2\ell}\phi^H,\quad \tilde{\gA}^H_4 = \gA^H_4 - \frac{x_3}{2\ell}\phi^H,
\end{equation}
has Q-exact curvature $\tilde{\mathscr{F}}^H_{34}=0 + \{\cQ^H,\dots\}$. Therefore, flatness of \eqref{effectiveA} is a BPS equation, and it also ensures that the curvature of $\gA^H_{3,4}$ is equivalent to $\frac1{\ell}\phi^H$ in the cohomology.

Additionally, the following combination is $\cQ^H$ closed:
\begin{equation}
\gA^H_{\tau} = x_1(A_2+i\Phi^1)-x_2 (A_1 - i\Phi^2)=A_\tau + i(x_1\Phi^1 + x_2\Phi^2),
\end{equation}
furthermore it is closed for any values of $x_{1,2,3,4}$. It can be used to construct a $\cQ^H$-supersymmetric circular Wilson loop, as mentioned in the main text.

\paragraph{The C case.} In the $\cQ^C$ cohomology, we only find a scalar operator
\begin{equation}
\phi^C(x_3, x_4) = \Phi^3 -i\Phi^1 \frac{4\ell^2-x_3^2 - x_4^2}{4\ell^2 + x_3^2 + x_4^2} - \Phi^2 \frac{4i\ell x_3}{4\ell^2 + x_3^2 + x_4^2} - \Phi^5 \frac{4i\ell x_4}{4\ell^2 +  x_3^2 + x_4^2}\Bigg|_{x_1=x_2=0},
\end{equation}
and a similar circular Wilson loop linking the $S^2$, which is constructed from the gauge field
\begin{equation}
\gA^C_\tau = x_1 A_2-x_2 A_1 - x_3 \Phi^2 - x_4\Phi^5 - \frac{1}{4\ell}(4\ell^2 -x_1^2-x_2^2-x_3^2-x_4^2)\Phi^1 - \frac{i}{4\ell}(4\ell^2 + x_1^2 + x_2^2 + x_3^2 + x_4^2)\Phi^3.
\end{equation}
As mentioned in the main text, S-duality implies that there must also exist magnetic observables.

\newpage

\bibliographystyle{JHEP}
\bibliography{traces}

\end{document}